\documentstyle[aps,epsf,floats,eqsecnum]{revtex} 
\begin{document}
\begin{center}
{\bf {\Large Computer Simulations of Friction, Lubrication and Wear}}\\
(To appear in the Handbook of Modern Tribology edited by Bharat Bhushan
(CRC Press))
\end{center}
\begin{tabbing}
1234567890123456789\=1234567890123456789012345678901234\=6789012345\=7890 \kill
\>Mark O. Robbins		\>  Martin H. M\"user\\
\>Dept. of Physics and Astronomy	\>
Institut f\"ur Physik, WA 311\\                          
\>The Johns Hopkins University \> Johannes Gutenberg-Universit\"at\\
\>3400 N. Charles St. \> 55 099 Mainz \\
\>Baltimore, MD 21218 \> GERMANY \\
\>USA \\
\end{tabbing}

\section{Introduction}

Computer simulations have played an important role in understanding
tribological processes.  They allow
controlled numerical "experiments" where the geometry, sliding conditions
and interactions between atoms can be varied at will to explore their effect
on friction, lubrication, and wear.
Unlike laboratory experiments, computer simulations enable scientists to follow
and analyze the full dynamics of all atoms.
Moreover, theorists have no other general approach to analyze processes
like friction and wear. There is no known principle 
like minimization of free energy that determines the steady state of 
non-equilibrium systems.
Even if there was, simulations would be needed to address
the complex systems of interest, just as 
in many equilibrium problems.

Tremendous advances in computing hardware and methodology have dramatically 
increased the ability of theorists to simulate tribological processes.
This has led to an explosion in the number of computational studies over
the last decade, and allowed increasingly sophisticated modeling of
sliding contacts.
Although it is not yet possible to treat all the length scales
and time scales that enter the friction coefficient of engineering
materials, computer simulations have revealed
a great deal of information about the 
microscopic origins of static and kinetic friction,
the behavior of boundary lubricants, and the 
interplay between molecular geometry and tribological properties.
These results provide valuable input to more traditional
macroscopic calculations. Given the rapid pace of developments,
simulations can be expected to play an expanding role in tribology.

In the following chapter we present an overview of the major results from the 
growing simulation literature.  The emphasis is on providing a coherent
picture of the field, rather than a historical review.
We also outline opportunities 
for improved simulations, and highlight unanswered questions.

We begin by presenting a brief overview of simulation techniques and focus on 
special features of simulations for tribological processes.  For example,
it is well known that the results of tribological experiments can be strongly 
influenced by the mechanical properties of the entire system that produces 
sliding.  In much the same way, the results from simulations depend on how 
relative motion of the surfaces is imposed, and how heat generated by sliding
is removed.  The different techniques that are used are described, so that
their influence on results can be understood in later sections.

The complexities of realistic three-dimensional systems can make it difficult
to analyze the molecular mechanisms that underly friction.
The third section focuses
on dry, wearless friction in less complex systems.  The discussion begins with 
simple one-dimensional models of friction between crystalline surfaces.  These 
models illustrate general results for the origin and trends of static and 
kinetic friction, such as the importance of metastability and the effect of 
commensurability.
Then two-dimensional studies are described, with an emphasis 
on the connection to atomic force microscope experiments and detailed 
measurements of the friction on adsorbed monolayers.

In the fourth section, simulations of the dry sliding of crystalline surfaces 
are addressed.  Studies of metal/metal interfaces, surfactant coated surfaces, 
and diamond interfaces with various terminations are described.  The results
can be understood from the simple pictures of the previous chapter.
However, the extra complexity of the interactions in these systems leads to a
richer variety of processes.
Simple examples of wear between metal surfaces are also discussed.

The fifth section describes how the behavior of lubricated systems begins to 
deviate from bulk hydrodynamics as the thickness of the lubricant decreases to 
molecular scales.  Deviations from the usual no-slip boundary condition are 
found in relatively thick films.  These are described, and correlated to 
structure induced in the lubricant by the adjoining walls.  As the film 
thickness decreases, the effective viscosity increases rapidly above the bulk 
value.  Films that are only one or two molecules thick typically exhibit solid 
behavior.  The origins of this liquid/solid transition are discussed, and the 
possibility that thin layers of adventitious carbon are responsible for the 
prevalence of static friction is explored.  The section concludes with 
descriptions of simulations using realistic models of hydrocarbon boundary 
lubricants between smooth and corrugated sufaces.

The sixth section describes work on the common phenomenon of stick-slip
motion, and microscopic models for its origins.
Atomic-scale ratcheting is contrasted 
with long-range slip events, and the structural changes that accompany
stick-slip transitions in simulations are described.

The seventh and final section describes work on friction at extreme conditions 
such as high shear rates or large contact pressures.  Simulations of 
tribochemical reactions, machining, and the evolution of microstructure in 
sliding contacts are discussed.
\section{Atomistic Computer Simulations}

The simulations described in this chapter all use an approach
called classical molecular dynamics (MD) that is described extensively
in a number of review articles and books,
including Allen and Tildesley (1987) and Frenkel and Smit (1996).
The basic outline of the method is straightforward.
One begins by defining the interaction potentials.
These produce forces on the individual particles whose dynamics will be
followed, typically atoms or molecules.
Next the geometry and boundary conditions are specified, and initial
coordinates and velocities are given to each particle.
Then the equations of motion for the particles are integrated
numerically, stepping forward in time by discrete steps of size $\Delta t$.
Quantities such as forces, velocities, work, heat flow,
and correlation functions are calculated as a function of time to
determine their steady-state values and dynamic fluctuations.
The relation between changes in these quantities and the motion of
individual molecules is also explored.

When designing a simulation, care must be taken to choose
interaction potentials and ensembles that capture the essential
physics that is to be addressed. 
The potentials may be as simple as ideal spring constants for
studies of general phenomena, or as complex as
electronic density-functional calculations 
in quantitative simulations.
The ensemble can also be tailored to the problem of interest.
Solving Newton's equations yields a constant energy and volume,
or microcanonical, ensemble.
By adding terms in the equation of motion that simulate heat baths or
pistons,
simulations can be done at constant temperature, pressure, lateral force,
or velocity.
Constant chemical potential can also be maintained by adding or removing
particles using Monte Carlo methods or explicit particle baths.

The amount of computational effort typically grows linearly with both
the number of particles, $N$, and the number of time-steps $M$.
The prefactor increases rapidly with the complexity of the interactions,
and substantial ingenuity is required to achieve linear scaling with $N$ for
long-range interactions or density-functional approaches.
Complex interactions also lead to a wide range of characteristic frequencies,
such as fast bond-stretching and slow bond-bending modes.
Unfortunately, the time step $\Delta t$ must be small ($\sim$ 2\%) compared to
the period of the fastest mode.
This means that many time steps are needed before one obtains information
about the slow modes.

The maximum feasible simulation size has increased continuously
with advances in computer technology, but remains relatively limited.
The product of $N$ times $M$ in the largest simulations described
below is about $10^{12}$.
A cubic region of $10^6$ atoms would have a side of about 50nm.
Such linear dimensions allow reasonable models of an atomic force
microscope tip, the boundary lubricant in a surface force apparatus,
or an individual asperity contact on a rough surface.
However $10^6$ time steps is only about 10 nanoseconds,
which is much smaller than experimental measurement times.
This requires intelligent choices in the problems that are attacked,
and how results are extrapolated to experiment.
It also limits sliding velocities to relatively high values,
typically meters per second or above.

A number of efforts are underway to increase the accessible time scale,
but the problem remains unsolved.
Current algorithms attempt to follow the deterministic equations of motion,
usually with
the Verlet or predictor-corrector algorithms (Allen and Tildesley, 1987).
One alternative approach is to make stochastic steps.
This would be a non-equilibrium generalization of the Monte Carlo
approach that is commonly used in equilibrium systems.
The difficulty is that there is no general principle for determining
the appropriate probability distribution of steps in a
non-equilibrium system.

In the following we describe some of the potentials that are commonly
used, and the situations where they are appropriate.
The final two subsections describe methods
for maintaining constant temperature and constant load.

\subsection{Model Potentials}
\label{sec:mod_pot}

A wide range of potentials has been employed in studies of tribology.
Many of the studies described in the next section use simple ideal
springs and sine-wave potentials.
The Lennard-Jones potential gives a more realistic representation
of typical inter-atomic interactions,
and is also commonly used in studies of general behavior.
In order to model specific materials, more detail must be built into
the potential.
Simulations of metals frequently use the embedded atom method,
while studies of hydrocarbons use potentials that include bond-stretching,
bending, torsional forces and even chemical reactivity.
In this section we give a brief definition of the most commonly used models.
The reader may consult the original literature for more detail.

The Lennard-Jones (LJ) potential is a two-body potential that is commonly used
for interactions between atoms or molecules with closed electron shells.
It is applied not only to the interaction between noble gases,
but also to the interaction between different segments on polymers.
In the latter case, one LJ particle
may reflect a single atom on the chain (explicit atom model),
a CH$_2$ segment (united atom model) or even a Kuhn's
segment consisting of several CH$_2$ units (coarse-grained model).
United atom models (Ryckaert and Bellemans, 1978)
have been shown by Paul et al. (1995)
to successfully reproduce explicit atom results for polymer melts, while
Tsch\"op et al. (1998a, 1998b)
have successfully mapped chemically detailed models of polymers
onto coarse-grained models and back.

The 12-6 LJ potential has the form
\begin{equation}
\label{eq:len_jones}
U(r_{ij}) = 4 \epsilon  
\left[ \left( \sigma \over r_{ij} \right)^{12} - 
\left( \sigma \over r_{ij} \right)^6 \right]
\end{equation}
where $r_{ij}$ is the distance between particles $i$ and $j$,
$\epsilon$ is the LJ interaction energy,
and $\sigma$ is the LJ interaction radius.
The exponents 12 and 6 used above are very common,
but depending on the system, other values may be chosen.
Many of the simulation results discussed in subsequent sections
are expressed in units derived from $\epsilon$, $\sigma$,
and a characteristic mass of the particles.
For example, the standard LJ time unit is defined as 
$t_{\rm LJ} = \sqrt{m \sigma^2/\epsilon}$,
and would typically correspond to a few picoseconds.
A convenient time step 
is $\Delta t = 0.005 t_{\rm LJ}$ for a LJ liquid or solid at
external pressures and temperatures that are not too large.

Most realistic atomic potentials can not be expressed as two-body
potentials.
For example, bond angle potentials in a polymer are effectively
three-body interactions and torsional potentials correspond to four-body
interactions.
Classical models of these interactions (Flory, 1988; Binder, 1995)
assume that a polymer is chemically inert and interactions between different
molecules are modeled by two-body potentials.
In the most sophisticated models, bond-stretching, bending and torsional
energies depend on the position of neighboring molecules
and bonds are allowed to rearrange (Brenner, 1990).
Such generalized model potentials are needed to model friction-induced
chemical interactions.

For the interaction between metals, a different approach has proven fruitful.
The embedded atom method (EAM), introduced
by Daw and Baskes (1984),
includes a contribution in the potential energy associated with the
cost of "embedding" an atom in the 
local electron density $\rho_i$ produced by surrounding atoms.
The total potential energy $U$ is approximated by
\begin{equation}
\label{eq:eam}
U = \sum_i \tilde{F}_i(\rho_i) +
\sum_i \sum_{j<i} \phi_{ij}(r_{ij}).
\end{equation}
where $\tilde{F}_i$ is the embedding energy,
whose functional form depends on the particular metal.
The pair potential
$\phi_{ij}(r_{ij})$
is a doubly-screened short-range potential reflecting core-core repulsion.
The computational cost of the EAM is not substantially greater than
pair potential calculations because the density $\rho_i$ is approximated
by a sum of independent atomic densities.
When compared to simple two-body potentials such as Lennard-Jones
or Morse potentials, the EAM has been particularly successful in
reproducing experimental vacancy formation energies and surface energies,
even though the potential parameters were only adjusted
to reproduce bulk properties.
This feature makes the EAM an important tool in tribological applications,
where surfaces and defects play a major role.

\subsection{Maintaining Constant Temperature}
\label{sec:thermo}

An important issue for tribological simulations is temperature regulation.
The work done when two walls slide past each other is
ultimately converted into random thermal motion.
The temperature of the system would increase indefinitely
if there was no way for this heat to flow out of the system.
In an experiment, heat flows away from the
sliding interface into the surrounding solid.
In simulations, the effect of the surrounding solid must be
mimicked by coupling the particles to a heat bath.

Techniques for maintaining constant temperature $T$ in equilibrium systems
are well-developed.
Equipartition guarantees that the average kinetic energy of each particle
along each Cartesian coordinate is $k_B T/2$ where $k_B$ is Boltzmann's
constant.\footnote{ This assumes that $T$ is above the Debye temperature
so that quantum statistics are not important.  The applicability of
classical MD decreases at lower $T$.}
To thermostat the system,
the equations of motion are modified so that the average kinetic energy
stays at this equilibrium value.

One class of approaches removes or adds kinetic energy to the system
by multiplying the velocities of all particles by the same global factor.
In the simplest version, velocity rescaling,
the factor is chosen to keep the kinetic energy exactly constant
at each time step.
However, in a true constant temperature ensemble there would be
fluctuations in the kinetic energy.
Improvements, such as the Berendsen
and Nos\'e-Hoover methods
(Nos\'e, 1991)
add equations of motion that gradually scale the velocities to
maintain the correct average kinetic energy over a longer time scale.

Another approach is to couple each atom to its own local thermostat
(Schneider and Stoll, 1978; Grest and Kremer, 1986). 
The exchange of energy with the outside world is modeled by a
Langevin equation that includes a
damping coefficient $\gamma$ and a random force
${\vec f}_{i}(t)$ on each atom $i$.
The equations of motion for the $\alpha$ component of the position
$x_{i\alpha}$ become:
\begin{equation}
m_i {{d^2 x_{i\alpha}}\over {dt^2}} =
- {\partial \over {\partial x_{i\alpha}}} U
- m_i \gamma {{dx_{i\alpha}} \over dt} + f_{i\alpha}(t) ,
\label{eq:langevin}
\end{equation}
where $U$ is the total potential energy and $m_i$ is the mass of the atom.
To produce the appropriate temperature,
the forces must be completely random, have zero mean, and
have a second moment given by
\begin{equation}
\label{eq:sec_mom_rand}
\langle \delta f_{i\alpha}(t)^2 \rangle =
2 k_{\rm B} T m_i \gamma / \Delta t.
\end{equation}
The damping coefficient $\gamma$ must be large enough that energy
can be removed from the atoms without substantial temperature increases.
However, it should be small enough that the trajectories
of individual particles are not perturbed too strongly.

The first issue in non-equilibrium simulations is what temperature means.
Near equilibrium, hydrodynamic theories define a local temperature
in terms of the equilibrium equipartition formula
and the kinetic energy relative to the local rest frame
(Sarman et al., 1998).
In $d$ dimensions, the definition is
\begin{equation}
k_B T = {1\over {dN}} \sum_i m_i\left[ \frac{d{\vec x}_i}{dt} -
\langle{\vec v}({\vec x})\rangle\right]^2,
\end{equation}
where the sum is over all $N$ particles and
$\langle{\vec v}({\vec x})\rangle$
is the mean velocity in a region around $\vec x$.
As long as the change in mean velocity is sufficiently slow,
$\langle{\vec v}({\vec x})\rangle$ is well-defined, and this definition of
temperature is on solid theoretical ground.

When the mean velocity difference between neighboring molecules 
becomes comparable to the random thermal velocities,
temperature is not well-defined.
An important consequence is that different strategies for defining
and controlling temperature give very different structural order and
friction forces (Evans and Morriss, 1986; Loose and Ciccotti, 1992;
Stevens and Robbins, 1993).
In addition, the distribution of velocities may become non-Gaussian,
and different directions $\alpha$ may have different effective
temperatures.
Care should be taken in drawing conclusions from simulations in
this extreme regime.
Fortunately, the above condition typically implies that the velocities
of neighboring atoms differ by something approaching 10\% of the speed
of sound.
This is generally higher than any experimental velocity, and would
certainly lead to heat buildup and melting at the interface.

In order to mimic experiments, the thermostat is often applied only
to those atoms that are at the outer boundary of the simulation
cell.  This models the flow of heat into surrounding material that
is not included explicitly in the simulation.
The resulting temperature profile is peaked
at the shearing interface
(e.g. Bowden and Tabor, 1986; Khare et al., 1996).
In some cases the temperature rise may lead to undesirable changes
in the structure and dynamics even at the 
lowest velocity that can be treated in the available simulation time.
In this case, a weak thermostat applied throughout the system may
maintain the correct temperature and yield the dynamics that would
be observed in longer simulations at lower velocities.
The safest course is to couple the thermostat only to those
velocity components that are perpendicular to the mean flow.
This issue is discussed further in Sec. \ref{sec:monolayer}.

There may be a marginal advantage to local Langevin methods in
non-equilibrium simulations because they remove heat only from
atoms that are too hot.
Global methods like Nos\'e-Hoover remove heat everywhere.  
This can leave high temperatures in the region where heat is
generated, while other regions are at an artificially low temperature.

\subsection{Imposing Load and Shear}

The magnitude of the friction that is measured in an experiment
or simulation may be strongly influenced by the way in which the normal
load and tangential motion are imposed (Rabinowicz, 1965).
Experiments almost always impose a constant normal load.
The mechanical system applying shear can usually be described as
a spring attached to a stage moving at controlled velocity.
The effective spring constant includes the compliance of all elements
of the loading device, including the material on either side of
the interface.
Very compliant springs apply a nearly constant force, while
very stiff springs apply a nearly constant velocity.

In simulations, it is easiest to treat the boundary of the system
as rigid, and to translate atoms in this region at a constant
height and tangential velocity.
However this does not allow atoms to move around (Sec. \ref{sec:2dAFM})
or up and over atoms on the opposing surface.
Even the atomic-scale roughness of a crystalline surface
can lead to order of magnitude variations in normal and tangential force 
with lateral position when sliding is imposed in this way
(Harrison et al., 1992b, 1993; Robbins and Baljon, 2000).
The difference between constant separation and pressure simulations
of thin films can be arbitrarily large, since they produce different
power law relations between viscosity and sliding velocity
(Sec. \ref{sec:viscosity}).

One way of minimizing the difference between constant separation and
pressure ensembles is to include a large portion of
the elastic solids that bound the interface.
This extra compliance allows the surfaces to slide at a more uniform
normal and lateral force.
However, the extra atoms greatly increase the computational effort.

To simulate the usual experimental ensemble more directly,
one can add equations of motion that describe the position of the boundary
(Thompson et al., 1990b, 1992, 1995),
much as equations are added to maintain constant temperature.
The boundary is given an effective mass that moves in response
to the net force from interactions with mobile
atoms and from the external pressure and lateral spring.
The mass of the wall should be large enough that its dynamics
are slower than those of individual atoms, but not too much slower,
or the response time will be long compared to the simulation time.
The spring constant should also be chosen to produce an appropriate
response time and ensemble.
\section{Wearless Friction in Low Dimensional Systems}
\subsection{Two Simple Models of Crystalline Surfaces in Direct Contact}
\label{sec:crystal}

Static and kinetic friction involve different
aspects of the interaction between surfaces.
The existence of static friction implies that the surfaces become
trapped in a local potential energy minimum.
When the bottom surface is held fixed, and
an external force is applied to the top surface, the system moves away from
the minimum until the derivative of the potential energy balances
the external force.
The static friction $F_s$ is the maximum force the
potential can resist, i.e. the maximum slope of the potential.
When this force is exceeded, the system begins to slide, and kinetic
friction comes in to play.
The kinetic friction $F_k(v)$ is the force required to
maintain sliding at a given velocity $v$.
The rate at which work is done on the system is $v \cdot F_k(v)$ and
this work must be dissipated as heat that flows away from
the interface.
Thus simulations must focus on the nature of potential energy minima
to probe the origins of static friction,
and must also address dissipation mechanisms
to understand kinetic friction.

\begin{figure}[tb]
\epsfxsize=7cm
\hfil\epsfbox{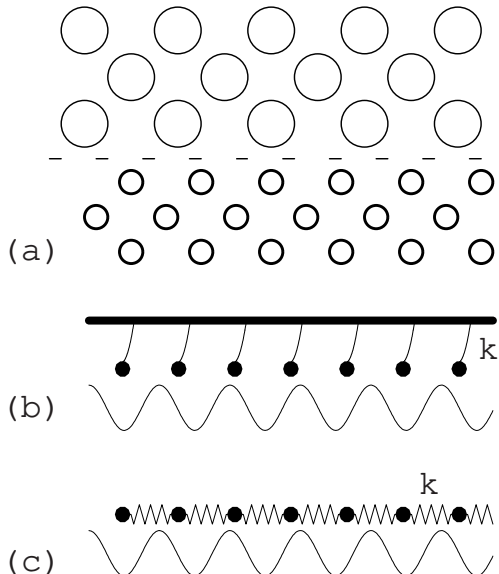}\hfill
\caption{
(a) Two ideal, flat crystals making contact at the plane indicated by
the dashed line.
The nearest-neighbor spacings in the bottom and top walls are $a$ and
$b$ respectively.
The Tomlinson model (b) and Frenkel-Kontorova model (c) replace the bottom
surface by a periodic potential.
The former model keeps elastic forces between atoms on the top surface and
the center of mass of the top wall, and the latter includes springs
of stiffness $k$ between neighbors in the top wall.
(From Robbins, 2000.)
\label{fig:models}
}
\end{figure}

Two simple ball and spring models are useful in illustrating the
origins of static and kinetic friction, and
in understanding the results of detailed simulations.
Both consider two clean, flat, crystalline surfaces in
direct contact
(Fig.~\ref{fig:models}a).
The bottom solid is assumed to be rigid, so that it can be
treated as a fixed periodic substrate potential acting on the top solid.
In order to make the problem analytically tractable, 
only the bottom layer of atoms from the top solid is retained,
and the interactions within the top wall are simplified.
In the Tomlinson model (Fig.~\ref{fig:models}b),
the atoms are coupled to
the center of mass of the top wall by springs of stiffness $k$, and coupling
between neighboring atoms is ignored
(Tomlinson, 1929; McClelland and Cohen, 1990).
In the Frenkel-Kontorova model (Fig.~\ref{fig:models}c),
the atoms are coupled to nearest-neighbors
by springs, and the coupling to the atoms above is ignored
(Frenkel and Kontorova, 1938).
Due to their simplicity, these models arise in a number of
different problems
and a great deal is known about their properties.
McClelland (1989) and 
McClelland and Glosli (1992)
have provided two early discussions of their relevance to friction.
Bak (1982)
has reviewed
the Frenkel-Kontorova model and the physical systems that it models in
different dimensions.

\subsection{Metastability and Static Friction in One Dimension} 

Many features of the Tomlinson and Frenkel-Kontorova models can be understood
from their one-dimensional versions.
One important parameter is the ratio $\eta$ between
the lattice constants of the two surfaces $\eta \equiv b/a$.
The other is the strength of the periodic potential from the
substrate relative to the spring stiffness $k$ that represents
interactions within the top solid.
If the substrate potential has a single Fourier component,
then the periodic force can be written as
\begin{equation}
f(x)= - f_1 \sin\left({2\pi \over a} x\right).
\label{eq:f(x)}
\end{equation}
The relative strength of the potential and springs
can be characterized by the dimensionless
constant $\lambda \equiv 2\pi f_1/ka$.

In the limit of infinitely strong springs ($\lambda \rightarrow 0$),
both models represent rigid solids.  The atoms of the top wall
are confined to lattice sites $x^0_l=x_{\rm CM} + lb$, where the integer $l$
labels successive atoms,
and $x_{\rm CM}$ represents a rigid translation
of the entire wall.
The total lateral or friction force is given by summing Eq.~\ref{eq:f(x)}
\begin{equation}
F=- f_1 \sum_{l=1}^N \sin \left[{{2\pi} \over a} (lb + x_{\rm CM})\right] \ \ ,
\label{eq:Fsum}
\end{equation}
where $N$ is the number of atoms in the top wall.
In the special case of equal lattice constants ($\eta = b/a=1$),
the forces on all atoms add in phase,
and $F=-Nf_1 \sin(2\pi x_{\rm CM}/a)$.
The maximum of this restraining force gives the static friction
$F_s=Nf_1$.

Unless there is a special reason for $b$ and $a$ to be related,
$\eta$ is most likely to be an irrational number.
Such surfaces are called incommensurate, while surfaces with a
rational value of $\eta$ are commensurate.
When $\eta$ is irrational, atoms on the top surface sample all phases
of the periodic force with equal probability
and the net force (Eq. \ref{eq:Fsum}) vanishes exactly.

When $\eta$ is a rational number, it can be expressed as $p/q$ where
$p$ and $q$ are integers with no common factors.
In this case, atoms only sample $q$ different phases.
The net force from Eq.~\ref{eq:Fsum} still vanishes because
the force is a pure sine wave and the phases are equally spaced.
However, the static friction is finite if the potential has higher
harmonics.
A Fourier component with wavevector
$q 2\pi/a$ and magnitude $f_q$ contributes $Nf_q$ to $F_s$.
Studies of surface potentials (Bruch et al., 1997)
show that $f_q$ drops exponentially with increasing $q$
and thus imply that
$F_s$ will only be significant for small $q$.

As the springs become weaker, the top wall is more able to
deform into a configuration that lowers the potential energy.
The Tomlinson model is the simplest case to consider, because
each atom can be treated as an independent oscillator
within the upper surface.
The equations of motion for the position $x_l$ of the $l^{th}$ atom
can be written as
\begin{equation}
m \ddot{x}_l = - \gamma {\dot{x}_l}
 -f_1 \sin\left({{2\pi} \over a}x_l\right) - k(x_l-x_l^0)
\label{eq:tomlinson}
\end{equation}
where $m$ is the atomic mass and $x_l^0$ is the position of the 
lattice site.
Here $\gamma$ is a phenomenological damping coefficient, like that in a
Langevin thermostat (Sec. \ref{sec:thermo}),
that allows work done on the atom to be dissipated as heat.
It represents the coupling to external degrees of freedom such
as lattice vibrations in the solids.

\begin{figure}[tb]
\epsfxsize=8cm
\hfil\epsfbox{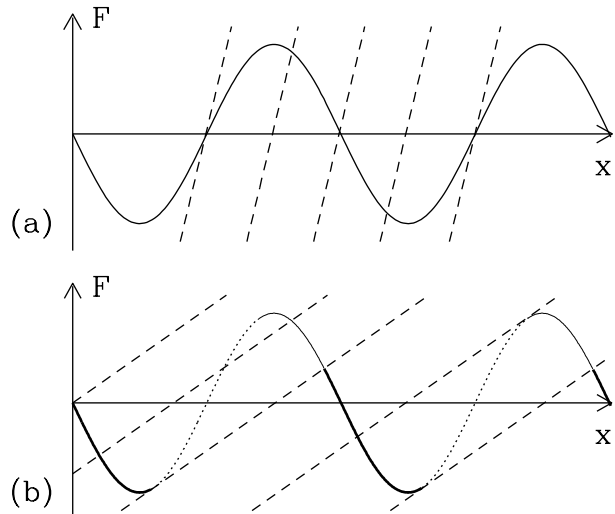}\hfill
\caption{
\label{fig:stab}
Graphical solution for metastability of an atom in the Tomlinson
model for (a) $\lambda=0.5$ and (b) $\lambda=3$.
The straight dashed lines show the force from the spring,
$k (x_l-x_l^0)$, at different $x_l^0$,
and the curved lines show the periodic substrate potential.
For $\lambda < 1$ there is a single intersection of the dashed lines
with the substrate potential for each value of $x_l^0$,
and thus a single metastable state.
For $\lambda >1$ there are multiple intersections with the substrate
potential.
Dotted portions of the potential curve indicate unstable maxima and
solid regions indicate metastable solutions.
As $x_l^0$ increases, an atom that started at $x_l=0$ gradually
moves through the set of metastable states indicated by a thick
solid line.
At the value of $x_l^0$ corresponding to the third dashed line,
this metastable state becomes unstable and the atom jumps
to the metastable state indicated by the continuation of the thick
region of the line.
It then follows this portion of the line until this state becomes
unstable at the $x_l^0$ corresponding to the rightmost dashed line.
}
\end{figure}

In any steady-state of the system,
the total friction can be determined either from the sum
of the forces exerted by the springs on the top wall,
or from the sum of the periodic potentials acting on the atoms
(Eq.~\ref{eq:Fsum}).
If the time average of these forces differed, there would
be a net force on the atoms and a steady acceleration
(Thompson and Robbins, 1990a; Matsukawa and Fukuyama, 1994).
The static friction is related to the force in metastable states
of the system where $\ddot{x_l}=\dot{x_l}=0$.
This requires that spring and substrate forces cancel for each $l$,
\begin{equation}
k(x_l-x_l^0)=-f_1 \sin\left({{2\pi} \over a}x_l\right)  .
\label{eq:stab}
\end{equation}
As shown graphically in Fig.~\ref{fig:stab}a,
there is only one solution for weak interfacial potentials
and stiff solids ($\lambda < 1$).
In this limit, 
the behavior is essentially the same as for infinitely rigid solids.
There is static friction for $\eta=1$, but not for incommensurate
cases.
Even though incommensurate potentials displace
atoms from lattice sites, there are
exactly as many displaced to the right as to the left, and the 
force sums to zero.

A new type of behavior sets in when $\lambda$ exceeds unity.
The interfacial potential is now strong enough
compared to the springs
that multiple metastable states are possible.
These states must satisfy both Eq.~\ref{eq:stab} and the condition
that the second derivative of the potential energy is positive:
$1 + \lambda \cos\left(2\pi x_l/a\right) > 0  $.
The number of metastable solutions increases as $\lambda$
increases. 

As illustrated in Fig.~\ref{fig:stab}b,
once an atom is in a given metastable minimum
it is trapped there until the center of mass moves far enough
away that the second derivative of the potential vanishes
and the minimum becomes unstable.
The atom then pops forward very rapidly to the nearest remaining
metastable state.
This metastability makes it possible to have a finite static
friction even when the surfaces are incommensurate.

If the wall is pulled to the right by an external force,
the atoms will only sample the metastable states corresponding to
the thick solid portion of the substrate potential in Fig.~\ref{fig:stab}b.
Atoms bypass other portions
as they hop to the adjacent metastable state.
The average over the solid portion of the curve is clearly negative
and thus resists the external force.
As $\lambda$ increases, the dashed lines in Fig. \ref{fig:stab}b become
flatter and the solid portion of the curve becomes confined
to more and more negative forces.
This increases the static friction which approaches $N f_1$ in
the limit $\lambda \rightarrow \infty$ (Fisher, 1985).

A similar analysis can be done for the one-dimensional
Frenkel-Kontorova model
(Frank et al., 1949; Bak, 1982; Aubry, 1979, 1983).
The main difference is that the static friction and ground state depend
strongly on $\eta$.
For any given irrational value of $\eta$ there is a threshold
potential strength $\lambda_c$.
For weaker potentials, the static friction vanishes.
For stronger potentials, metastability produces a finite static friction.
The transition to the onset of static friction
was termed a breaking of analyticity by Aubry (1979)
and is often called the Aubry transition.
The metastable states for $\lambda> \lambda_c$
take the form of locally commensurate
regions that are separated by domain walls where the two crystals
are out of phase.
Within the locally commensurate regions the ratio of the periods is a rational
number $p/q$ that is close to $\eta$.
The range of $\eta$ where locking occurs grows with increasing potential
strength ($\lambda$) until it spans all values.
At this point there is an infinite number of different metastable
ground states that form a fascinating
``Devil's staircase'' as $\eta$ varies (Aubry, 1979, 1983; Bak, 1982).

Weiss and Elmer (1996)
have performed a careful study of the 1D Frenkel-Kontorova-Tomlinson
model where both types of springs are included.
Their work illustrates how one can have a finite static friction
at all rational $\eta$ and an Aubry at all irrational $\eta$.
They showed that magnitude of the static friction is a
monotonically increasing function of $\lambda$ and 
decreases with decreasing commensurability.
If $\eta=p/q$ then the static friction rises with corrugation
only as $\lambda^q$. 
Successive approximations to an irrational number involve progressively
larger values of $q$.
Since $\lambda_c <1$, the value of $F_s$  at $\lambda < \lambda_c $
drops closer and closer to zero as the irrational number is approached.
At the same time, the value of $F_s$ rises more and more rapidly with
$\lambda$ above $\lambda_c$.
In the limit $q \rightarrow \infty$ one has the discontinuous rise
from zero to finite values of $F_s$ described by Aubry.
Weiss and Elmer also considered the connection between
the onsets of static friction,
of metastability, and of a finite kinetic friction as $v \rightarrow 0$
that is discussed in the next section.
Their numerical results showed that all these transitions coincide.

Work by Kawaguchi and Matsukawa (1998) shows that varying the strengths
of competing elastic interactions can lead to even more
complex friction transitions.
They considered a model proposed by Matsukawa and Fukuyama (1994) that
is similar to the one-dimensional
Frenkel-Kontorova-Tomlinson model.
For some parameters the static friction oscillated several
times between zero and finite
values as the interaction between surfaces increased.
Clearly the transitions from finite to vanishing static friction
continue to pose a rich mathematical challenge.

\subsection{Metastability and Kinetic Friction}

The metastability that produces static friction in these simple
models is also important in determining the kinetic friction.
The kinetic friction between two solids is usually fairly
constant at low center of mass velocity differences $v_{\rm CM}$.
This means that the same amount of work must be done to 
advance by a lattice constant no matter how slowly the system moves.
If the motion were adiabatic, this irreversible work would vanish as
the displacement was carried out more and more slowly.
Since it does not vanish, some atoms must remain very far from equilibrium
even in the limit $v_{\rm CM} \rightarrow 0$.

The origin of this non-adiabaticity is most easily illustrated
with the Tomlinson model.
In the low velocity limit, atoms stay near to the metastable
solutions shown in Fig.~\ref{fig:stab}.
For $\lambda < 1$ there is a unique metastable solution that
evolves continuously.
The atoms can move adiabatically, and the kinetic friction vanishes
as $v_{\rm CM} \rightarrow 0$.
For $\lambda > 1$ each atom is trapped in a
metastable state.
As the wall moves, this state becomes unstable and the
atom pops rapidly to the next metastable state.
During this motion the atom experiences very large forces
and accelerates to a peak velocity $v_{\rm p}$ that is
independent of $v_{\rm CM}$.
The value of $v_{\rm p}$
is typically comparable to the sound and thermal
velocities in the solid and thus can not be treated as a
small perturbation from equilibrium.
Periodic pops of this type are seen in many of the realistic
simulations described in Sec. \ref{sec:chap4}.
They are frequently referred
to as atomic-scale stick-slip motion (Secs. \ref{bar_com_che}
and \ref{sec:stick}),
because of the oscillation
between slow and rapid motion (Sec. \ref{sec:stick}).

\begin{figure}[tb]
\epsfxsize=10cm
\hfil\epsfbox{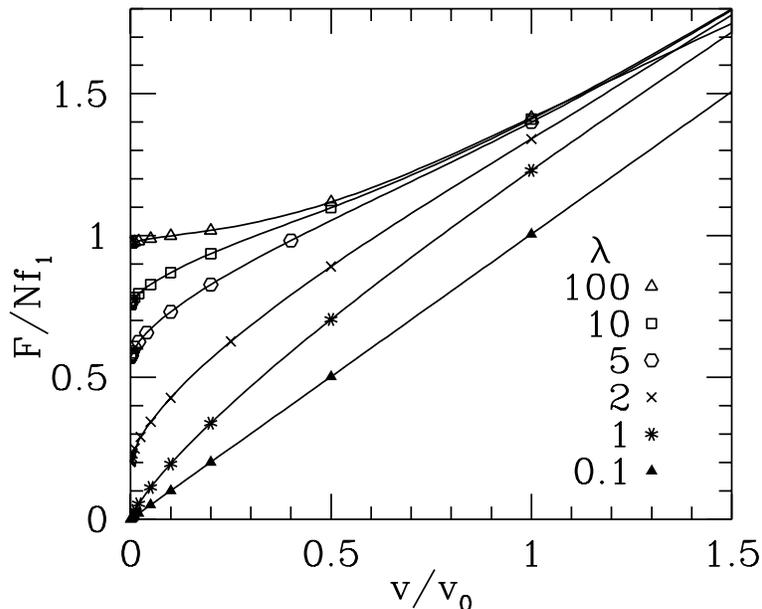}\hfill
\caption{
Force vs. velocity for the Tomlinson model at the indicated
values of $\lambda$.
The force is normalized by the static friction $N f_1 $ and the
velocity is normalized by $v_0 \equiv f_1/\gamma$ where
$\gamma$ is the phenomenological damping rate.
(Data from Joanny and Robbins, 1990).
\label{fig:fvsvTom}
}
\end{figure}

The dynamic equation of motion for the Tomlinson model
(Eq.~\ref{eq:tomlinson})
has been solved in several different contexts.
It is mathematically identical
to simple models for Josephson junctions
(McCumber, 1968),
to the single-particle model of charge-density wave depinning
(Gr\"uner et al., 1981),
and to the equations of motion for a contact line on a periodic
surface
(Raphael and deGennes, 1989; Joanny and Robbins, 1990).
Fig.~\ref{fig:fvsvTom} shows the time-averaged force as a function of
wall velocity for several values of the interface potential strength
in the overdamped limit.
(Since each atom acts as an independent oscillator, these curves
are independent of $\eta$.)
When the potential is much weaker than the springs ($\lambda < 1$),
the atoms can not deviate significantly from their equilibrium
positions.
They go up and down over the periodic potential at
constant velocity in an adiabatic manner.
In the limit $v_{\rm CM} \rightarrow 0$ the periodic potential
is sampled uniformly and the kinetic friction vanishes, just
as the static friction did for incommensurate walls.
At finite velocity the
kinetic friction is just due to the drag force on each atom
and rises linearly with velocity.
The same result holds for all spring constants in the Frenkel-Kontorova
model with equal lattice constants ($\eta=1$).

As the potential becomes stronger, the periodic force begins to
contribute to the kinetic friction of the Tomlinson model.
There is a transition at $\lambda = 1$, and at larger $\lambda$
the kinetic friction remains finite in the limit of zero velocity.
The limiting $F_k(v=0)$
is exactly equal to the static friction for
incommensurate walls.
The reason is that as $v_{\rm CM} \rightarrow 0$ atoms spend 
almost all of their time in metastable states.
During slow sliding,
each atom samples all the metastable states that contribute to
the static friction and with exactly the same
weighting.

The solution for commensurate walls has two different
features.
The first is that the static friction is higher than
$F_k(0)$.
This difference is greatest for the case $\lambda < 1$
where the kinetic friction vanishes, while the static
friction is finite.
The second difference is that the force/velocity curve
depends on whether the simulation is done at constant
wall velocity (Fig.~\ref{fig:fvsvTom}) or constant
force.
The constant force solution is independent of $\lambda$
and equals the constant velocity solution in the limit
$\lambda \rightarrow \infty$.

The only mechanism of dissipation in the Tomlinson
model is through the phenomenological
damping force, which is proportional to the velocity of the atom.
The velocity is essentially zero except in the rapid pops that
occur as a state becomes unstable and the atom pops to the next
metastable state.
In the overdamped limit, atoms pop with peak velocity
$v_p \sim f_1/\gamma$ --
independent of the average velocity of the center of mass.
Moreover, the time of the pop is nearly independent of $v_{\rm CM}$,
and so the total energy dissipated per pop is independent of $v_{\rm CM}$.
This dissipated energy is of course consistent
with the limiting force determined from arguments based on the
sampling of metastable states given above
(Fisher, 1985; Raphael and DeGennes, 1989; Joanny and Robbins, 1990).
The basic idea that kinetic friction is due to dissipation during
pops that remain rapid as $v_{\rm CM} \rightarrow 0$ is very general,
although the phenomenological damping used in the model is far from 
realistic.
A constant dissipation during each displacement by a lattice constant
immediately implies a velocity independent $F_k$, and vice versa.

\subsection{Tomlinson Model in Two-Dimensions: Atomic Force Microscopy}
\label{sec:2dAFM}

Gyalog et al. (1995)
have studied a generalization of
the Tomlinson model where the atoms can move in two dimensions
over a substrate potential.
Their goal was to model the motion of an atomic-force microscope
(AFM) tip over a surface.
In this case the spring constant $k$ reflects the elasticity of the
cantilever, the tip, and the substrate.
It will in general be different along the scanning direction than
along the perpendicular direction.

The extra degree of freedom provided by the second dimension means
that the tip will not follow the nominal scanning direction, but
will be deflected to areas of lower potential energy.
This distorts the image and also lowers the measured friction force.
The magnitude of both effects decreases with increasing stiffness.

As in the one-dimensional model there is a transition from smooth
sliding to rapid jumps with decreasing spring stiffness.
However, the transition point now depends on sliding direction and
on the position of the scan line along the direction normal to the
nominal scan direction.
Rapid jumps tend to occur first near the peaks of the potential,
and extend over greater distances as the springs soften.
The curves defining the unstable points can have very complex,
anisotropic shapes.

H\"olscher et al. (1997)
have used a similar model to simulate scans of MoS$_2$.
Their model also includes kinetic and damping terms in order
to treat the velocity dependence of the AFM image.
They find marked anisotropy in the friction as a function of sliding
direction, and also discuss deviations from the nominal scan direction
as a function of the position and direction of the scan line. 

Rajasekaran et al. (1997)
considered a simple elastic solid of varying stiffness that
interacted with a single atom at the end of an AFM tip with
Lennard-Jones potentials.
Unlike the other calculations mentioned above, this paper explicitly includes
variations in the height of the atom and maintains a constant
normal load.
The friction rises linearly with load in all cases,
but the slope depends strongly on sliding direction, scan
position and the elasticity of the solid.

The above papers and related work show the complexities that can
enter from treating detailed surface potentials and the full
elasticity of the materials and machines that drive sliding.
All of these factors can influence the measured friction and
must be included in a detailed model of any experiment.
However, the basic concepts derived from 1D models carry forward.
In particular, 
1) static friction results when there is sufficient
compliance to produce multiple metastable states,
and
2) a finite $F_k(0)$ arises when 
energy is dissipated during rapid pops between metastable states.

All of the above work considers a single atom or tip in a
two-dimensional potential.
However, the results can be superimposed to
treat a pair of two-dimensional surfaces in contact,
because the oscillators are independent in the Tomlinson model.
One example of such a system is the work by Glosli and McClelland (1993)
that is described in Sec. \ref{bar_com_che}.
Generalizing the Frenkel-Kontorova model to two dimensions is
more difficult.

\subsection{Frenkel-Kontorova Model in Two Dimensions: Adsorbed Monolayers}
\label{sec:monolayer}
 
The two-dimensional Frenkel-Kontorova model provides a simple model
of a crystalline layer of adsorbed atoms (Bak, 1982).
However, the behavior of adsorbed layers can be much richer because
atoms are not connected by fixed springs, and thus can rearrange to form
new structures in response to changes in equilibrium conditions
(i.e. temperature) or due to sliding.
Overviews of the factors that determine the wide variety
of equilibrium structures, including fluid, incommensurate and commensurate
crystals, can be found in
Bruch et al., (1997) and Taub et al. (1991).
As in one-dimension, both the structure and the strength of
the lateral variation or ``corrugation'' in the substrate potential 
are important in determining the friction.
Variations in potential normal to the substrate are
relatively unimportant
(Persson and Nitzan, 1996; Smith et al., 1996).

Most simulations of the friction between adsorbed layers and substrates
have been motivated by the pioneering Quartz Crystal Microbalance (QCM)
experiments of Krim et al. (1988, 1990, 1991).
The quartz is coated with metal electrodes that are used
to excite resonant shear oscillations in the crystal.
When atoms adsorb onto the electrodes, the increased mass causes a
decrease in the resonant frequency.
Sliding of the substrate under the adsorbate leads to friction
that broadens the resonance.
By measuring both quantities, the friction per atom can be calculated.
The extreme sharpness of the intrinsic resonance in the crystal
makes this a very sensitive technique.

In most experiments the electrodes were the noble metals Ag or Au.
Deposition produces fcc crystallites with close-packed (111) surfaces.
Scanning tunneling microscope studies show that the surfaces are
perfectly flat and ordered over regions at least 100nm across.
At larger scales there are grain boundaries and other defects.
A variety of molecules have been physisorbed onto these surfaces,
but most of the work has been on noble gases.

The interactions within the noble metals are typically much stronger
than the van der Waals interactions between the adsorbed molecules.
Thus, to a first approximation, the substrate remains unperturbed 
and can be replaced by a periodic potential
(Smith et al., 1996; Persson et al., 1998).
However, the mobility of substrate atoms is important in allowing heat
generated by the sliding adsorbate to flow into the substrate.
This heat transfer into substrate lattice vibrations or phonons
can be modeled by a Langevin thermostat (Eq. \ref{eq:langevin}).
If the surface is metallic, the Langevin damping should also include
the effect of energy dissipated to the
electronic degrees of freedom
(Schaich and Harris, 1981; Persson, 1991; Persson and Volokitin, 1995).

With the above assumptions, the equation of motion for an
adsorbate atom can be written as
\begin{equation}
m \ddot{x}_{\alpha} = - \gamma_\alpha \dot{x}_\alpha
+ F_{\alpha}^{ext}-{\partial \over {\partial x_\alpha}}
U +f_\alpha(t)
\label{eq:adsorb}
\end{equation}
where $m$ is the mass of an adsorbate atom,
$\gamma_\alpha$ is the damping rate from
the Langevin thermostat in the $\alpha$ direction,
$f_\alpha (t)$ is the corresponding
random force, ${\vec F}^{ext}$ is an external force applied to the particles,
and U is the total energy from the interactions of the adsorbate atoms
with the substrate and with each other.

Interactions between noble gas adsorbate atoms
have been studied extensively, and are reasonably well described by
a Lennard-Jones potential (Bruch et al., 1997).
The form of the substrate interaction is less well-known.  However,
if the substrate is crystalline, its potential can be expanded as a Fourier
series in the reciprocal lattice vectors $\vec Q$ of the surface layer
(Bruch et al., 1997).
Steele (1973) has considered Lennard-Jones interactions with substrate
atoms and shown that the higher Fourier components drop off exponentially
with increasing $| \vec Q |$ and height $z$ above the substrate.
Thus most simulations have kept only the shortest wavevectors,
writing:
\begin{equation}
U_{sub}(\vec r, z) = U_0(z) + U_1(z) \sum_l \cos [\vec Q_l \cdot \vec r]
\label{eq:usub}
\end{equation}
where $\vec r$ is the position within the plane of the surface,
and the sum is over symmetrically equivalent $\vec Q$.
For the close-packed (111) surface of fcc crystals there are 6 equivalent
lattice vectors of length $4 \pi/(\sqrt{3}a)$ where $a$ is the nearest
neighbor spacing in the crystal.
For the (100) surface there are 4 equivalent lattice vectors of length
$2\pi/a$.
Cieplak et al. (1994)
and Smith et al. (1996)
used Steele's potential with
an additional 4 shells of symmetrically equivalent wavevectors in
their simulations.
However, they found that their results were almost unchanged when only
the shortest reciprocal lattice vectors were kept.

Typically the Lennard-Jones $\epsilon$, $\sigma$ and $m$ are used to
define the units of energy, length and time, as described in
Sec. \ref{sec:mod_pot}.
The remaining parameters in Eq.~\ref{eq:adsorb} are the damping rates,
external force, and the substrate potential which is characterized 
by the strength of the adsorption potential $U_0(z)$ and the corrugation
potential $U_1(z)$.
The Langevin damping for the two directions in the plane of the
substrate is expected to be the same and will be denoted by $\gamma_\parallel$.
The damping along $z$, $\gamma_\perp$, may be different
(Persson and Nitzan, 1996).
The depth of the minimum in the adsorption potential
can be determined from the energy needed to desorb an atom,
and the width is related to the frequency of vibrations along $z$.
In the cases of interest here, the adsorption energy is much larger 
than the Lennard-Jones interaction or the corrugation.
Atoms in the first adsorbed layer
sit in a narrow range of $z$ near the minimum $z_0$.
If the changes in $U_1$ over this range are small, then the effective
corrugation for the first monolayer is $U_1^0 \equiv U_1(z_0)$.
As discussed below, the calculated friction in most simulations varies
rapidly with $U_1^0$ but is insensitive to 
other details in the substrate potential.

The simplest case is the limit of weak corrugation and a fluid or
incommensurate solid state of the adsorbed layer.
As expected based on results from 1D models, such layers
experience no static friction, and the kinetic friction is proportional
to velocity: $F_k = - \Gamma v$
(Persson, 1993a; Cieplak et al., 1994).
The constant of proportionality $\Gamma$ gives the
"slip-time" $t_s  \equiv m/\Gamma$ that is
reported by Krim and coworkers (1988, 1990, 1991).
This slip time
represents the time for the transfer of momentum
between adsorbate and substrate.
If atoms are set moving with an initial velocity, the velocity will decay
exponentially with time constant $t_s$.
Typical measured values are of order nanoseconds for rare gases.
This is surprisingly large when compared to the picosecond time
scales that characterize momentum transfer in a bulk fluid of the
same rare gas.  (The latter is directly related to the viscosity.)

The value of $t_s$ can be determined from simulations in several
different ways.
All give consistent results in the cases where they have been compared,
and should be accurate if used with care.
Persson (1993a),
Persson and Nitzan (1996),
and Liebsch et al.  (1999)
have calculated the average
velocity as a function of ${\vec F}^{ext}$ and obtained $\Gamma$ from the
slope of this curve.
Cieplak et al. (1994)
and Smith et al. (1996)
used this approach and also mimicked experiments by finding
the response
to oscillations of the substrate.
They showed $t_s$ was constant over a wide range of frequency
and amplitude.
The frequency is difficult to vary in experiment,
but Mak and Krim (1998)
found that $t_s$ was independent
of amplitude in both fluid and crystalline phases of Kr on Au.
Tomassone et al. (1997)
have used two additional techniques
to determine $t_s$.
In both cases they used no thermostat ($\gamma_\alpha=0$).
In the first method all atoms were given an initial velocity and
the exponential decay of the mean velocity was used to determine
$t_s$.
The second method made use of the fluctuation-dissipation theorem,
and calculated $t_s$ from equilibrium velocity fluctuations.

A coherent picture has emerged for the relation between
$t_s$ and the damping and corrugation in Eqs.~\ref{eq:adsorb}
and \ref{eq:usub}.
In the limit where the corrugation vanishes, the substrate potential
is translationally invariant and can not exert any friction on the
adsorbate.
The value of $\Gamma$ is then just equal to $\gamma_\parallel$.
In his original 2D simulations Persson (1993a)
used relatively large values of $\gamma_\parallel$ and reported
that $\Gamma$ was always proportional to $\gamma_\parallel$.
Later work by Persson and Nitzan (1996)
showed that this proportionality
only held for large
$\gamma_\parallel$.
Cieplak et al. (1994),
Smith et al. (1996),
and Tomassone et al. (1997)
considered the opposite
limit, $\gamma_\parallel=0$ and found a nonzero $\Gamma_{ph}$ that reflected
dissipation due to phonon excitations in the adsorbate film.
Smith et al.
(1996)
found that including a Langevin damping
along the direction of sliding produced a simple additive shift in $\Gamma$.
This relation has been confirmed in extensive simulations
by Liebsch et al. (1999).
All of their data can be fit to the relation 
\begin{equation}
\Gamma = \gamma_\parallel + \Gamma_{ph} = \gamma_\parallel + C (U_1^0)^2
\label{eq:Gamma}
\end{equation}
where the constant $C$ depends on temperature, coverage, and other factors.

Cieplak et al. (1994)
and Smith et al. (1996)
had previously shown that the damping increased quadratically
with corrugation and
developed a simple perturbation theory for the prefactor $C$
in Eq. \ref{eq:Gamma}.
Their approach follows that of Sneddon et al. (1982)
for charge-density waves, and of Sokoloff (1990)
for friction between two semi-infinite incommensurate solids.
It provides the simplest illustration of how dissipation occurs
in the absence of metastability, and is directly relevant to studies
of flow boundary conditions discussed in Sec. \ref{sec:flowbc}. 

The basic idea is that the adsorbate monolayer acts like an elastic sheet.
The atoms are attracted to regions of low corrugation potential
and repelled from regions of high potential.
This produces density modulations $\rho(\vec Q_l)$ in the adsorbed layer
with wavevector $\vec Q_l$.
When the substrate moves underneath the adsorbed layer, the
density modulations attempt to follow the substrate potential.
In the process, some of the energy stored in the modulations
leaks out into other phonon modes of the layer due to anharmonicity.
The energy dissipated to these other modes eventually flows into
the substrate as heat.
The rate of energy loss can be calculated to lowest order in
a perturbation theory in the strength of the corrugation if
the layer is fluid or incommensurate.
Equating this to the average energy dissipation rate given by 
the friction relation,
gives an expression for the phonon contribution to dissipation.

The details of the calculation can be found in Smith et al. (1996).
The final result is that the damping rate is proportional to
the energy stored in the density modulations and to the rate
of anharmonic coupling to other phonons.
To lowest order in perturbation theory the energy is proportional
to the square of the density modulation and thus the square of
the corrugation as in Eq.~\ref{eq:Gamma}.
This quantity is experimentally accessible by measuring the
static structure factor $S(\vec Q)$
\begin{equation}
{{S(\vec Q)} \over{ N_{ad}}} \equiv | \rho ( \vec Q ) |^2
\label{eq:Sofk}
\end{equation}
where $N_{ad}$ is the number of adsorbed atoms.
The rate of anharmonic coupling is the inverse of an effective
lifetime for acoustic phonons, $t_{\rm phon}$, that could also be measured in
scattering studies.
One finds 
\begin{equation}
\Gamma_{ph}/m = {{c S(Q)}\over {N_{ad}}} {1 \over {t_{\rm phon}}}
\label{eq:phon}
\end{equation}
where $c$ is half of the number of symmetrically equivalent
$\vec Q_l$.
For an fcc crystal $c = 3$ on the (111) surface and $c=2$ on
the (100) surface.
In both cases the damping is independent of the direction of
sliding, in agreement with simulations by Smith et al. (1996).
Smith et al. performed a quantitative test of Eq.~\ref{eq:phon}
showing that values of $S(Q)$ and $t_{\rm phon}$ from equilibrium
simulations were consistent with non-equilibrium determinations of
$\Gamma_{ph}$.
The results of Liebsch et al. (1999) provide the first comparison
of (111) and (100) surfaces.
Data for the two surfaces collapse on to a single curve when
divided by the values of $c$ given above.
Liebsch et al. (1999) noted that the barrier for motion between local
minima in the substrate potential is much smaller for (111) than
(100) surfaces and thus it might seem surprising that $\Gamma_{ph}$
is 50\% higher on (111) surfaces.
As they state, the fact that the corrugation is weak means that
atoms sample all values of the potential and
the energy barrier plays no special role.

The major controversy between different theoretical groups concerns
the magnitude of the substrate damping $\gamma_\parallel$ that should
be included in fits to experimental systems.
A given value of $\Gamma$ can be obtained with an infinite number
of different combinations of $\gamma_\parallel$ and corrugation
(Robbins and Krim, 1998; Liebsch et al., 1999).
Unfortunately both quantities are difficult to calculate and to measure.

Persson (1991, 1998) has discussed the relation between electronic
contributions to $\gamma_\parallel$ and changes in surface resistivity with
coverage.
The basic idea is that adsorbed atoms exert a drag on electrons that
increases resistivity.
When the adsorbed atoms slide, the same coupling produces a drag on
them.
The relation between the two quantities is somewhat more complicated
in general because of disorder and changes in electron density due
to the adsorbed layer.
In fact adsorbed layers can decrease the resistivity in certain
cases.
However, there is a qualitative agreement between changes in surface
resistivity and the measured friction on adsorbates (Persson, 1998).
Moreover, the observation of a drop in friction at the superconducting
transition of lead substrates is clear evidence that electronic damping
is significant in some systems (Dayo et al., 1998).

There is general agreement that the electron damping is relatively
insensitive to the number of adsorbed atoms per unit area or coverage.
This is supported by experiments that show the variation of surface
resistivity with coverage is small (Dayo and Krim, 1998).
In contrast, the phonon friction varies dramatically with increasing
density (Krim et al., 1988, 1990, 1991).
This makes fits to measured values of friction as a function of coverage
a sensitive test of the relative size of electron and phonon friction.

\begin{figure}[tb]
\epsfxsize=10cm
\hfil\epsfbox{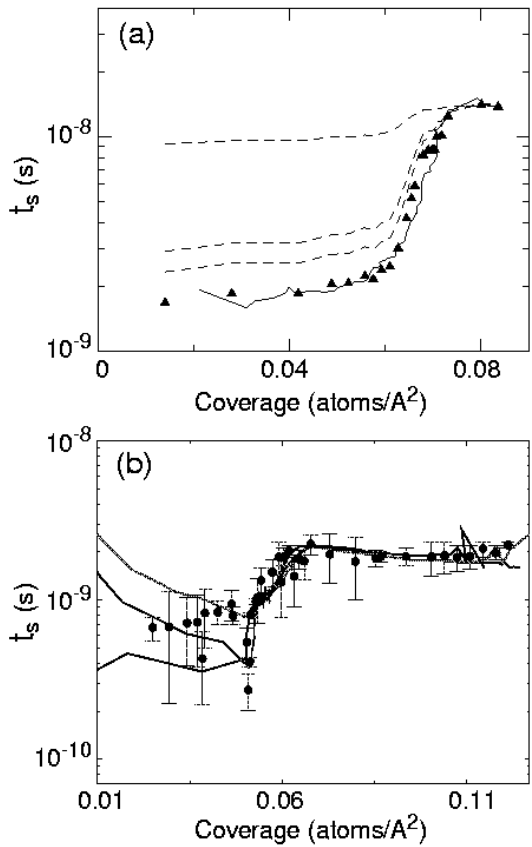}\hfill
\caption{
Slip times vs. coverage for (a) Kr on Au
(Cieplak et al., 1994, Krim et al., 1991)
and (b) Xe on Ag (Tomassone et al., 1997).
Calculated values are indicated by symbols, and experimental results
by solid lines.
Experimental data for three different runs are shown in (b).
The dashed lines in (a) indicate theoretical values obtained
by fitting experiments with 1/3, 1/2, or 9/10 (from bottom to top)
of the friction at high coverage coming from electronic damping.
(From Robbins and Krim, 1998.)
\label{fig:cover}
}
\end{figure}

Two groups have found that calculated values of
$\Gamma$ with $\gamma_\parallel=0$
can reproduce experiment.
Calculations for Kr on Au by Cieplak et al. (1994)
are compared to data from Krim et al. (1991)
in Fig.~\ref{fig:cover} (a).
Fig.~\ref{fig:cover}(b) shows the comparison between fluctuation-dissipation
simulations and experiments for Xe on Ag from Tomassone et al. (1997).
In both cases there is a rapid rise in slip time with increasing coverage
$n_{ad}$.
At liquid nitrogen temperatures
krypton forms islands of uncompressed fluid for $n_{ad} < 0.055$\AA$^{-2}$
and the slip time is relatively constant.
As the coverage increases from 0.055 to 0.068 \AA$^{-2}$, the monolayer
is compressed into an incommensurate crystal.
Further increases in coverage lead to an increasingly dense crystal.
The slip time increases by a factor of seven during  the compression
of the monolayer.

For low coverages, Xe forms solid islands on Ag at T=77.4K.
The slip time drops slightly with increasing coverage, presumably
due to increasing island size (Tomassone et al., 1997).
There is a sharp rise in slip time as the islands merge into a complete
monolayer that is gradually compressed with increasing coverage.
Fig.~\ref{fig:cover} shows that the magnitude of the rise in $t_s$ varies
from one experiment to the next.
The calculated rise is consistent with the larger measured increases.

The simulation results of
the two groups can be extended to nonzero values of $\gamma_\parallel$,
using Eq.~\ref{eq:Gamma}.
This would necessarily change the ratio between the slip times of
the uncompressed and compressed layers.
The situation is illustrated for Kr on Au in Fig.~\ref{fig:cover}(a).
The dashed lines were generated by fitting the damping of the
compressed monolayer with different ratios of
$\gamma_\parallel$ to $\Gamma_{ph}$.
As the importance of $\gamma_\parallel$ increases, the change in slip
time during compression of the monolayer decreases substantially.
The comparison between theory and experiment suggests that $\gamma_\parallel$
is likely to contribute less than 1/3 of the friction in the compressed
monolayer, and thus less than 5\% in the uncompressed fluid.
The measured increase in slip time for Xe on Ag is smaller
and the variability noted in Fig.~\ref{fig:cover}b makes it
harder to place bounds on $\gamma_\parallel$.
Tomassone et al. (1997)
conclude that their results are
consistent with no contribution from $\gamma_\parallel$.
When they included a value of $\gamma_\parallel$ suggested by Persson and
Nitzan (1996)
they still found that phonon friction provided 75\%
of the total.
Persson and Nitzan had concluded that phonons contributed only 2\%
of the friction in the uncompressed monolayer.

Liebsch et al. (1999)
have reached an intermediate conclusion.
They compared calculated results for different corrugations to a set
of experimental data and chose the corrugation that matched the change
in friction with coverage.
They conclude that most of the damping at high coverages
is due to $\gamma_\parallel$ and most of the damping at low coverages
is due to phonons.
However, the data they fitted had only a factor of 3 change with increasing
coverage and some of the data in Fig.~\ref{fig:cover}b change by a factor
of more than 5.
Fitting to these sets would decrease their estimate of the size
of $\gamma_\parallel$.

The behavior of commensurate monolayers is very different than
that of the incommensurate and fluid layers described so far.
As expected from studies of one dimensional models,
simulations of commensurate monolayers show that they
exhibit static friction.
Unfortunately, no experimental results have been obtained
because the friction is too high for the QCM technique to measure.

In one of the earliest simulation studies,
Persson (1993a)
considered a two-dimensional model of Xe on the (100) surface of Ag.
Depending on the corrugation strength he found
fluid, 2x2 commensurate, and incommensurate phases.
He studied $1/\Gamma$ as the commensurate phase was approached
by lowering temperature in the fluid phase, or decreasing coverage
in the incommensurate phase.
In both cases he found that $1/\Gamma$ went to zero at the boundary
of the commensurate phase, implying that there was no flow in
response to small forces.

When the static friction is exceeded, the dynamics of adsorbed layers
can be extremely complicated.
In the model just described, Persson (1993a, 1993b, 1995)
found that sliding caused a transition from a commensurate crystal
to a new phase.
The velocity was zero until the static friction was exceeded.
The system then transformed into a sliding fluid layer.
Further increases in force caused a first order transition
to the incommensurate structure that would be stable in the absence
of any corrugation.
The velocity in this phase was also what would be calculated for
zero corrugation $F = \gamma_\parallel v$ (dashed line).
Decreasing the force led to a transition back to the fluid
phase at essentially the same point.
However, the layer did not return to the initial commensurate phase
until the force dropped well below the static friction.

The above hysteresis in the transition between commensurate and fluid states
is qualitatively similar
to that observed in the underdamped Tomlinson model or the equivalent
case of a Josephson junction (McCumber, 1968).
As in these cases,
the magnitude of the damping effects the range of the hysteresis.
The major difference is the origin of the hysteresis.
In the Tomlinson model, hysteresis arises solely because the
inertia of the moving system allows it to overcome potential barriers
that a static system could not.
This type of hysteresis would disappear at finite temperature
due to thermal excitations (Braun et al., 1997a).
In the adsorbed layers, the change in the physical state of the system
has also changed the nature of the potential barriers.
Similar sliding induced phase transitions were observed
earlier in experimental and simulation studies of shear
in bulk crystals (Ackerson et al., 1986; Stevens et al., 1991, 1993)
and in thin films (Gee et al., 1990; Thompson and Robbins, 1990b).
The relation between such transitions and stick-slip motion
is discussed in Section \ref{sec:stick}.

Braun and collaborators have considered the transition from
static to sliding states at coverages near to a commensurate value.
They studied one
(Braun et al., 1997b; Paliy et al., 1997)
and two
(Braun et al., 1997a, 1997c)
dimensional Frenkel-Kontorova models
with different degrees of damping.
If the corrugation is strong,
the equilibrium state consists of locally commensurate regions
separated by domain walls or kinks.
The kinks are pinned because of the discreteness of the lattice,
but this Peierls-Nabarro pinning potential is smaller than the substrate
corrugation.
In some cases there are different types of kinks with different
pinning forces.
The static friction corresponds to the force needed to initiate motion
of the most weakly pinned kinks.
As a kink moves through a region,
atoms advance between adjacent local minima in the substrate potential.
Thus the average velocity depends on both the kink velocity and the
density of kinks.
If the damping is strong, there may be a series of sudden transitions
as the force increases.
These may reflect depinning of more strongly pinned kinks,
or creation of new kink-antikink pairs that lead to faster and faster
motion.
At high enough velocity the kinks become unstable, and
a moving kink generates a cascade of new kink-antikink pairs
that lead to faster and faster motion.
Eventually the layer decouples from the substrate and there are
no locally commensurate regions.
As in Persson (1995),
the high velocity state looks like
an equilibrium state with zero corrugation.
The reason is that the atoms move over the substrate so quickly
that they can not respond.
Although this limiting behavior is interesting, it would only
occur in experiments between flat crystals
at velocities comparable to the speed of sound.

\section{Dry Sliding of Crystalline Surfaces}
\label{sec:chap4}

The natural case of interest to tribologists is the sliding
interface between two three-dimensional objects.
In this section we consider sliding of bare surfaces.
We first discuss general issues related to the effect of commensurability,
focusing on strongly adhering surfaces such as clean metal surfaces.
Then simulations of chemically-passivated surfaces of practical
interest are described.
The section concludes with studies of
friction, wear and indentation in single-asperity contacts.

\subsection{Effect of Commensurability}

\begin{figure}[tb]
\epsfxsize=12cm
\hfil\epsfbox{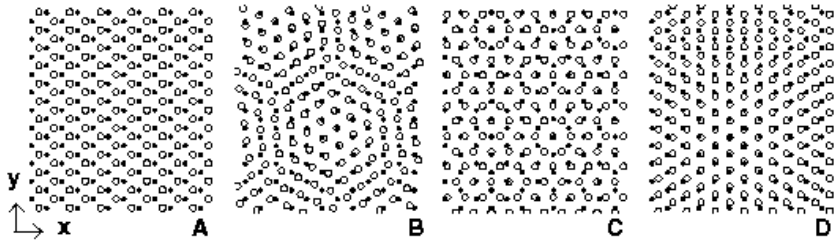}\hfill
\caption{
The effect of crystalline alignment and lattice constant
on commensurability is illustrated
by projecting atoms from the bottom (filled circles) and top (open
circles) surfaces into the plane of the walls.
In A-C the two walls have the same structure and lattice constant but
the top wall has been rotated by 0$^o$, 8.2$^o$ or 90$^o$, respectively.
In D the walls are aligned, but the lattice constant of the top wall
has been reduced by a small amount.
Only case A is commensurate.
The other cases are incommensurate and atoms from the two
walls sample all possible relative positions with equal probability.
(From He et al., 1999.)
\label{fig:walls}
}
\end{figure}

The effect of commensurability in three dimensional systems has
been studied by Hirano and Shinjo (1990, 1993).
They noted that even two identical surfaces are likely to be
incommensurate.
As illustrated in Fig.~\ref{fig:walls},
unless the crystalline surfaces are perfectly aligned, the periods
will no longer match up.
Thus one would expect almost all contacts between surfaces to
be incommensurate.

Hirano and Shinjo (1990)
calculated the condition for static
friction between high symmetry surfaces of fcc and bcc metals.
Many of their results are consistent with the conclusions described
above for lower dimensions.
They first showed that the static friction between incommensurate
surfaces vanishes exactly if the solids are perfectly rigid.
They then allowed the bottom layer of the top surface to relax
in response to the atoms above and below.
The relative strength of the interaction between the two surfaces
and the stiffness of the top surface plays the same role as $\lambda$
in the Tomlinson model.
As the interaction becomes stronger, there is an Aubry transition to a 
finite static friction.
This transition point was related to the condition for
multi-stability of the least stable atom.

To test whether realistic potentials would be strong enough to produce
static friction between incommensurate surfaces,
Hirano and Shinjo (1990) applied their theory to
noble and transition metals.
Contacts between various surface orientations of the same
metal (i.e. (111) and (100) or (110) and (111)) were tested.
In all cases the interactions were too weak to produce static friction.

Shinjo and Hirano (1993)
extended this line of work
to dynamical simulations of sliding.
They first considered the undamped Frenkel-Kontorova model with ideal
springs between atoms.
The top surface was given an initial velocity and the evolution
of the system was followed.
When the corrugation was small,
the kinetic friction vanished, and
the sliding distance increased linearly with time.
This "superlubric" state disappeared above a threshold corrugation.
Sliding stopped because of energy transfer from the
center of mass motion into vibrations within the surface.
The transition point depended on the initial velocity, since that set
the amount of energy that needed to be converted into lattice vibrations.
Note that the {\em kinetic}
friction only vanishes in these simulations because
atoms are connected by ideal harmonic springs
(Smith et al., 1996).
The damping due to energy transfer between
internal vibrations (e.g. Eq.~\ref{eq:phon}) is zero
because the phonon lifetime is infinite.
More realistic anharmonic potentials always lead to
an exponential damping of the velocity at long times.

Simulations for two dimensional surfaces were also described
(Shinjo and Hirano, 1993; Hirano and Shinjo, 1993).
Shinjo and Hirano noted that static friction
is less likely in higher dimensions because of the ability of
atoms to move around maxima in the substrate potential, as
described in Sec. \ref{sec:2dAFM}.
A particularly interesting feature of their results is that
the Aubry transition
to finite static friction depends on the relative orientation of
the surfaces (Hirano and Shinjo, 1993).
Over an intermediate range of corrugations,
the two surfaces slide freely in some alignments and are pinned
in others.
This extreme dependence on relative alignment has not been seen
in experiments,
but strong orientational variations in friction have been
seen between mica surfaces (Hirano et al., 1991)
and between a crystalline AFM tip and substrate
(Hirano et al., 1997).

Hirano and Shinjo's conclusion 
that two flat, strongly adhering 
but incommensurate surfaces are likely to have zero static friction
has been supported by two other studies.
As described in more detail in Section~\ref{bare_sin_asp_con},
S$\o$rensen et al. (1996)
found that there was no static friction between a sufficiently large copper tip
and an incommensurate copper substrate.
M\"user and Robbins (1999)
studied a simple model system and found that interactions within
the surfaces needed to be much smaller than the interactions between
surfaces in order to get static friction.

M\"user and Robbins (1999) considered two identical but orientationally
misaligned triangular surfaces similar to Fig.~\ref{fig:walls}C.
Interactions within each surface were represented by coupling atoms
to ideal lattice sites with a spring constant $k$.
Atoms on opposing walls interacted through a Lennard-Jones potential.
The walls were pushed together by an external force ($\sim 3$MPa) that
was an order of magnitude less than the adhesive pressure from the LJ 
potential.
The bottom wall was fixed, and the free diffusion of the top wall was
followed at a low temperature
($T=0.1 \epsilon/k_B$).
For $k \le  10 \epsilon \sigma^{-2}$, 
the walls were pinned by static friction for all system sizes
investigated.
For $k \ge 25 \epsilon \sigma^{-2}$, $F_s$ vanished, and
the top wall diffused freely in the long time limit.
By comparison,
Lennard-Jones interactions between atoms within the walls
would give rise to $k \approx 200 \epsilon \sigma^{-2}$.
Hence, the adhesive interactions between atoms on different surfaces
must be an order of magnitude stronger than the cohesive
interactions within each surface in order to produce static friction between
the flat, incommensurate walls that were considered.

The results described above make it clear that the static friction between
ideal crystals can be expected to vanish in many cases.
This raises the question of why static friction is observed so universally
in experiments.
One possibility is that roughness or chemical disorder pins the
two surfaces together.
Theoretical arguments indicate that disorder will always pin low
dimensional objects
(e.g. Gr\"uner et al., 1988).
However, the same arguments show that the pinning between three-dimensional
objects is exponentially weak
(Caroli and Nozieres, 1996;
Persson and Tosatti, 1996; Volmer and Natterman, 1997).
This suggests that other effects like mobile atoms between the surfaces
may play a key role in creating static friction.
This idea is discussed below in Sec.~\ref{sec:eff_adsorbed}.

\subsection{Chemically Passivated Surfaces}
\label{bar_com_che}

The simulations just described aimed at revealing
general aspects of friction.
There is also a need to understand the tribological
properties of specific materials on the nanoscale.
Advances in the chemical vapor deposition of diamond
hold promise for producing
hard protective diamond coatings on a variety of materials.
This motivated Harrison et al. (1992b) to perform
molecular-dynamics simulations of atomic-scale friction between diamond
surfaces.

Two orientationally-aligned, hydrogen-terminated 
diamond (111) surfaces were placed in sliding contact.
Potentials based on the work of Brenner (1990) were used.
As discussed in Section~\ref{tribochem}, these potentials
have the ability to account for chemical reactions, but none occurred
in the work described here.
The lattices contained ten layers of carbon atoms and two layers of hydrogen
atoms, and each layer consisted of 16 atoms.
The three outermost layers were treated as rigid units,
and were displaced relative to each other at constant
sliding velocity and constant separation. 
The atoms of the next five layers were coupled to a thermostat.

Energy dissipation mechanisms were investigated as a function
of load, temperature, sliding velocity, and sliding direction.
At low loads, the top wall moved almost rigidly over the potential
from the bottom wall, and the average friction was nearly zero.
At higher loads, colliding hydrogen atoms on opposing surfaces
locked into a metastable
state before suddenly slipping past each other.
As in the Tomlinson model, energy was dissipated during these rapid pops.
The kinetic friction was smaller for sliding along the grooves between
nearest neighbor hydrogen terminations, [1$\bar{1}$0],
than in the orthogonal direction, [11$\bar{2}$],
because hydrogen atoms on different surfaces could remain farther apart.

In a subsequent study, Harrison et al. (1993)
investigated the effect of atomic scale roughness by randomly replacing
one eighth of the hydrogen atoms on one surface
with methyl, ethyl or n-propyl groups.
Changing hydrogen to methyl had little effect on the friction at a given
load.
However a new
type of pop between metastable states was observed: Methyl groups rotated
past each other in a rapid turnstile motion.
Further increases in the length of the substituted molecules
led to much smaller $F_k$ at high loads.
These molecules were flexible enough to be pushed into the grooves between hydrogen
atoms on the opposing surface, reducing the number of collisions.

Note that Harrison et al. (1992b, 1993) and 
Perry and Harrison (1996, 1997)
might have obtained somewhat
different trends using
a different ensemble and/or incommensurate walls.
Their case of constant separation and velocity corresponds to a system that is
much stiffer than even the stiffest AFM.
Because they used commensurate walls and constant velocity,
the friction depended on the relative displacement of the lattices
in the direction normal to the velocity.
The constant separation also led to variations in normal load
by up to an order of magnitude
with time and lateral displacement.
To account for these effects,
Harrison et al. (1992b, 1993) and
Perry and Harrison (1996, 1997)
presented values for friction and load that were
averaged over both time and lateral displacement.
Studies of hydrogen-terminated silicon surfaces (Robbins and Mountain)
indicate that changing to a constant load and lateral force ensemble
allows atoms to avoid each other more easily.
Metastability sets in at higher loads than in a constant separation
ensemble, the friction is lower, and variations with sliding
direction are reduced.

Glosli and coworkers have
investigated the sliding motion between two ordered monolayers of
longer alkane chains bound to commensurate walls (McClelland and Glosli,
1992; Glosli and McClelland, 1993; Ohzono et al., 1998).
Each chain contained six alkane monomers with fixed bond lengths.
Next-nearest neighbors and third-nearest
neighbors on the chain interacted via bond bending and torsional
potentials, respectively.
One end of each chain was harmonically coupled
to a site on the $6 \times 6$ triangular lattices that made up each wall.
All other
interactions were Lennard Jones (LJ) potentials between CH$_3$ and CH$_2$
groups (the united atom model of Sec. \ref{sec:mod_pot}).
The chain density was high enough that chains pointed away from the
surface they were anchored to.
A constant vertical separation of the walls was maintained,
and the sliding velocity $v$ was well below the sound velocity.
Friction was studied as a function of
$T$, $v$, and the ratio of the LJ interaction energies
between endgroups on opposing surfaces, $\epsilon_1$,
to that within each surface, $\epsilon_0$.

Many results of these simulations correspond to the predictions of the
Tomlinson model.
Below a threshold value of $\epsilon_1/\epsilon_0$
(0.4 at $k_BT/\epsilon_0=0.284$),
molecules moved smoothly, and the force decreased to zero with velocity.
When the interfacial interactions became stronger than this threshold
value, ``plucking motion''
due to rapid pops between metastable states was observed.
Glosli and McClelland (1993) showed that at each pluck,
mechanical energy was converted to kinetic energy
that flowed away from the interface as heat.
Ohzono et al. (1998) showed that a generalization of the Tomlinson
model could quantitatively describe the sawtooth shape of the shear
stress as a function of time.
The instantaneous lateral force did not vanish in any of 
Glosli and McClelland's (1993) or Ohzono et al.'s (1998) simulations.
This shows that there was always a finite static friction, as expected
between commensurate surfaces.

For both weak ($\epsilon_1/\epsilon_0 = 0.1$)
and strong ($\epsilon_1/\epsilon_0 = 1.0$)
interfacial interactions,
Glosli and McClelland (1993)
observed an interesting maximum in the $T$-dependent friction force.
The position of this maximum coincided with the rotational
``melting'' temperature $T_M$
where orientational order at the interface was lost.
It is easy to understand that $F$ drops at $T>T_M$ because thermal
activation helps molecules move past each other.
The increase in $F$ with $T$ at low $T$ was attributed to increasing
anharmonicity that allowed more of the plucking energy to be dissipated.

\subsection{Single Asperity Contacts}
\label{bare_sin_asp_con}

Engineering surfaces are usually rough,
and friction is generated between
contacting asperities on the two surfaces.
These contacts typically have diameters of order a $\mu$m or more
(e.g. Dieterich and Kilgore, 1996).
This is much larger than atomic scales, and the models above may provide
insight into the behavior within a representative portion of these contacts.
However, it is important to determine how finite contact area and
surface roughness effect friction.
Studies of atomic-scale asperities can address these issues, and
also provide direct models of the small contacts typical of AFM tips.

\begin{figure}[tb]
\epsfxsize=10cm
\hfil\epsfbox{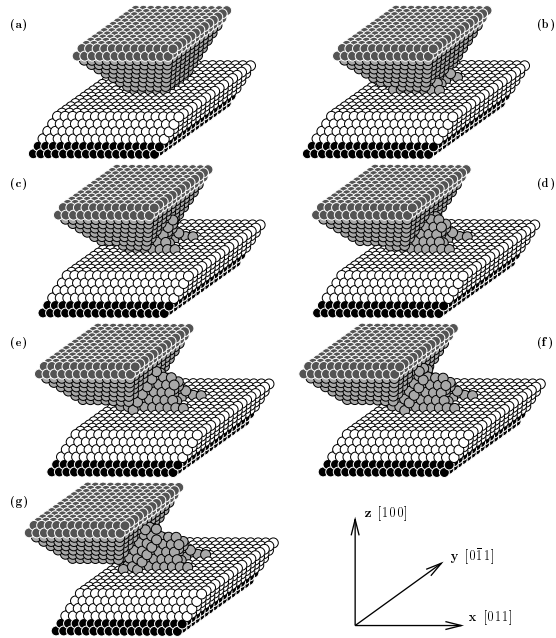}\hfill
\caption{
Snapshots showing the evolution of a Cu(100) tip on a Cu(100)
substrate during sliding to the left.
(From S$\o$rensen et al., 1996.)
\label{sorensen96_15acf}
}
\end{figure}

S$\o$rensen et al.
performed simulations of sliding tip-surface and surface-surface
contacts consisting of copper atoms (S$\o$rensen et al., 1996).
Flat, clean tips with (111) or (100) surfaces were brought into
contact with corresponding crystalline substrates
(Fig. \ref{sorensen96_15acf}).
The two exterior layers of tip and surface 
were treated as rigid units,
and the dynamics of the remaining mobile layers was followed.
Interatomic forces and
energies were calculated using semiempirical potentials derived from
effective medium theory (Jacobsen et al., 1987).
At finite temperatures, the outer mobile layer of both tip and surface
was coupled to a Langevin thermostat.
Zero temperature simulations gave similar results.
To explore the effects of commensurability,
results for crystallographically aligned and
misoriented tip-surface configurations
were compared.

In the commensurate Cu(111) case, S$\o$rensen et al. observed
atomic-scale stick-slip motion of the tip.
The trajectory was of zig-zag form which could be
related to jumps of the tip's surface between fcc and hcp positions.
Similar zig-zag motion is seen in shear along (111) planes
of bulk fcc solids (Stevens and Robbins, 1993).
Detailed analysis of the slips
showed that they occurred via a dislocation mechanism.
Dislocations were nucleated at the corner of the interface, and then
moved rapidly through the contact region.
Adhesion led to a large static friction at zero load:
The static friction per unit area, or critical yield stress,
dropped from 3.0GPa to 2.3GPa as $T$ increased from 0 to 300K.
The kinetic friction increased linearly with load with a surprisingly small
differential friction coefficient
$\tilde{\mu}_{\rm k} \equiv \partial F_{\rm k}/ \partial L \approx .03$.
In the load regime investigated,
$\tilde{\mu}_{\rm k}$ was independent of temperature and load. 
No velocity dependence was detectable up
to sliding velocities of $v = 5$~m/s.
At higher velocities,
the friction decreased.
Even though the interactions between the surfaces are identical
to those within the surfaces, no wear was observed.
This was attributed to the fact that (111) surfaces are the preferred
slip planes in fcc metals.

Adhesive wear was observed between a commensurate (100) tip and substrate
(Fig.~\ref{sorensen96_15acf}).
Sliding in the (011) direction at either constant height or constant load
led to inter-plane sliding between (111) planes inside the tip.
As shown in Fig. \ref{sorensen96_15acf}, this plastic deformation
led to wear of the tip, which left a trail of atoms in its wake.
The total energy was an increasing function of sliding distance due
to the extra surface area.
The constant evolution of the tip kept the motion from being periodic,
but the saw-toothed variation of force with displacement
that is characteristic of atomic-scale stick-slip was still observed.

Nieminen et al. (1992) observed a different mechanism of plastic
deformation in essentially the same geometry, but at higher velocities
(100m/s vs. 5m/s) and with Morse potentials between Cu atoms.
Sliding took place place between (100) layers inside the tip.
This led to a reduction of the tip by two layers
that was described as the climb of two successive edge dislocations,
under the action of the compressive load.
Although wear covered more of the surface with material
from the tip, the friction remained constant at constant normal load.
The reason was that the portion of the surface where the tip
advanced had a constant area.
While the detailed mechanism of plastic mechanism is very different
than in S$\o$rensen et al. (1996), the main conclusions of both papers
are similar: When two commensurate surfaces with strong adhesive
interactions are slid against each other, wear is obtained
through formation of dislocations that nucleate at the corners of the 
moving interface. 

S$\o$rensen et al. (1996) also examined the effect of incommensurability.
An incommensurate Cu(111) system was obtained by rotating the tip
by 16.1$^o$ about the axis perpendicular to the substrate.
For a small tip (5x5 atoms) they observed an Aubry transition from
smooth sliding with no static friction at low loads, to atomic-scale
stick-slip motion at larger loads.
Further increases in load led to sliding within the tip and plastic
deformation.
Finite systems are never truly incommensurate,
and pinning was found to occur at the corners of the contact, suggesting
it was a finite-size effect.
Larger tips (19x19) slid without static friction at all loads.
Similar behavior was observed for incommensurate Cu(100) systems.
These results confirm the conclusions of Hirano and Shinjo (1990)
that even bare metal surfaces of the same material will not
exhibit static friction if the surfaces are incommensurate.
They also indicate that contact areas as small as a few hundred
atoms are large enough to exhibit this effect.

Many other tip-substrate simulations of bare metallic surfaces have been
carried out.  Mostly, these simulations concentrated on indentation,
rather than on sliding or scraping (see Sec. \ref{wear}).
Among the indentation studies of metals
are simulations of a Ni tip indenting
Au(100) (Landman et al., 1990),
a Ni tip coated with an
epitaxial gold monolayer indenting Au(100) (Landman et al., 1992),
an Au tip indenting Ni(001) (Landman and Luedtke, 1989, 1991),
an Ir tip indenting a soft Pb substrate (Raffi-Tabar et al., 1992),
and an Au tip indenting Pb(110)
(Tomagnini et al., 1993).
These simulations have been reviewed in detail
within this series by Harrison et al. (1999).

\begin{figure}[tb]
\epsfxsize=7cm
\hfil\epsfbox{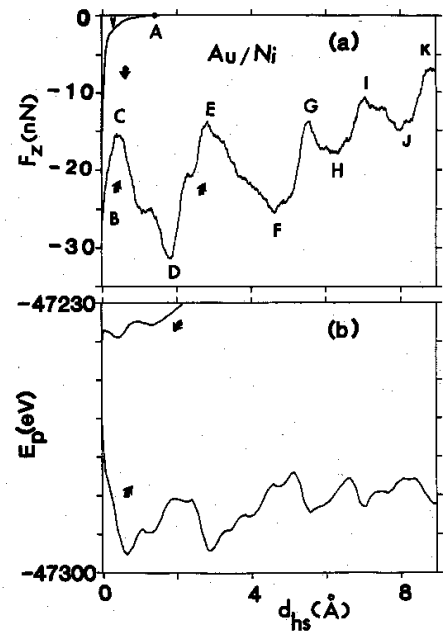}\hfill
\caption{Normal force $F_{\rm z}$ and potential energy $E_{\rm p}$
of an Au tip lowered
toward a Ni (001) surface as a function of separation $d_{\rm hs}$,
which is defined
to be zero after the jump to contact has occurred.
Upper lines were obtained by lowering the tip, lower lines are obtained
by raising the tip.
The points marked C,E,G,I, and K
correspond to ordered configurations of the tip, each containing
an additional layer.
(From Landman et al., 1991.)
\label{landman91_1}
}
\end{figure}

In general, plastic deformation occurs mainly in the softer of
the two materials, typically Au or Pb in the cases above.
Fig. \ref{landman91_1} shows the typical evolution of the
normal force and potential energy
during an indentation at high enough loads to produce
plastic deformation (Landman and Luedtke, 1991).
As the Au tip approaches the Ni surface (upper line),
the force remains nearly zero until a separation of about 1 \AA.
The force then becomes extremely attractive and there is a jump
to contact (A).
During the jump to contact, Au atoms in the tip displace
by 2 \AA\ within a short time span of 1~ps.
This strongly adhesive contact produces reconstruction of the
Au layers through the fifth layer of the tip.
When the tip is withdrawn, a neck is pulled out of the
substrate.
The fluctuations in force seen in Fig. \ref{landman91_1}
correspond to periodic increases in the
number of layers of gold atoms in the neck.

Nanoscale investigations of indentation, adhesion and fracture
of non-metallic diamond (111) surfaces have been carried out by 
Harrison et al. (1992a).
A hydrogen-terminated diamond tip was brought in contact with a
(111) surface that was either bare or hydrogen-terminated.
The tip was constructed by removing atoms from a (111) crystal
until it looked like an inverted pyramid with a flattened apex.
The model for the surface was similar to that
described in Section~\ref{bar_com_che}, but one layer contained 64 atoms.
The indentation was performed by moving the rigid layers of the tip
in steps of 0.15 \AA. The system was then equilibrated before
observables were calculated.

\begin{figure}[tb]
\epsfxsize=8cm
\hfil\epsfbox{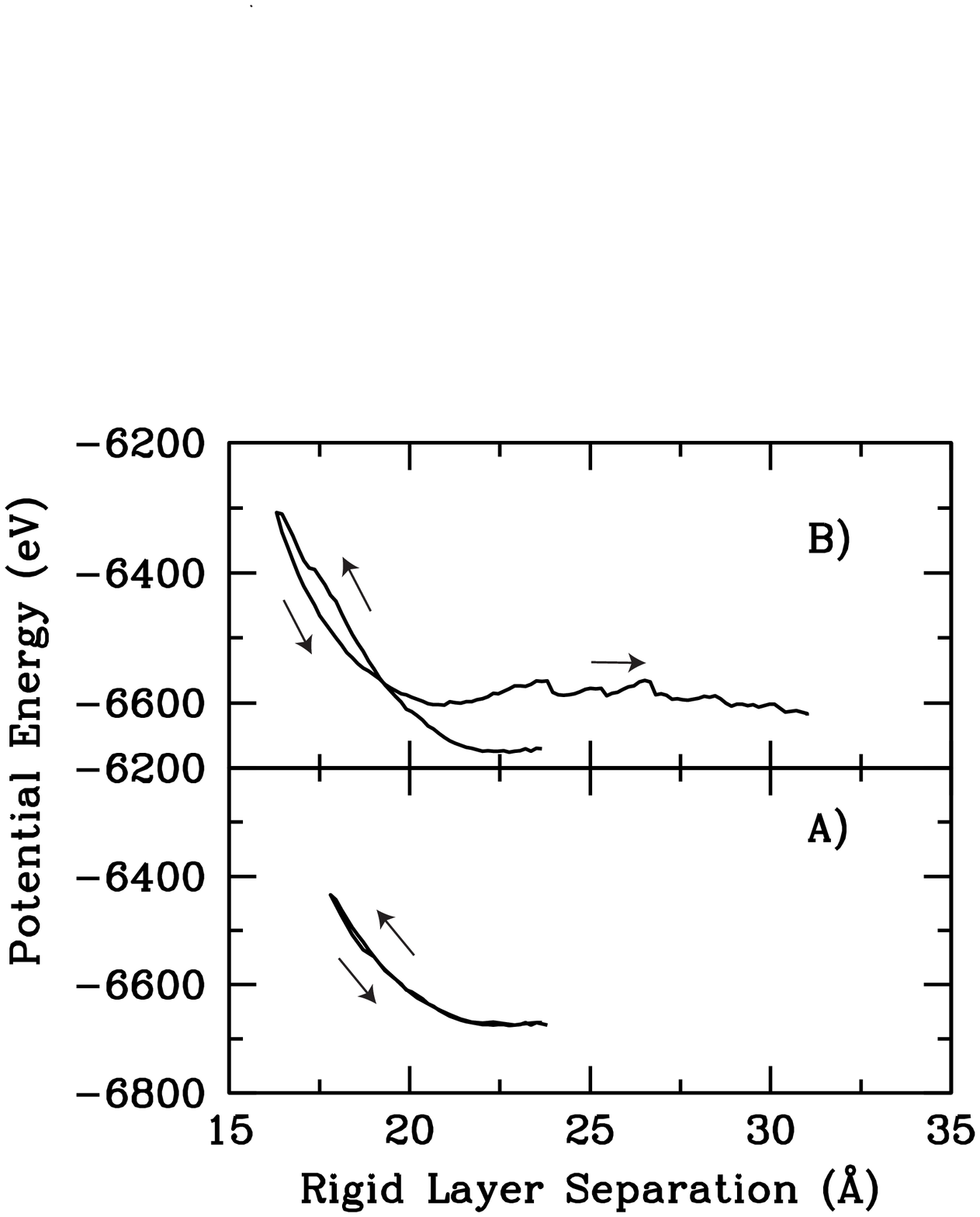}\hfill
\caption{Total potential energy of a hydrogen-terminated diamond-diamond
system as a function of separation during indentation to a
maximum load of 200nN (A) or 250nN (B).
Arrows indicate the direction of the motion.
(From Harrison et al., 1992a.)
\label{harrison92_2ab}
}
\end{figure}

Unlike the metal/metal systems
(Fig. \ref{landman91_1}),
the diamond/diamond systems
(Fig. \ref{harrison92_2ab})
did not show a pronounced jump to contact (Harrison et al., 1992a).
This is because the adhesion between diamond (111) surfaces is
quite small if at least one is hydrogen-terminated 
(Harrison et al., 1991).
For effective normal loads up to 200~nN (i.e. small indentations),
the diamond tip and surface deformed elastically and the force
distance curve was reversible (Fig. \ref{harrison92_2ab} (A)).
A slight increase to 250nN, led to plastic deformation
that produced hysteresis and steps in the force distance curve
(Fig. \ref{harrison92_2ab} (B))
(Harrison et al., 1992a).

\section{Lubricated Surfaces}
Hydrodynamics and elasto-hydrodynamics have been very successful in
describing lubrication by micron-thick films (Dowson and Higginson, 1968).
However, these continuum theories begin to break down as atomic
structure becomes important.
Experiments
and simulations
reveal a sequence of dramatic changes in the static and
dynamic properties of fluid films as
their thickness decreases from microns down to molecular scales.
These changes have important implications for the function of
boundary lubricants.
This section describes simulations of these changes, beginning
with changes in flow boundary conditions for relatively thick
films, and concluding with simulations of submonolayer films
and corrugated walls.

\subsection{Flow boundary conditions}
\label{sec:flowbc}

Hydrodynamic theories of lubrication need to assume a boundary condition
(BC) for the fluid velocity at solid surfaces.
Macroscopic experiments are generally well-described by a "no-slip" BC;
that is that the tangential component of the fluid velocity equals
that of the solid at the surface.
The one prominent exception is contact line motion, where
an interface between two fluids moves along a solid surface.
This motion would require an infinite force in hydrodynamic theory,
unless slip occurred near the contact line
(Huh and Scriven, 1971; Dussan, 1979).

The experiments on adsorbed monolayers described in Section \ref{sec:monolayer}
suggest that slip may occur more generally on solid surfaces.
As noted, the kinetic friction between the first monolayer and
the substrate can be orders of magnitude
lower than that between two layers in a fluid.
The opposite deviation from no-slip is seen in some Surface Force
Apparatus experiments -- a layer of fluid molecules becomes immobilized
at the solid wall (Chan and Horn, 1985; Israelachvili, 1986).

In pioneering theoretical work,
Maxwell (1867)
calculated the deviation from
a no-slip boundary condition for an ideal gas.
He assumed that at each collision molecules were either
specularly reflected or exchanged momentum to emerge with a
velocity chosen at random from a thermal distribution.
The calculated flow velocity at the wall was non-zero, and increased
linearly with the velocity gradient near the wall.
Taking $z$ as the direction normal to the wall and $u_\parallel$ as the
tangential component of the velocity relative to the wall, Maxwell found
\begin{equation}
u_\parallel (z_0) = {\cal S}
\left( {{\partial u_\parallel} \over {\partial z}}\right)_{z_0}
\label{eq:maxwell}
\end{equation}
where $z_0$ is the position of the wall.
The constant of proportionality, ${\cal S}$, has units of
length and is called the slip length.
It represents the distance into the wall at which the velocity gradient
would extrapolate to zero.
Calculations with a fictitious wall at this position and no-slip boundary
conditions would reproduce the flow in the region far from the wall.
${\cal S}$ also provides a measure of the kinetic friction per unit area
between the wall and the
adjacent fluid.
The shear stress must be uniform in steady state, because any
imbalance would lead to
accelerations.
Since the velocity gradient times the viscosity $\mu$ gives the stress in
the fluid, the kinetic friction per unit area is
$u_\parallel (z_0) \mu/{\cal S}$.
Maxwell found that
${\cal S}$ increased linearly with the mean free path
and that it
also increased with the probability of specular reflection.

Early simulations used mathematically flat walls and phenomenological
reflection rules
like those of Maxwell.
For example, Hannon et al. (1988)
found that the slip length was reduced to molecular scales
in dense fluids.
This is expected from Maxwell's result, since the mean free path becomes
comparable to an atomic separation at high densities.
However work of this type does not address how the atomic structure
of realistic walls is related to collision probabilities and 
whether Maxwell's reflection rules are relevant.
This issue was first addressed in simulations of moving contact lines
where deviations from no-slip boundary conditions have their most
dramatic effects
(Koplik et al., 1988, 1989; Thompson and Robbins, 1989).
These papers found that even when no-slip boundary conditions held
for single fluid flow, the large stresses near moving contact lines
led to slip within a few molecular diameters of the contact line.
They also began to address the relation between the flow boundary
condition and structure induced in the fluid by the solid wall.

The most widely studied type of order is layering in planes
parallel to the wall.
It is induced by the sharp cutoff in fluid density at the wall
and the pair correlation function $g(r)$ between fluid atoms
(Abraham, 1978; Toxvaerd, 1981; Nordholm and Haymet, 1980;
Snook and van Megen, 1980;
Plischke and Henderson, 1986).
An initial fluid layer forms at the preferred wall-fluid spacing.
Additional fluid molecules tend to lie in a second layer,
at the preferred fluid-fluid spacing.
This layer induces a third, and so on.

\begin{figure}[tb]
\epsfxsize=12cm
\hfil\epsfbox{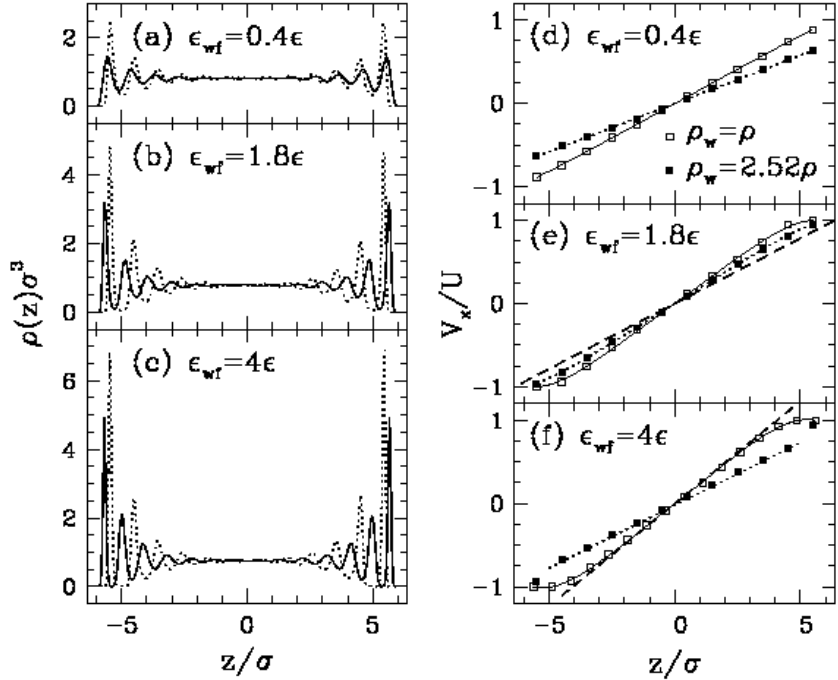}\hfill
\caption{
Panels (a-c) show density as a function of position $z$ relative to
two walls whose atoms are centered at $z/\sigma=\pm 6.4$.
Values of $\epsilon_{\rm wf}$ are indicated.
Solid lines are for equal wall and fluid densities and dotted
lines are for $\rho_w =2.52 \rho$.
Squares in panels (d-f) show the average velocity in each layer
as a function of $z$, and solid and dotted
lines are fits through these values for low and high density
walls, respectively.
For this data the walls were moved in opposite directions at
speed $U=1\sigma/t_{LJ}$.
The dashed line in (e) represents the flow expected from hydrodynamics
with a no-slip BC (${\cal S} =0$).
As shown in (f), the slope (dashed line) far from the walls
is used to define $\cal S$.
(Panels d-f from Thompson and Robbins, 1990a).
\label{fig:layers}
}
\end{figure}

Some of the trends that are observed in the degree and extent
of layering are illustrated in Fig. \ref{fig:layers}
(Thompson and Robbins, 1990a).
The fluid density is plotted as a function of
the distance between walls for a model considered in almost all
the studies of flow BC's described below.
The fluid consists of spherical molecules interacting with
a Lennard-Jones potential.
They are confined by crystalline walls containing discrete atoms.
In this case the walls were planar (001) surfaces of an fcc crystal.
Wall and fluid atoms also interact with a
Lennard-Jones potential,
but with a different binding energy $\epsilon_{\rm wf}$.

The net adsorption potential from the walls (Eq. \ref{eq:usub}) can
be increased by raising $\epsilon_{\rm wf}$ or by increasing the
density of the walls $\rho_w$ so that more wall atoms attract the fluid.
Fig. \ref{fig:layers} shows that both increases lead to increases
in the height of the first density peak.
The height also increases with the pressure in the fluid
(Koplik et al., 1989; Barrat and Bocquet, 1999a)
since that forces atoms into steeper regions of the adsorption potential.
The height of subsequent density peaks decreases smoothly
with distance from the wall, and only four or five well-defined layers
are seen near each wall in Fig. \ref{fig:layers}.
The rate at which the density oscillations decay is determined
by the decay length of structure in the bulk pair-correlation function
of the fluid.
Since all panels of Fig. \ref{fig:layers} have the same conditions
in the bulk, the decay rate is the same.
The adsorption potential only determines the initial height of the peaks.

The pair correlation function usually decays over a few molecular
diameters except under special conditions, such as near a critical
point.
For fluids composed of simple spherical molecules, the oscillations
typically extend out to a distance of order 5 molecular diameters
(Magda et al., 1985; Schoen et al., 1987; Thompson and Robbins, 1990a).
For more complex systems containing fluids with chain or branched
polymers, the oscillations are usually negligible beyond $\sim$ 3
molecular diameters
(Bitsanis and Hadziianou, 1990; Thompson et al., 1995;
Gao et al. 1997a, 1997b).
Some simulations with realistic potentials for alkanes show more pronounced
layering near the wall
because the molecules adopt a rod-like conformation in the first layer
(Ribarsky et al., 1992; Xia et al., 1992).

Solid surfaces also induce density modulations within the plane of
the layers
(Landman et al., 1989; Schoen et al., 1987, 1988, 1989;
Thompson and Robbins, 1990a, 1990b).
These correspond directly to the modulations induced in adsorbed
layers by the corrugation potential
(Sec. \ref{sec:monolayer}),
and can also be quantified
by the two-dimensional static structure factor at the
shortest reciprocal lattice vector $Q$ of the substrate.
When normalized by the number of atoms in the layer, $N_l$,
this becomes an intensive variable
that would correspond to the Debye-Waller factor in a crystal.
The maximum possible value, $S(Q)/N_l = 1$, corresponds to fixing
all atoms exactly at crystalline lattice sites.
In a strongly ordered case such as $\rho_w=\rho$ in Fig. \ref{fig:layers}(c),
the small oscillations about lattice sites in
the first layer only decrease $S(Q)/N_l$ to 0.71.
This is well above the value of 0.6 that is typical of bulk
3D solids at their melting point and indicates that the
first layer has crystallized onto the wall.
This was confirmed by computer simulations of diffusion and flow
(Thompson and Robbins, 1990a).
The values of $S(Q)/N_l$ in the second and third layers
are 0.31 and 0.07, respectively, and atoms in these layers exhibit typical
fluid diffusion.

There is some correlation between the factors that produce strong layering
and those that produce strong in-plane modulations.
For example,
chain molecules have several conflicting length scales that
tend to frustrate both layering and in-plane order
(Thompson et al., 1995; Gao et al., 1997a, 1997b; Koike and Yoneya, 1998, 1999).
Both types of order
also have a range that is determined by $g(r)$ and a magnitude
that decreases with decreasing $\epsilon_{\rm wf}/\epsilon$.
However,
the dependence of in-plane order on the density of substrate atoms
is more complicated than for layering.
When $\rho_w=\rho$,
the fluid atoms can naturally sit on the sites of a
commensurate lattice, and $S(Q)$ is large.
When the substrate density $\rho_w$ is increased by a factor of 2.52,
the fluid atoms no longer fit easily into the corrugation potential.
The degree of induced in-plane order drops sharply, although the layering
becomes stronger (Fig. \ref{fig:layers}). 
Sufficiently strong adsorption potentials may eventually lead to
crystalline order in the first layers, and stronger layering.
However, this may actually increase slip, as shown below.

Fig. \ref{fig:layers} also illustrates the range of flow boundary
conditions that have been seen in many studies
(Heinbuch and Fischer, 1989; Koplik et al., 1989; Thompson and Robbins, 1990a;
Bocquet and Barrat, 1994; Mundy et al., 1996; Khare et al., 1996;
Barrat and Bocquet, 1999a).
Flow was imposed by displacing the walls in opposite
directions along the $x$-axis with speed $U$ (Thompson and Robbins, 1990a).
The average velocity $V_x$
was calculated within each of the layers defined by
peaks in the density (Fig. \ref{fig:layers}),
and normalized by $U$.
Away from the walls, all systems exhibit
the characteristic Couette flow profile expected for 
Newtonian fluids.
The value of $V_x$ rises linearly with $z$,
and the measured shear stress divided by $\partial V_x/\partial z$
equals the bulk viscosity.
Deviations from this behavior occur within the first few
layers, in the region where layering and in-plane order
are strong.
In some cases the fluid velocity remains substantially less
than $U$, indicating slip occurs.
In others, one or more layers move at the same velocity as
the wall, indicating they are stuck to it.

Applying Maxwell's definition of slip length Eq. \ref{eq:maxwell}
to these systems is complicated by uncertainty in where the
plane of the solid surface $z_0$ should be defined.
The wall is atomically rough,
and the fluid velocity can not be evaluated too near
to the wall because of the pronounced layering.
In addition, the curvature evident in some flow profiles represents
a varying local viscosity whose effect must be included
the boundary condition.

One approach is to fit the linear flow profile in the
central region and extrapolate to the value of $z^*$ where
the velocity would reach $+U$.
The slip length can then be defined as ${\cal S} = z^*-z_{tw}$
where $z_{tw}$ is the height of the top wall atoms.
This is equivalent to applying Maxwell's definition (Eq. \ref{eq:maxwell})
to the extrapolated flow profile at $z_{tw}$.
The no-slip condition corresponds to a flow profile that extrapolates
to the wall velocity at $z_{tw}$ as illustrated by the dashed
line in Fig. \ref{fig:layers}(e).
Slip produces a smaller velocity gradient and a positive value of
$\cal S$.
Stuck layers lead to a larger velocity gradient and a negative value of
$\cal S$.

The dependence of slip length on many parameters has been studied.
All the results are consistent with a decrease in slip as the
in-plane order increases.
Numerical results for $\rho_w=\rho$ and $\epsilon_{\rm wf} =0.4\epsilon$
(Fig. \ref{fig:layers}(d)) are very close to the no-slip condition.
Increasing $\epsilon_{\rm wf}$ leads to stuck layers
(Koplik et al., 1988, 1989; Thompson and Robbins, 1989, 1990a;
Heinbuch and Fischer, 1989),
and decreasing $\epsilon_{\rm wf}$
can produce large slip lengths
(Thompson and Robbins, 1990a; Barrat and Bocquet, 1999a)
Increasing pressure (Koplik et al., 1989; Barrat and Bocquet, 1999a)
or decreasing temperature
(Heinbuch and Fischer, 1989; Thompson and Robbins, 1990a)
increases structure in $g(r)$ and leads to less slip.
These changes could also be attributed to increases in layering.
However, increasing the wall density $\rho_w$ from $\rho$ to 2.52 $\rho$
increases slip in Fig. \ref{fig:layers}.
This correlates with the drop in in-plane order, while the layering
actually increases.
Changes in in-plane order also explain the pronounced increase
in slip when $\epsilon_{\rm wf}/\epsilon$ is increased to 4 in
the case of dense walls (Fig. \ref{fig:layers}(f)).
The first layer of fluid atoms becomes crystallized with
a very different density than the bulk fluid.
It becomes the ``wall'' that produces order in the second layer,
and it gives an adsorption potential characterized by $\epsilon$
rather than $\epsilon_{\rm wf}$.
The observed slip is consistent with that for dense walls
displaced to the position of the first layer and interacting with $\epsilon$.

Thompson and Robbins (1990a)
found that all of their
results for $\cal S$ collapsed on to a universal curve
when plotted against the structure factor $S(Q)/N_l$.
When one or more layers crystallized onto the wall, the same collapse
could be applied as long as the effective wall position was shifted
by a layer
and the $Q$ for the outer wall layer was used.
The success of this collapse at small $S(Q)/N_l$
can be understood from the perturbation theory for the kinetic
friction on adsorbed monolayers (Eq. \ref{eq:phon}).
The slip length is determined by the friction between the outermost
fluid layer and the wall.
This depends only on $S(Q)/N_1$ and the phonon lifetime for
acoustic waves.
The latter changes relatively little over the temperature range
considered, and hence ${\cal S}$ is a single-valued function of $S(Q)$.
The perturbation theory breaks down at large $S(Q)$, but the
success of the collapse indicates that there is still a 
one-to-one correspondence between it and the friction.

In a recent paper Barrat and Bocquet (1999b) have derived an
expression relating ${\cal S}$ and $S(Q)$ that is equivalent to
Eq. \ref{eq:phon}.
However, in describing their numerical results\footnote{They considered
almost the same parameter range as Thompson
and Robbins (1990a), but at a lower temperature
($k_BT/ \epsilon=0.7$ vs. 1.1 or 1.4) and at a single wall density
($\rho = \rho_w$).}
they emphasize the
correlation between increased slip and decreased wetting
(Barrat and Bocquet, 1999a, 1999b).
In general the wetting properties of fluids are determined
by the adsorption term of the substrate potential $U_0$
(Eq. \ref{eq:usub}).
This correlates well with the degree of layering, but has
little to do with in-plane order.
In the limit of a perfectly structureless wall one may
have complete wetting and yet there is also complete slip.
The relation between wetting and slip is very much like
that between adhesion and friction.
All other things being equal, a greater force of attraction
increases the effect of corrugation in the potential and
increases the friction.
However, there is no one-to-one correspondence between them.

In earlier work Bocquet and Barrat (1993, 1994)
provided a less ambiguous resolution to the definition of
the slip length than that of Thompson and Robbins (1990a).
They noted that the shear rate in the central region of
Couette flow depended only on the sum of the wall position
and the slip length.
Thus one must make a somewhat arbitrary choice of wall position to
fix the slip length.
However, if one also fits the flow profile for Poiseuille flow,
unique values of slip length and the effective distance between
the walls $h$ are obtained (Barrat and Bocquet, 1999a).
Bocquet and Barrat (1993, 1994)
also suggested and implemented
an elegant approach for determining both ${\cal S}$ and $h$
using equilibrium simulations and the fluctuation-dissipation theorem.
This is one of the first applications of the fluctuation-dissipation
theorem to boundary conditions.
It opens up the possibility of
calculating flow boundary conditions directly from equilibrium
thermodynamics, and Bocquet and Barrat (1994)
were able to derive Kubo relations for $z_0$ and $\cal S$.
Analytic results for these relations are not possible in
general, but in the limit of weak interactions
they give an expression equivalent to Eq. \ref{eq:phon} for the drag
on the wall as noted above
(Barrat and Bocquet, 1999b).
Mundy et al. (1996)
have proposed a non-equilibrium
simulation method for calculating these quantities directly.

In all of the work described above, care was taken to ensure
that the slip boundary condition was independent of the
wall velocity.
Thus both the bulk of the fluid and the interfacial
region were in the linear response
regime where the fluctuation-dissipation theorem holds.
This linear regime usually extends to very high shear rates
($> 10^{10} {\rm s}^{-1}$ for spherical molecules).
However,
Thompson and Troian (1997)
found that under
some conditions the interfacial region exhibits nonlinear
behavior at much lower shear rates than the bulk fluid.
They also found a universal form for the deviation from
a linear stress/strain-rate relationship at the interface.
The fundamental origin for this non-linearity is that there
is a maximum stress that the substrate can apply to the fluid.
This stress roughly corresponds to the maximum of the force
from the corrugation potential times the areal density of fluid atoms.
The stress/velocity relation at the interface starts out
linearly and then flattens as the maximum stress is approached.
The shear rate in the fluid saturates at the value corresponding
to the maximum shear stress and the amount of slip at the wall grows
arbitrarily large with increasing wall velocity.
Similar behavior was observed for more realistic potentials by
Koike and Yoneya (1998, 1999).

\subsection{Phase Transitions and Viscosity Changes in Molecularly Thin Films}
\label{sec:viscosity}

One of the surprising features of Fig. \ref{fig:layers}
is that the viscosity remains the same even in regions near the wall where 
there is pronounced layering.
Any change in viscosity would produce a change in the
velocity gradient since the stress is constant.
However, the flow profiles in panel (d) remain linear throughout the cell.
The profile for the dense walls in panel (f) is linear up to the last
layer, which has crystallized onto the wall.
From Fig. \ref{fig:layers} it is apparent that density variations
by at least a factor of seven can be accommodated without a viscosity
change.

The interplay between layering and viscosity was studied in detail by
Bitsanis et al. (1987),
although their use of artificial
flow reservoirs kept them from addressing flow BC's.
They were able to fit detailed flow profiles using only the bulk viscosity
evaluated at the average local density.
This average was taken over a distance of order $\sigma$ that smeared
out the rapid density modulations associated with layering, but
not slower variations due to the adsorption potential of the walls
or an applied external potential.

In subsequent work, Bitsanis et al. (1990)
examined the change in viscosity with film thickness.
They found that results for films thicknesses $h > 4\sigma$ could
be fit using the bulk viscosity for the average density.
However, as $h$ decreased below $4\sigma$, the viscosity diverged much
more rapidly than any model based on bulk viscosity could explain.

These observations were consistent with experiments
on nanometer thick films of a wide variety of small molecules.
These experiments used the Surface Force Apparatus (SFA) which
measures film thickness with \AA\ resolution using optical interferometry.
Films were confined between atomically flat mica sheets at a fixed normal
load, and sheared with a steady (Gee et al., 1990)
or oscillating (Granick, 1992)
velocity.
Layering in molecularly thin films gave rise to oscillations in
the energy, normal force and effective viscosity 
(Horn and Israelachvili, 1981; Israelachvili, 1991; Georges et al., 1993)
as the film thickness decreased.
The period of these oscillations was a characteristic molecular
diameter.
As the film thickness decreased below 7 to 10 molecular diameters,
the effective viscosity increased dramatically
(Gee et al., 1990; Granick, 1992; Klein and Kumacheva, 1995).
Most films of one to three molecular layers exhibited a yield stress
characteristic of solid-like behavior, even though the molecules form
a simple Newtonian fluid in the bulk.

Pioneering grand canonical Monte Carlo simulations by
Schoen et al. (1987)
showed crystallization of spherical molecules between commensurate walls
separated by up to 6 molecular diameters.
However, the crystal was only stable when the thickness was near to
an integral number of crystalline layers.
At intermediate $h$, the film transformed to a fluid state.
Later work (Schoen et al., 1988, 1989), showed that translating the
walls could also destabilize the crystalline phase and lead to periodic
melting and freezing transitions as a function of displacement.
However, these simulations were carried out at equilibrium and
did not directly address the observed changes in viscosity.

SFA experiments can not determine the flow profile within
the film, and this introduces ambiguity in the meaning of the
viscosity values that are reported.
Results are typically expressed as an effective viscosity
$\mu_{\rm eff}\equiv \tau_s / {\dot \gamma}_{\rm eff}$
where $\tau_s$ is the measured shear
stress and ${\dot \gamma}_{\rm eff} \equiv v/h$
represents the effective shear rate that would be present if 
the no-slip condition held and walls were displaced at relative
velocity $v$.
Deviations from the no-slip condition might cause $\mu_{\rm eff}$
to differ from the bulk viscosity by an order of magnitude.
However, they could not explain the observed changes of $\mu_{\rm eff}$
by more than five orders of magnitude
or the even more dramatic changes by 10 to 12 orders of magnitude
in the characteristic viscoelastic relaxation time determined
from the shear rate dependence of $\mu_{\rm eff}$
(Gee et al., 1990; Hu et al., 1991).

\begin{figure}[tb]
\epsfxsize=9cm
\hfil\epsfbox{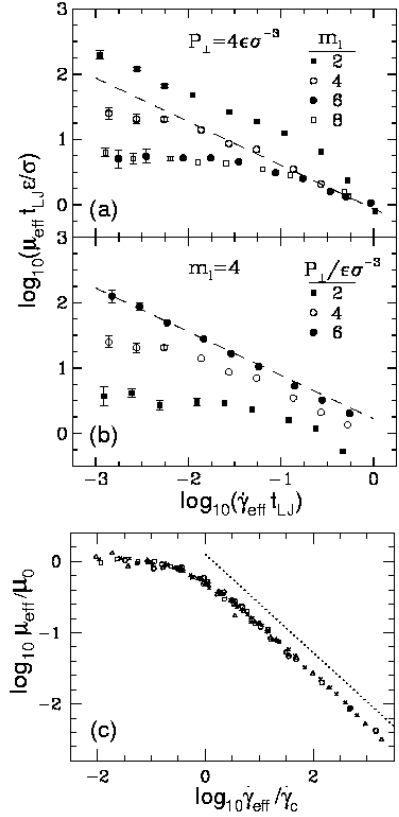}\hfill
\caption{
Plots of $\mu_{\rm eff}$ vs. ${\dot \gamma}_{\rm eff}$
at (a) fixed
normal pressure $P_{\bot} = 4\epsilon \sigma^{-3}$
and varying numbers of layers $m_l$,
and (b) fixed $m_l=4$ and varying $P_{\bot}$
(Thompson et al., 1992).
Dashed lines have slope -2/3.
Panel (c) shows that this and other data can be collapsed onto
a universal response function (Baljon and Robbins, 1999).
Results are for decreasing temperature in bulk systems (circles),
increasing normal pressure at fixed number of fluid layers
(triangles), or decreasing film thickness at fixed pressure with
two different sets of interaction potentials (squares and crosses).
The dashed line has a slope of -0.69.
(Panels a and b from Thompson et al., 1992; panel c from
Baljon and Robbins, 1999.)
\label{fig:viscos}
}
\end{figure}

Thompson et al. (1992)
found very similar changes in viscosity and relaxation time in
simulations of a simple bead-spring model of linear molecules
(Kremer and Grest, 1990).
Some of their results for the effective viscosity vs. effective shear
rate are shown in Fig. \ref{fig:viscos}.
In (a), the normal pressure $P_\bot$ was fixed and the number of atomic
layers, $m_l$, was decreased from 8 to 2.
In (b), the film was confined by increasing the pressure
at fixed particle number.
Both methods of increasing confinement lead to dramatic changes
in the viscosity and relaxation time.

The shear rate dependence of $\mu_{\rm eff}$ in Fig. \ref{fig:viscos}
has the same form as in experiment (Hu et al., 1991).
A Newtonian regime with constant viscosity $\mu_0$ is seen at
the lowest shear rates in all but the uppermost curve in each panel.
Above a characteristic shear rate
${\dot \gamma}_c$ the viscosity begins to decrease rapidly.
This shear thinning is typical of viscoelastic media and indicates
that molecular rearrangements are too slow to respond to the sliding
walls at ${\dot \gamma_{\rm eff}} > {\dot \gamma}_c$.
As a result, the structure of the fluid begins to change in a way
that facilitates shear.
A characteristic time for molecular rearrangements in the film can
be associated with $1/{\dot \gamma}_c$.
Increasing confinement
by decreasing $m_l$ or increasing
pressure, increases the Newtonian viscosity and relaxation time.
For the uppermost curve in each panel, the relaxation time is
longer than the longest simulation runs ($> 10^6$ time steps) and
the viscosity continues to increase at the lowest accessible shear rates.
Note that the ranges of shear rate covered in experiment and simulations
differ by orders of magnitude. 
However, the scaling discussed below suggests that the same behavior
may be operating in both.

For the parameters used in Fig. \ref{fig:viscos}, studies
of the flow profile in four layer films showed that one layer of molecules
was stuck to each wall and shear occurred in the middle layers.
Other parameter sets produced varying degrees of slip at the wall/film
interface, yet the viscoelastic response curves showed the same behavior.
Later work by Baljon and Robbins (1996, 1997)
shows that lowering
the temperature through the bulk glass transition also produces similar
changes in viscoelastic response.
This suggests that the same glass transition is being produced by
changes in thickness, pressure or temperature.

Following the analogy to bulk glass transitions, Thompson et al.
(1993, 1995)
have shown that changes in equilibrium
diffusion constant and ${\dot \gamma}_c$ can be fit to a free volume
theory.
Both vanish as $\exp (-h_0/(h-h_c))$ where $h_c$ is the film
thickness at the glass transition.
Moreover,
at $h<h_c$,
they found behavior characteristic of a solid.
Films showed a yield stress and no measurable diffusion.
When forced to slide, shear localized at the wall/film interface 
and $\mu_{\rm eff}$ dropped as $1/{\dot \gamma}$, implying that
the shear stress is independent of sliding velocity.
This is just the usual form of kinetic friction between solids.

The close relation between bulk glass transitions and those
induced by confinement can perhaps best be illustrated by
using a generalization of time-temperature scaling.
In bulk systems
it is often possible to collapse the viscoelastic
response onto a universal curve by dividing the viscosity by
the Newtonian value, $\mu_0$, and dividing the shear rate by
the characteristic rate, ${\dot \gamma}_c$.
Demirel and Granick (1996a)
found that this approach
could be used to collapse data for the real and imaginary
parts of the elastic moduli of confined films at different thicknesses.
Fig. \ref{fig:viscos}(c) shows that simulation results
for the viscosity of thin films can also be collapsed
using Demirel and Granick's approach (Robbins and Baljon, 2000).
Data for different thicknesses, normal pressures,
and interaction parameters taken from all parameters considered
by Thompson et al. (1992, 1995)
collapse onto a universal curve.
Also shown on the plot (circles) are data for different temperatures that were
obtained for longer chains in films that are thick enough to exhibit bulk
behavior
(Baljon and Robbins, 1996, 1997).
The data fit well on to the same curve, providing a strong indication
that a similar glass transition occurs whether thickness,
normal pressure, or temperature is varied.

The high shear rate region of the universal curve shown in Fig. 
\ref{fig:viscos}(c) exhibits power law shear thinning:
$\mu_{\rm eff} \propto {\dot \gamma}^{-x}$ with a best fit exponent
$x =-0.69\pm .02$.
In SFA experiments, Hu et al. (1991)
found shear thinning of many molecules was consistent with $x$ near -2/3.
However, the response of some fluids followed power laws closer to -.5
as they became less confined (Carson et al., 1992).
One possible explanation for this is that these
measurements fall onto the crossover region of a universal
curve like Fig. \ref{fig:viscos}(c).
The apparent exponent obtained from the slope of this log-log
plot varies from 0 to -2/3 as $\mu_{\rm eff}$ drops from $\mu_0$
to about $\mu_0/30$.
Many of the experiments that found smaller exponents were
only able to observe a drop in $\mu$ by an order of magnitude.
As confinement was increased, and a larger drop in viscosity
was observed, the slope increased toward -2/3.
It would be interesting to attempt a collapse of experimental
data on a curve like Fig. \ref{fig:viscos}(c) to test this
hypothesis.

Another possibility is that the shear thinning exponent depends
on some detail of the molecular structure.
Chain length alone does not appear to affect the exponent, since
results for chains of length $16$ and $6$ are combined in Fig.
\ref{fig:viscos}(c).
Manias et al. (1996) find that changing the geometry from
linear to branched has little effect on shear thinning.
In most simulations of simple spherical molecules, crystallization
occurs before the viscosity can rise substantially above the
bulk value.
However, Hu et al. (1996)
have found a set of conditions where spherical molecules
follow a $-2/3$ slope over two-decades in shear rate.

Stevens et al. (1997)
have performed simulations of confined films of hexadecane
using a detailed model of the molecular structure and interactions.
They found that films crystallized before the viscosity rose
much above bulk values.
This prevented them from seeing large power law scaling regimes,
but the apparent exponents were consistently less than those
for bead spring models.
It is not clear whether the origin of this discrepancy is the
inability to approach the glass transition, or whether structural
changes under shear lead to different behavior.
Hexadecane molecules have some tendency to adopt a linear
configuration and become aligned with the flow.
This effect is not present in simpler bead-spring models.

Shear thinning typically reflects changes in structure that
facilitate shear.
Experiments and the above simulations were done at constant normal
load.
In bead-spring models the dominant structural change is a dilation
of the film that creates more room for molecules to slide past
each other.
The dilations are relatively small, but have been detected in
some experiments (Dhinojwala and Granick).
When simulations are done at constant wall spacing, the
shear-thinning exponent drops to $x=-0.5$
(Thompson et al., 1992, 1995; Manias et al., 1996).
Kr\"oger et al. (1993)
find the same exponent in bulk simulations of these
molecules at constant volume, indicating that a universal
curve like that in Fig. \ref{fig:viscos}(c) might also
be constructed for constant volume instead of constant pressure.

Several analytic models have been developed to explain the power law
behavior observed in experiment and simulations.
All the models find an exponent of -2/3 in certain limits. 
However, they start from very different sets of assumptions and it
is not clear if any of these correspond to the simulations and experiments.
Two of the models yield an exponent of -2/3 for constant film
thickness (Rabin and Hersht, 1993; Urbakh et al., 1995)
where simulations give $x=-1/2$.
Urbakh et al. (1995) also find that the exponent depends on the
velocity profile, while simulations do not.
The final model (deGennes)
is based on scaling results for
the stretching of polymers under shear.
While it may be appropriate for thick films, it can not describe
the behavior of films which exhibit plug-like flow.
It remains to be seen if the -2/3 exponent has a single explanation
or arises from different mechanisms in different limits.

The results described in this section have interesting implications for
the function of macroscopic bearings.
Some bearings may operate in the boundary lubrication regime where
the separation between asperities decreases to molecular dimensions.
The dramatic increase in viscosity due to confinement may play
a key role in preventing squeeze-out of the lubricant and
direct contact between asperities.
Although the glassy lubricant layer would not have a low frictional
force, the yield stress would be lower than that for asperities
in contact.
More importantly, the amount of wear would be greatly reduced by
the glassy film.
Studies of confined films may help to determine what factors control
a lubricant's ability to form a robust protective layer at atomic scales.
As we now discuss, they may also help explain the pervasive observation
of static friction.

\subsection{Submonolayer Lubrication}
\label{sec:eff_adsorbed}

Physisorbed molecules, such as the short hydrocarbon chains considered
above, can be expected to sit on any surface exposed to atmospheric conditions.
Even in ultra-high-vacuum, special surface treatments are needed
to remove strongly physisorbed species from surfaces.
Recent work shows that the presence of these physisorbed molecules
qualitatively alters the tribological
behavior between two incommensurate walls (He et al., 1999; M\"user and
Robbins, 1999)
or between two disordered
walls (M\"user and Robbins)

As noted in Sec. \ref{sec:chap4},
the static friction is expected to vanish between most incommensurate
surfaces.
Under similar conditions,
the static friction between most amorphous, but flat, interfaces
vanishes in the thermodynamic limit (M\"user and Robbins).
In both cases, the reason is that the density modulations on two bare
surfaces can not lock into phase with each other unless the surfaces
are unrealistically compliant.
However, a sub-monolayer of molecules
that form no strong covalent bonds with the walls can
simultaneously lock to the density modulations of both walls.
This gives rise to a finite static friction
for all surface symmetries: commensurate, incommensurate,
and amorphous (M\"user and Robbins).

A series of simulations were performed in order to elucidate the influence
of such ''between-sorbed'' particles on tribological properties
(He et al., 1999; M\"user and Robbins, 1999).
A layer of spherical or short ``bead-spring'' (Kremer and Grest, 1990)
molecules
was confined between two fcc (111) surfaces.
The walls had the orientations and lattice spacings
shown in Fig. \ref{fig:walls},
and results are labeled by the letters in this figure.
Wall atoms were bound to their lattice
sites with harmonic springs as in the Tomlinson model.
The interactions between atoms on opposing walls,
as well as fluid-fluid and fluid-wall interactions,
had the LJ form.
Unless noted, the potential parameters for all three interactions
were the same.

\begin{figure}[tb]
\epsfxsize=7cm
\hfil\epsfbox{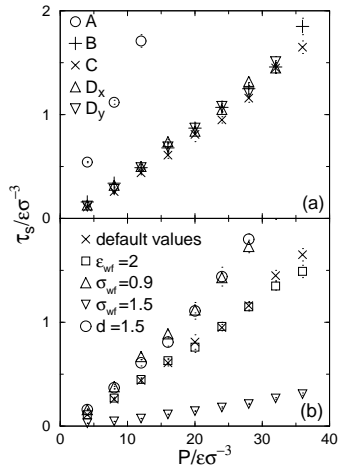}\hfill
\caption{Shear pressure as a function of normal pressure for various systems.
In (a) the letters correspond to the labels of wall geometries in
Fig. 5, and system D was slid in both $x$ and $y$ directions.
Panel (b) shows the effect of increasing the wall/fluid coupling
$\epsilon_{wf}$, decreasing or increasing the wall/fluid length $\sigma_{wf}$,
and increasing the nearest-neighbor spacing $d$ on the friction of system B.
The unit of pressure and stress, $\epsilon \sigma^{-3}$ corresponds to
about 30-50MPa.  (from He et al., 1999.)
\label{fig:he99_2}
}
\end{figure}

The static friction per unit contact area, or yield stress
$\tau_{\rm s}$, was determined from the lateral force
needed to initiate steady sliding of the surfaces at fixed pressure.
When there were no molecules between the surfaces, there was no
static friction unless the surfaces were commensurate.
As illustrated in Fig. \ref{fig:he99_2}(a), introducing a thin film led
to static friction in all cases.
Moreover, all incommensurate
\footnote{Although perfectly incommensurate walls are not consistent
with periodic boundary conditions, the effect of residual commensurability
was shown to be negligible (Muser and Robbins, 1999).}
cases (B-D)
showed nearly the same static friction, and $\tau_s$ was independent
of the direction of sliding relative to crystalline axes
(e.g. along $x$ or $y$ for case D).

Most experiments do not control the crystallographic orientation
of the walls relative to each other or to the sliding direction,
yet the friction is fairly reproducible.
This is hard to understand based on models of bare surfaces
which show dramatic variations in friction with orientation
(Hirano and Shinjo, 1993; S$\o$rensen et al., 1996; Robbins and Smith, 1996).
Fig. \ref{fig:he99_2} shows that a thin layer of molecules eliminates
most of this variation.
In addition, the friction is insensitive to chain length, coverage,
and other variables that are not well controlled in experiments
(He et al., 1999).
The kinetic friction is typically about 10 to 20\% lower than the
static friction in all cases (He and Robbins).

Of course experiments do observe changes in friction with
surface material.
The main factor that changed $\tau_s$ in this simple model
was the ratio of the characteristic length for wall-fluid interactions
$\sigma_{\rm wf}$ to the nearest-neighbor spacing on the walls, $d$.
As shown in Fig. \ref{fig:he99_2}(b), increasing $\sigma_{\rm wf}/d$
decreases the friction.  The reason is that larger fluid atoms are less
able to penetrate between wall atoms and thus feel less surface
corrugation.
Using amorphous, but flat, walls
also produced a somewhat larger static friction.

Note that $\tau_s$ rises linearly with the imposed pressure
in all cases shown in Fig. \ref{fig:he99_2}.
This provides a microscopic basis for the phenomenological explanation
of Amontons' laws that was proposed by Bowden and Tabor (1986).
The total static friction is given by the integral of the yield stress
over areas of the surface that are in molecular contact. 
If $\tau_s = \tau_0 + \alpha P$, then the total force is
$F_s = \alpha L + \tau_0 A_{\rm real}$ where $L$ is the load and
$A_{\rm real}$ is the total contact area.
The coefficient of friction is then $\mu_s = \alpha + \tau_0/{\overline P}$
where ${\overline P} = L/A_{\rm real}$ is the mean contact pressure.
Amontons' laws say that
$\mu_s$ is independent of load and the apparent area of the surfaces
in contact.
This condition is satisfied if $\tau_0$ is small or if ${\overline P}$
is constant.
The latter condition is expected to hold for both ideal elastic
(Greenwood and Williamson, 1966) and plastic (Bowden and Tabor, 1986)
surfaces.

The above results suggest that adsorbed molecules and other ``third-bodies''
may prove key to understanding macroscopic friction measurements.
It will be interesting to extend these studies to more realistic
molecular potentials and to rough surfaces.
To date, realistic potentials have only been used between commensurate
surfaces, and we now describe some of this work.

The effect of small molecules injected between two sliding
hydrogen-terminated (111) diamond surfaces on kinetic friction
was investigated by Perry and Harrison (1996, 1997).
The setup of the simulation was similar to the one described in
Section~\ref{bar_com_che}.
Two sets of simulations were performed.
In one set, bare surfaces were considered.
In another set, either two methane (CH$_4$) molecules, one 
ethane (C$_2$H$_6$) molecule,
or one isobutane (CH$_3$)$_3$CH molecule was
introduced into the interface between the sliding diamond surfaces.
Experiments show that the friction between diamond surfaces goes down
as similar molecules are formed in the contact due to wear
(Hayward, 1991).

Perry and Harrison found that
these third bodies also reduced the calculated frictional force between
commensurate diamond surfaces.
The reduction of the frictional force with respect
to the bare hydrogen-terminated case was most pronounced for
the smallest molecule, methane.
The molecular motions were analyzed in detail to determine how
dissipation occurred.
Methane produced less friction because it was small enough to roll
in grooves between the terminal hydrogen atoms without collisions.
The larger ethane and isobutane collided frequently.

As for the simple bead-spring model described above, the friction increased
roughly linearly with load. 
Indeed, Perry and Harrison's data for all third bodies corresponds to
$\alpha \approx 0.1$, which is close
to that for commensurate surfaces in Fig. \ref{fig:he99_2}(a).
One may expect that the
friction would decrease if incommensurate
walls were used.
However, the rolling motion of methane might be prevented in such
geometries.

Perry and Harrison (1996, 1997) compared their results to earlier simulations
(Harrison et al., 1993) where hydrogen terminations
on one of the two surfaces were replaced by
chemisorbed methyl (-CH$_3$), ethyl (-C$_2$H$_5$),
or $n$-propyl (-C$_3$H$_7$) groups
(Sec. \ref{bar_com_che}).
The chemisorbed molecules were seen to give a considerably smaller reduction
of the friction with respect to the physisorbed molecules.
Perry and Harrison note that the chemisorbed molecules have fewer
degrees of freedom, and so are less able to avoid collisions that dissipate
energy.

\subsection{Corrugated Surfaces}

\begin{figure}[tb]
\epsfxsize=14cm
\hfil\epsfbox{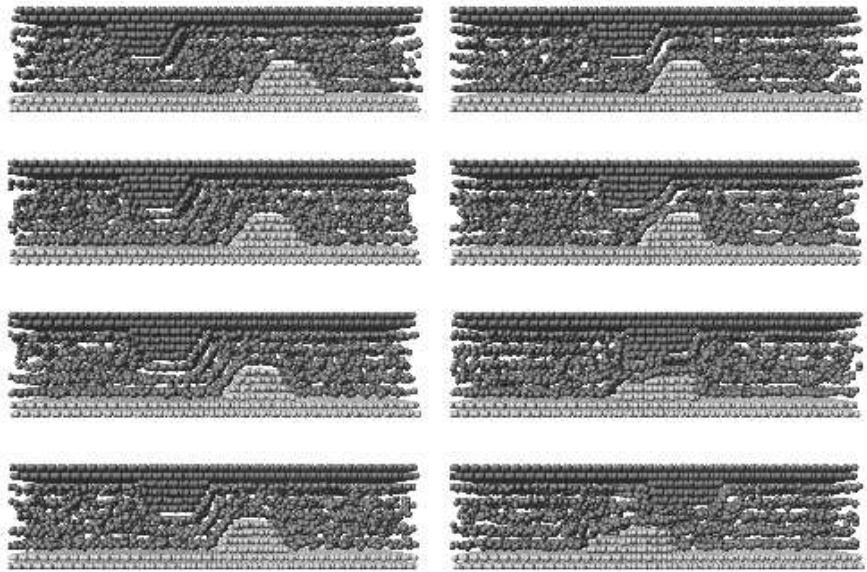}\hfill
\caption{Atomic and molecular configurations obtained in a slice
through the near-overlap system at times
156, 250, 308, 351, 400, 461, 614 and 766ps.
Time increases from top to bottom and then left to right.
(From Gao et al., 1995.)
\label{gao96_3}
}
\end{figure}

The presence of roughness can be expected to alter the behavior
of lubricants, particularly when the mean film thickness is
comparable to the surface roughness.
Gao et al. (1995, 1996)
and Landman et al. (1996)
used molecular dynamics to investigate this thin film limit.
Hexadecane ($n$-C$_{16}$H$_{34}$) was confined between two gold substrates
exposing only (111) surfaces
(Fig. \ref{gao96_3}).
The two outer layers of the substrates were completely rigid and were
displaced laterally
with a constant relative velocity of 10 to 20m/s
at constant separation.
The asperities on both walls were modeled by flat-topped pyramidal ridges
with initial heights of 4 to 6 atomic layers.
The united atom model (Sec. \ref{sec:mod_pot})
was used for the interactions within the film,
and the embedded atom method for Au-Au interactions.
All other interactions were modeled with
suitable 6-12 Lennard-Jones potentials. 
The alkane molecules and the gold atoms
in the asperities were treated dynamically using the Verlet 
algorithm.
The temperature was kept constant at $T = 350$~K by rescaling the
velocities every fiftieth time step.

Simulations were done in three different regimes, which can be
categorized according to the separation $\Delta h_{\rm aa}$ that the
outer surfaces of the
asperities would have if they were placed on top of one another without
deforming elasticly or plasticly.
The cases were
(1) large separation of the asperities $\Delta h_{\rm aa} = 17.5$ \AA,
(2) a near-overlap regime with $\Delta h_{\rm aa} = 4.6$ \AA, and
(3) an asperity-overlap regime  with $\Delta h_{\rm aa} = -6.7$ \AA.
Some selected atomic and molecular configurations obtained in a slice through
the near-overlap system are shown in Fig. \ref{gao96_3}. In all cases,
the initial separation of the walls was chosen such that the normal pressure
was zero for large lateral asperity separations.

\begin{figure}[tb]
\epsfxsize=8cm
\hfil\epsfbox{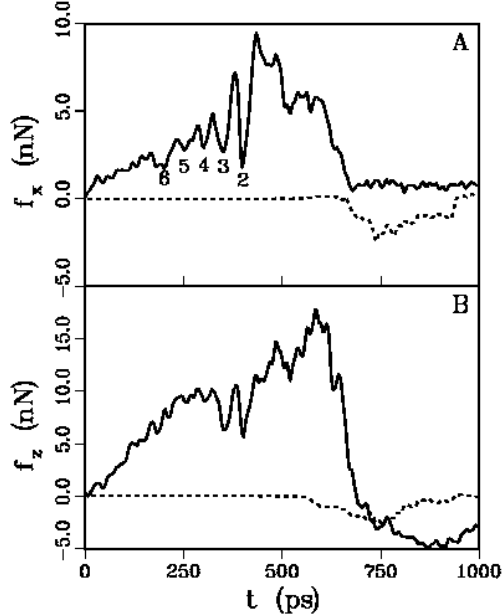}\hfill
\caption{Total force in driving direction $f_x$ and normal to the walls $f_z$
plotted vs. time.
Solid lines correspond to the net force between lubricant and
gold surface, dashed lines to direct forces between opposing gold surfaces.
The numbers in panel A give the number of layers between the asperity ridges.
(From Gao et al., 1995.)
\label{gao96_2}
}
\end{figure}

One common feature of all simulations was the formation of lubricant layers
between the asperities as they approached each other (Fig. \ref{gao96_3}).
This is just like the layering observed in equilibrium between flat walls
(e.g. Fig. \ref{fig:layers}), but the layers form dynamically.
The number of layers decreased with decreasing
lateral separation between the asperities,
and the lateral force showed strong oscillations
as successive layers were pushed out.
This behavior is shown in Fig. \ref{gao96_2} for the near-overlap case.

For large separation, case (1), four lubricant layers 
remained at the point of closest approach between the
asperities, and no plastic deformation occurred.
In case (2), severe plastic deformation occurred 
after local shear and normal stresses
exceeded a limiting value of close to 4~GPa.
This deformation led to direct
intermetallic junctions, which were absent in simulations under
identical conditions but with no lubricant molecules in the interface.
The junctions eventually broke upon continued sliding,
resulting in transfer of some metal atoms between the asperities.
As in the overlap case (3), great densification and pressurization of the
lubricant in the asperity region occurred, accompanied by a significant
increase in the effective viscosity in that region. For the near-overlap
system, local rupture of the film in the region between the departing
asperities was seen. A nanoscale cavitated zone of length scale 
$\approx 30$ \AA\ was observed that persisted for about 100~ps.
The Deborah number $D$ was studied as well.
$D$ can be defined as the ratio of the relaxation time of the fluid
to the time of passage of the fluid through a characteristic distance $l$.
$D \approx 0.25$ was observed for the near-overlap system, which corresponds
to a viscoelastic response of the lubricant. The increased confinement in
the overlap system, resulted in $D = 2.5$, which can be associated
with highly viscoelastic behavior, perhaps even elasto-plastic or waxy
behavior.

Gao et al. (1996)
observed extreme pressures of 150~GPa 
in the near-overlap case when the asperities were treated as rigid units.
Allowing only the asperities to deform reduces the peak pressure
by almost two orders of magnitude.
However, even these residual pressures are still quite large, and
one may expect that a full treatment of the elastic response of
the substrates might lead to further dramatic decreases in pressure
and damage.
Tutein et al. (2000) have recently compared friction between
monolayers of anchored hydrocarbon molecules and
rigid or flexible nanotubes.
Studies of elasticity effects in larger asperities
confining films that are several molecules thick
are currently in progress (Persson and Ballone).
\section{Stick-Slip Dynamics}

\label{sec:stick}

The dynamics of sliding systems can be very complex and
depend on many factors, including the types of
metastable states in the system, the times needed to transform between
states, and the mechanical properties of the device that imposes the stress.
At high rates or stresses, systems usually slide smoothly.
At low rates the motion often becomes intermittent,
with the system alternately sticking and slipping forward
(Rabinowicz, 1965; Bowden and Tabor, 1986).
Everyday examples of such stick-slip motion include
the squeak of hinges and the music of violins.

The alternation between stuck and sliding states of the system
reflects changes in the way energy is stored.
While the system is stuck, elastic energy is pumped into the system
by the driving device.
When the system slips, this elastic energy is released into kinetic energy,
and eventually dissipated as heat.
The system then sticks once more, begins to store elastic energy, 
and the process continues.
Both elastic and kinetic energy can be stored in all the mechanical
elements that drive the system.
The whole coupled assembly must be included in any analysis of the dynamics.

\begin{figure}[tb]
\epsfxsize=8cm
\hfil\epsfbox{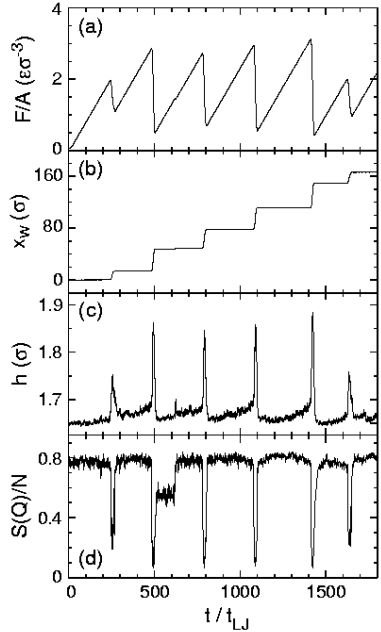}\hfill
\caption{
\label{fig:stkslp}
Time profiles of the (a) frictional force per unit area $F/A$,
(b) displacement of the top wall $x_W$, (c) wall spacing $h$,
and (d) Debye-Waller factor $S(Q)/N$
during stick-slip motion of spherical
molecules that form two crystalline layers in the static state.
Note that the system dilates (c) during each slip event.
The coinciding drop in Debye-Waller factor shows a dramatic decrease
from a typical crystalline value to a characteristic value for a fluid.
(From Robbins, 2000)
}
\end{figure}

The simplest type of intermittent motion is the atomic-scale
stick-slip that occurs in the 
multistable regime ($\lambda >1$) of the Tomlinson model
(Fig. \ref{fig:stab}(b)).
Energy is stored in the springs while atoms are trapped in a metastable
state, and converted to kinetic energy as they pop to the next metastable
state.
This phenomenon is quite general and has been observed in several of the
simulations
of wearless friction described in Sec. \ref{sec:chap4}
as well as
in the motion of atomic force microscope tips (e.g. Carpick and Salmeron, 1997).
In these cases, motion involves a simple ratcheting 
over the surface potential through a regular series of hops between
neighboring metastable states.
The slip distance is determined entirely by the periodicity of the
surface potential.
Confined films and adsorbed layers have a much 
richer potential energy landscape
due to their many internal degrees of freedom.
One consequence is that stick-slip motion between neighboring metastable
states can involve microslips by distances much less than a lattice constant
(Thompson and Robbins, 1990b; Baljon and Robbins, 1997;
Robbins and Baljon, 2000).
An example is seen at $t/t_{\rm LJ} = 620$ in Fig. \ref{fig:stkslp}(b).
Such microslips involve atomic-scale rearrangements within a small fraction
of the system.
Closely related microslips have been studied in granular media
(Nasuno et al., 1997; Veje et al., 1999) and foams (Gopal and Durian, 1995).

Many examples of stick-slip involve a rather different type of motion
that can lead to intermittency and chaos (Ruina, 1983; Heslot et al., 1994).
Instead of jumping between neighboring metastable states, the 
system slips for very long distances before sticking.
For example, Gee et al. (1990) and Yoshizawa et al. (1993)
observed slip distances
of many microns in their studies of confined films.
This distance is much larger than any characteristic periodicity in
the potential, and varied with velocity, load, and the
mass and stiffness of the SFA.
The fact that the SFA does not stick after moving by a lattice
constant indicates that sliding has changed the state of the
system in some manner, so that it can continue sliding even
at forces less than the yield stress.
Phenomenological theories of stick-slip
often introduce an unspecified
"state" variable to model the evolving properties of the system
(Dieterich, 1979; Ruina, 1983; Batista and Carlson, 1998).

One simple property that depends on past history is the 
amount of stored kinetic energy.
This can provide enough inertia to carry a system over potential energy
barriers even when the stress is below the yield stress.
Inertia is
readily included in the Tomlinson model and has been thoroughly
studied in the mathematically equivalent case of an underdamped
Josephson junction (McCumber, 1968).
One finds a hysteretic response function
where static and moving steady-states coexist over
a range of forces between $F_{\rm min}$ and the static friction $F_s$.
There is a minimum stable steady-state velocity $v_{\rm min}$ corresponding 
to $F_{\rm min}$.
At lower velocities, the only steady state is linearly unstable because
$\partial v /\partial F < 0$ -- pulling harder slows the system.
It is well-established that this type of instability can lead to
stick-slip motion (Bowden and Tabor, 1986; Rabinowicz, 1965).
If the top wall of the Tomlinson model is pulled at an average velocity
less than $v_{\rm min}$ by a sufficiently compliant system,
it will exhibit large-scale stick-slip motion.

Confined films have structural degrees of freedom that can
change during sliding, and this provides an alternative
mechanism for stick-slip motion
(Thompson and Robbins, 1990b).
Some of these structural changes are illustrated in
Fig. \ref{fig:stkslp}
which shows stick-slip motion of a two layer film of simple
spherical molecules.
The bounding walls were held together by a constant normal load.
A lateral force was applied to the top wall through a spring $k$
attached to a stage that moved with fixed velocity $v$ in the $x$
direction.
The equilibrium configuration of the film at $v=0$
is a commensurate crystal that resists shear.
Thus at small times, the top wall remains pinned at $x_W=0$.
The force grows linearly with time, $F=k v$, as the stage advances ahead
of the wall.
When $F$ exceeds $F_s$, the wall slips forward.
The force drops rapidly because the slip velocity ${\dot x}_W$
is much greater than $v$.
When the force drops sufficiently, the film recrystallizes,
the wall stops, and the force begins to rise once more.

One structural change that occurs during each slip event is
dilation by about 10\% (Fig. \ref{fig:stkslp}(c)).
Dhinojwala and Granick have recently confirmed that
dilation occurs during slip in SFA experiments.
The increased volume makes it easier for atoms to slide past
each other and is part of the reason that the sliding friction
is lower than $F_s$.
The system may be able to keep sliding in this dilated state
as long as it takes more time for the volume to contract
than for the wall to advance by a lattice constant.
Dilation of this type plays a crucial role in the yield, flow
and stick-slip dynamics of granular media
(Thompson and Grest, 1991; Jaeger et al., 1996; Nasuno et al., 1997).

The degree of crystallinity also changes during sliding.
As in Secs. \ref{sec:monolayer} and \ref{sec:flowbc},
deviations from an ideal crystalline structure
can be quantified by the Debye-Waller factor $S(Q)/N$
(Fig. \ref{fig:stkslp}d),
where $Q$ is one of the shortest reciprocal lattice vectors and $N$
is the total number of atoms in the film.
When the system is stuck, $S(Q)/N$ has a large value that is
characteristic of a 3D crystal.
During each slip event, $S(Q)/N$ drops dramatically.
The minimum values are characteristic of simple fluids that would show
a no-slip boundary condition (Sec. \ref{sec:flowbc}).
The atoms also exhibit rapid diffusion that is characteristic of
a fluid.
The periodic melting and freezing transitions that occur during
stick-slip are
induced by shear and not by the negligible changes in temperature.
Shear-melting transitions at constant temperature have been
observed in both theoretical and experimental studies of bulk
colloidal systems (Ackerson et al., 1986; Stevens and Robbins, 1993).
While the above simulations of confined
films used a fixed number of particles,
Lupowski and van Swol (1991)
found equivalent results at fixed chemical potential.

Very similar behavior has been observed in simulations of
sand (Thompson and Grest, 1991),
chain molecules (Robbins and Baljon, 2000),
and incommensurate or amorphous walls (Thompson and Robbins, 1990b).
These systems transform between glassy and fluid
states during stick-slip motion.
As in equilibrium, the structural differences between 
glass and fluid states are small.
However, there are strong changes in the self-diffusion and
other dynamic properties when the film goes from the static glassy
to sliding fluid state.

In the cases just described, the entire film transforms to a new
state, and shear occurs throughout the film.
Another type of behavior is also observed.
In some systems shear is confined to a single plane - either a wall/film
interface, or a plane within the film (Baljon and Robbins, 1997;
Robbins and Baljon, 2000).
There is always some dilation at the shear plane to facilitate sliding.
In some cases there is also in-plane ordering of the film to enable it to
slide more easily over the wall.
This ordering remains after sliding stops, and provides a mechanism
for the long-term memory seen in some
experiments (Gee et al., 1990; Yoshizawa et al., 1993; Demirel and Granick,
1996b).
Buldum and Ciraci (1997)
found stick-slip motion due to periodic structural transformations
in the bottom layers of a pyramidal Ni(111) tip sliding on an incommensurate
Cu(110) surface.

The dynamics of the transitions between stuck and sliding states
are crucial in determining the range of velocities where stick-slip motion is
observed, the shape of the stick-slip events, and whether stick-slip
disappears in a continuous or discontinuous
manner.
Current models are limited to energy balance arguments
(Robbins and Thompson, 1991; Thompson and Robbins, 1993)
or phenomenological models of the nucleation and growth of "frozen" regions
(Yoshizawa et al., 1993; Heslot et al., 1994;
Batista and Carlson, 1998; Persson, 1998).
Microscopic models and detailed experimental data on the sticking and
unsticking process are still lacking.

Rozman et al. (1996, 1997, 1998) have taken an interesting approach to
unraveling this problem.
They have performed detailed studies of stick-slip in a simple model
of a single incommensurate chain between two walls.
This model reproduces much of the complex dynamics seen in experiments
and helps to elucidate what can be learned about the nature of
structural changes within a contact using only the measured
macroscopic dynamics.
\section{Strongly Irreversible Tribological Processes}

Sliding at high pressures, high rates, or for long times can produce
more dramatic changes in the structure and even chemistry
of the sliding interface than those discussed so far.
In this concluding chapter, we describe some of the more strongly
irreversible tribological processes that have been studied with
simulations.
These include grain boundary formation and mixing, machining
and tribochemical reactions.

\subsection{Plastic Deformation}

For ductile materials, plastic deformation is likely to occur 
throughout a region of some characteristic width about the
nominal sliding interface (Rigney and Hammerberg, 1998).
Sliding induced mixing of material from the two surfaces 
and sliding induced grain boundaries are
two of the experimentally observed processes
that lack microscopic theoretical explanations.
In an attempt to get insight into the microscopic dynamics of these phenomena,
Hammerberg et al. (1998)
performed large-scale 
simulations of a two-dimensional model for copper.
The simulation cell contained $256 \times 256$
Cu atoms that were subject to a constant normal pressure $P_{\bot}$.
Two reservoir regions at the upper and lower boundaries of the cell were
constrained to move at opposite lateral velocities $\pm u_{\rm p}$.
The initial interface was midway between the two reservoirs.

The friction was measured at $P_\bot = 30$GPa
as a function of the relative sliding velocity $v$.
Different behavior was seen at velocities above and below
about 10\% of the speed of transverse sound.
At low velocities, the interface welded together and the
system formed a single workhardened object.
Sliding took place at the artificial boundary with one of the
reservoirs.
At higher velocities the friction was smaller, and decreased steadily
with increasing velocity.
In this regime, intense plastic deformation occurred at the interface.
Hammerberg et al. (1998) found that the early time-dynamics of the
interfacial structure could be reproduced with a
Frenkel-Kontorova model.
As time increased,
the interface was unstable to the formation of a fine-grained
polycrystalline microstructure,
which coarsened with distance away from the interface as a function of time.
Associated with this microstructure was the mixing of material across the
interface.

\begin{figure}[tb]
\epsfxsize=9cm
\hfil\epsfbox{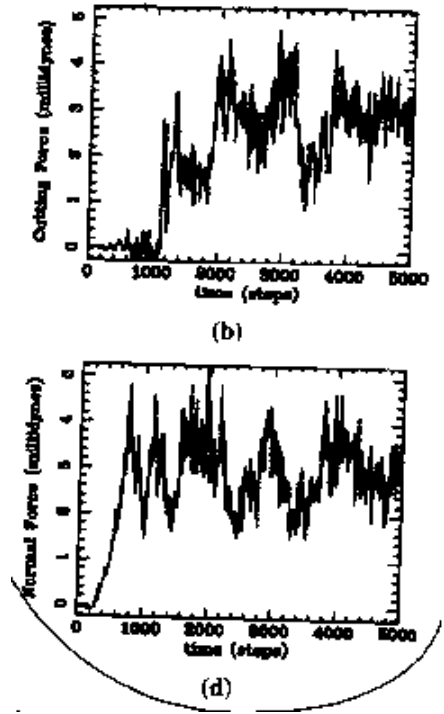}\hfill
\caption{Normal (bottom) and lateral (top) force on a three dimensional,
pyramidal Si tip on a copper surface as a function of time.
No plastic flow was reported up to 1,000 time steps. The indentation stopped
at about 5 layers after 2,000 time steps. (From Belak and Stowers, 1992.)
\label{belak92_6bd}
}
\end{figure}
\subsection{Wear}
\label{wear}

Large scale, two and three-dimensional molecular dynamics simulations of
the indentation and scraping of metal surfaces were carried out by Belak and
Stowers (1992).
Their simulations show that tribological properties
are strongly affected by wear or the generation of debris.
A blunted carbon tip was first indented into a copper
surface and then pulled over the surface.
The tip was treated as a rigid unit.
Interactions within the metal
were modeled with an embedded atom potential
and Lennard-Jones potentials were used between Si and Cu atoms.

In the two-dimensional simulation, indentation was
performed at a constant velocity of about 1~m/s.
The contact followed Hertzian behavior
up to a load $L \approx 2.7$~nN and an indentation of about 3.5 Cu layers.
The surface then yielded on one side of the tip,
through the creation of a single dislocation edge
on one of the easy slip planes.
The load needed to continue indenting
decreased slightly until an indentation of about five layers.
Then the load began to rise again as stress built up on the
side that had not yet yielded.
After an indentation of about six layers, this side yielded,
and further indentation could be achieved without a considerable increase
in load.
The hardness, defined as the ratio of load to contact length (area), slightly
decreased with increasing load once plastic deformation had occurred.

After indentation was completed, the carbon tip was slid parallel to the
original Cu surface. The work to scrape off material was determined as a
function of the tip radius. A power law dependence was found at small tip
radii that did not correspond to experimental findings for micro-scraping.
However, at large tip radii, the power law approached the experimental
value.
Belak and Stowers found that this change in power law was due to a change
in the mechanism
of plastic deformation from intragranular to intergranular plastic deformation.

In the three-dimensional (3D) simulations, the substrate contained as
many as 36 layers or 72,576 atoms.
Hence long-range elastic deformations were included.
The surface yielded plastically after an indentation of only 1.5 layers,
through the creation of a small dislocation loop.
The accompanying release of load was much bigger than in 2D.
Further indentation to about 6.5 layers produced several of these
loading-unloading events.
When the tip was pulled out of the substrate, both elastic and plastic
recovery was observed.
Surprisingly, the plastic deformation in the 3D studies
was confined to a region within a few lattice spacings of
the tip, while dislocations spread several hundred lattice spacings
in the 2D simulations.
Belak and Stowers concluded that dislocations
were not a very efficient mechanism for accommodating strain at the
nanometer length scale in 3D.

When the tip was slid laterally at $v=100$m/s during indentation,
the friction or ``cutting'' force fluctuated around zero as long as the
substrate did not yield (Fig.~\ref{belak92_6bd}).
This nearly frictionless sliding can be attributed to
the fact that the surfaces were incommensurate and the adhesive
force was too small to induce locking.
Once plastic deformation occurred,
the cutting force increased dramatically.
Fig. \ref{belak92_6bd} shows that the lateral and normal forces
are comparable, implying a friction coefficient of about one.
This large value was expected for cutting by a conical asperity
with small adhesive forces (Suh, 1986).

\subsection{Tribochemistry}
\label{tribochem}

The extreme thermomechanical conditions in sliding contacts can
induce chemical reactions.
This interaction of chemistry and friction is known as 
tribochemistry (Rabinowicz, 1965).
Tribochemistry plays important roles in many processes,
the best known example being the generation of fire through sliding friction.
Other examples include the formation of wear debris and adhesive
junctions which can have a major impact on friction.

Harrison and Brenner (1994) were the first to observe tribochemical reactions
involving strong covalent bonds
in molecular dynamics simulations.
A key ingredient of their work is the use of reactive potentials
that allow breaking and formation of chemical bonds.
Two (111) diamond surfaces terminated with
hydrogen atoms were brought into contact as in Sec.~\ref{bar_com_che}.
In some simulations, two hydrogen atoms from the upper surface were
removed, and replaced with ethyl (-CH$_2$CH$_3$) groups. The simulations were 
performed for 30~ps at an average normal pressure of about 33~GPa.
The sliding velocity was 100m/s along either the
$(1\bar{1}0)$ or $(11\bar{2})$ crystallographic direction

\begin{figure}[tb]
\epsfxsize=7cm
\hfil\epsfbox{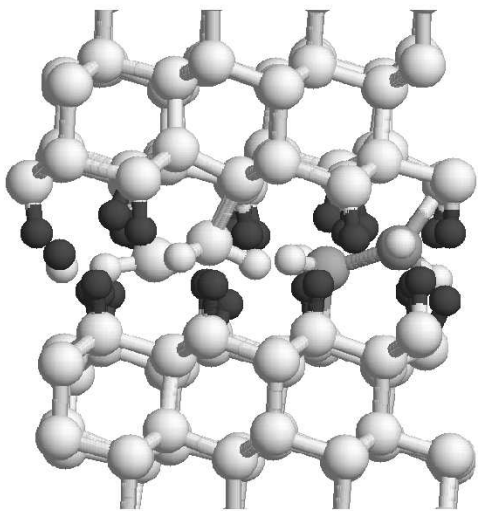}\hfill
\caption{Adhesion of the two diamond surfaces via the formation of a
carbon-carbon bond. The C atoms forming the bond originated from an ethyl
group chemisorbed on the upper diamond surface. (From Harrison et al., 1994.)}
\label{harrison94_2b}
\end{figure}

Sliding did not produce any chemical changes in
the hydrogen-terminated surfaces.
However, wear and chemical reactions were observed when ethyl groups
were present.
For sliding along the $(11\bar{2})$ direction, wear
was initiated by the shearing of hydrogen atoms from the tails
of the ethyl groups.
The resulting free hydrogen atoms reacted at the interface by
combining with an existing radical site or abstracting
a hydrogen from either a surface or a radical.
If no combination with a free hydrogen atom occurred,
the reactive radicals left on the tails
of the chemisorbed molecules abstracted a hydrogen from the opposing
surface, or they formed a chemical bond with existing radicals
on the opposing surface (Fig. \ref{harrison94_2b}). 
In the latter case, the two surface bonds sometimes broke simultaneously,
leaving molecular wear debris trapped at the interface. 

It is interesting to note that the wear debris, in the form of an ethylene
molecule CH$_2$CH$_2$, did not undergo another chemical reaction for the
remainder of the simulation. Similarly, methane CH$_4$, ethane C$_2$H$_6$,
and isobutane (CH$_3$)$_3$CH were not seen to 
undergo chemical reactions when introduced into a similar interface composed
of hydrogen terminated (111) diamond surfaces (Sec. \ref{sec:eff_adsorbed}) at
normal loads up to about 0.8~nN/atom
(Perry and Harrison, 1996, 1997;
Harrison and Perry, 1998).
At higher loads, only ethane reacted.
In some cases a hydrogen broke off of the ethane.
The resulting free H atom then reacted with an H atom from one surface
to make an H$_2$ molecule.
The remaining C$_2$H$_5$ could then form a carbon-carbon bond with
that surface when dragged close enough to the nascent radical site.
The C-C bond of the ethane was also
reported to break occasionally. However, due to the proximity
of the nascent methyl radicals and the absence of additional reactive species,
the bond always reformed.

Sliding along the $(1\bar{1}0)$ direction produced other types
of reaction between surfaces with ethyl terminations.
In some cases,
tails of the ethyl groups became caught between hydrogen atoms on the
lower surface. Continued sliding
sheared the entire tail from the rest of the ethyl group, leaving a 
chemisorbed CH${_2^{\;\bullet}}$ group and a free CH${_3^{\;\bullet}}$ species.
The latter group could form a bond with an existing radical site, it could
shear a hydrogen from a chemisorbed ethyl group, or it could be recombined with
the chemisorbed CH${_2^{\;\bullet}}$.

\section{Acknowledgement}
Support from the National Science Foundation through Grant No.
DMR-9634131 and from the German-Israeli Project Cooperation
``Novel Tribological Strategies from the Nano to Meso Scales''
is greatfully acknowledged.
We thank Gang He, Marek Cieplak, Miguel Kiwi, Jean-Louis Barrat,
Patricia McGuiggan and especially
Judith Harrison for providing comments on the text.
We also thank Jean-Louis Barrat for help in improving the
density oscillation data in Fig. 9, and Peter A. Thompson for
many useful conversations and for his role in creating Fig. 14.
\begin{description}
\itemsep .1ex
\item[]
Abraham, F.~F. (1978),  ``The interfacial density profile of a Lennard-Jones
  fluid in contact with a (100) Lennard-Jones wall and its relationship to
  idealized fluid/wall systems: A Monte Carlo simulation'', {\em J. Chem.
  Phys.}  {\bf 68}, 3713--3716.
\item[]
Ackerson, B.~J., Hayter, J.~B., Clark, N.~A., and Cotter, L. (1986),  ``Neutron
  scattering from charge stabilized suspensions undergoing shear'', {\em J.
  Chem. Phys.}  {\bf 84}, 2344--2349.
\item[]
Allen, M.~P. and Tildesley, D.~J. (1987), {\em Computer Simulation of Liquids},
  Clarendon Press, Oxford.
\item[]
Aubry, S. (1979),  ``The New Concept of Transitions by Breaking of Analyticity
  in a Crystallographic Model'', in {\em Solitons and Condensed Matter Physics}
  (Bishop, A.~R. and Schneider, T., eds.), pp. 264--290, Springer-Verlag,
  Berlin.
\item[]
Aubry, S. (1983),  ``The Twist Map, The Extended Frenkel-Kontorova Model and
  the Devil's Staircase'', {\em Physica D}  {\bf 7}, 240--258.
\item[]
Bak, P. (1982),  ``Commensurate phases, incommensurate phases and the devil's
  staircase'', {\em Rep. Progr. Phys.}  {\bf 45}, 587--629.
\item[]
Baljon, A. R.~C. and Robbins, M.~O. (1996),  ``Energy Dissipation During
  Rupture of Adhesive Bonds'', {\em Science}  {\bf 271}, 482--484.
\item[]
Baljon, A. R.~C. and Robbins, M.~O. (1997),  ``Stick-Slip Motion, Transient
  Behavior, and Memory in Confined Films'', in {\em Micro/Nanotribology and Its
  Applications} (Bhushan, B., ed.), pp. 533--553, Kluwer, Dordrecht.
\item[]
Barrat, J.-L. and Bocquet, L. (1999a),  ``Large Slip Effect at a Nonwetting
  Fluid-Solid Interface'', {\em Phys. Rev. Lett.}  {\bf 82}, 4671--4674.
\item[]
Barrat, J.-L. and Bocquet, L. (1999b),  ``Influence of wetting properties on
  hydrodynamic boundary conditions at a fluid/solid interface'', {\em Faraday
  Discuss.}  {\bf 112}, 1--9.
\item[]
Batista, A.~A. and Carlson, J.~M. (1998),  ``Bifurcations from steady sliding
  to stick slip in boundary lubrication'', {\em Phys. Rev. E}  {\bf 57},
  4986--4996.
\item[]
Belak, J. and Stowers, I.~F. (1992),  ``The Indentation and Scraping of a Metal
  Surface: A Molecular Dynamics Study'', in {\em Fundamentals of Friction:
  Macroscopic and Microscopic Processes} (Singer, I.~L. and Pollock, H.~M.,
  eds.), pp. 511--520, Kluwer Academic Publishers, Dordrecht.
\item[]
Binder, K. (1995), {\em Monte Carlo and Molecular Dynamics Simulations in
  Polymer Science}, Oxford University Press, New York.
\item[]
Bitsanis, I. and Hadziioannou, G. (1990),  ``Molecular dynamics simulations of
  the structure and dynamics of confined polymer melts'', {\em J. Chem. Phys.}
  {\bf 92}, 3827--3847.

\item[]
Bitsanis, I., Magda, J.~J., Tirrell, M., and Davis, H.~T. (1987),  ``Molecular
  dynamics of flow in micropores'', {\em J. Chem. Phys.}  {\bf 87}, 1733--1750.

\item[]
Bitsanis, I., Somers, S.~A., Davis, H.~T., and Tirrell, M. (1990),
  ``Microscopic dynamics of flow in molecularly narrow pores'', {\em J. Chem.
  Phys.}  {\bf 93}, 3427--3431.

\item[]
Bocquet, L. and Barrat, J.-L. (1993),  ``Hydrodynamic Boundary Conditions and
  Correlation Functions of Confined Fluids'', {\em Phys. Rev. Lett.}  {\bf 70},
  2726--2729.

\item[]
Bocquet, L. and Barrat, J.-L. (1994),  ``Hydrodynamic boundary conditions,
  correlation functions, and Kubo relations for confined fluids'', {\em Phys.
  Rev. E}  {\bf 49}, 3079--3092.

\item[]
Bowden, F.~P. and Tabor, D. (1986), {\em The Friction and Lubrication of
  Solids}, Clarendon Press, Oxford.

\item[]
Braun, O.~M., Bishop, A.~R., and R\"oder, J. (1997b),  ``Hysteresis in the
  Underdamped Driven Frenkel-Kontorova Model'', {\em Phys. Rev. Lett.}  {\bf
  79}, 3692--3695.

\item[]
Braun, O.~M., Dauxois, T., Paliy, M.~V., and Peyrard, M. (1997a),  ``Dynamical
  Transitions in Correlated Driven Diffusion in a Periodic Potential'', {\em
  Phys. Rev. Lett.}  {\bf 78}, 1295--1298.

\item[]
Braun, O.~M., Dauxois, T., Paliy, M.~V., and Peyrard, M. (1997c),  ``Nonlinear
  mobility of the generalized Frenkel-Kontorova model'', {\em Phys. Rev. E}
  {\bf 55}, 3598--3612.

\item[]
Brenner, D.~W. (1990),  ``Empirical Potentials for Hydrocarbons for Use in
  Simulating the Chemical Vapor Deposition of Diamond Films'', {\em Phys. Rev.
  B}  {\bf 42}, 9458--9471.

\item[]
Bruch, L.~W., Cole, M.~W., and Zaremba, E. (1997), {\em Physical Adsorption:
  Forces and Phenomena}, Oxford, New York.

\item[]
Buldum, A. and Ciraci, S. (1997),  ``Interplay between stick-slip motion and
  structural phase transitions in dry sliding friction'', {\em Phys. Rev. B}
  {\bf 55}, 12892--12895.

\item[]
Caroli, C. and Nozieres, P. (1996),  ``Dry Friction as a Hysteretic Elastic
  Response'', in {\em Physics of Sliding Friction} (Persson, B. N.~J. and
  Tosatti, E., eds.), pp. 27--49, Kluwer, Dordrecht.

\item[]
Carpick, R.~W. and Salmeron, M. (1997),  ``Scratching the Surface: Fundamental
  Investigations of Tribology with Atomic Force Microscopy'', {\em Chem. Rev.}
  {\bf 97}, 1163--1194.

\item[]
Carson, G.~A., Hu, H., and Granick, S. (1992),  ``Molecular Tribology of Fluid
  Lubrication: Shear Thinning'', {\em Tribol. Trans.}  {\bf 35}, 405--410.

\item[]
Chan, D. Y.~C. and Horn, R.~G. (1985),  ``The drainage of thin liquid films
  between solid surfaces'', {\em J. Chem. Phys.}  {\bf 83}, 5311--5324.

\item[]
Cieplak, M., Smith, E.~D., and Robbins, M.~O. (1994),  ``Molecular Origins of
  Friction: The Force on Adsorbed Layers'', {\em Science}  {\bf 265},
  1209--1212.

\item[]
Daw, M.~S. and Baskes, M.~I. (1984),  ``Embedded-Atom Method: Derivation and
  Application to Impurities, Surfaces, and Other Defects in Metals'', {\em
  Phys. Rev. B}  {\bf 29}, 6443--6453.

\item[]
Dayo, A., Alnasrallay, W., and Krim, J. (1998),  ``Superconductivity-Dependent
  Sliding Friction'', {\em Phys. Rev. Lett.}  {\bf 80}, 1690--1693.

\item[]
de~Gennes, P.~G., unpublished.

\item[]
Demirel, A.~L. and Granick, S. (1996a),  ``Glasslike Transition of a Confined
  Simple Fluid'', {\em Phys. Rev. Lett.}  {\bf 77}, 2261--2264.

\item[]
Demirel, A.~L. and Granick, S. (1996b),  ``Friction Fluctuations and Friction
  Memory in Stick-Slip Motion'', {\em Phys. Rev. Lett.}  {\bf 77}, 4330--4333.

\item[]
Dhinojwala, A. and Granick, S., unpublished.

\item[]
Dieterich, J.~H. (1979),  ``Modeling of Rock Friction. 2. Simulation of
  Pre-Seismic Slip'', {\em J. Geophys. Res.}  {\bf 84}, 2169--2175.

\item[]
Dieterich, J.~H. and Kilgore, B.~D. (1996),  ``Imaging surface contacts: power
  law contact distributions and contact stresses in quartz, calcite, glass and
  acrylic plastic'', {\em Tectonophysics}  {\bf 256}, 219--239.

\item[]
Dowson, D. and Higginson, G.~R. (1968), {\em Elastohydrodynamic Lubrication},
  Pergamon, Oxford.

\item[]
Dussan, E.~B. (1979),  ``On the Spreading of Liquids on Solid Surfaces: Static
  and Dynamic Contact Lines'', {\em Ann. Rev. Fluid Mech.}  {\bf 11}, 371--400.

\item[]
Evans, D.~J. and Morriss, G.~P. (1986),  ``Shear Thickening and Turbulence in
  Simple Fluids'', {\em Phys. Rev. Lett.}  {\bf 56}, 2172--2175.

\item[]
Fisher, D.~S. (1985),  ``Sliding charge-density waves as a dynamic critical
  phenomenon'', {\em Phys. Rev B}  {\bf 31}, 1396--1427.

\item[]
Flory, P. (1988), {\em Statistical Mechanics of Chain Molecules}, Hanser
  Publishers, M\"unchen.

\item[]
Frank, F.~C. and van~der Merwe, J.~H. (1949),  ``One-dimensional dislocations.
  I. Static theory'', {\em Proc. R. Soc. A}  {\bf 198}, 205--225.

\item[]
Frenkel, D. and Smit, B. (1996), {\em Understanding Molecular Simulation: From
  Algorithms to Applications}, Academic Press, San Diego.

\item[]
Frenkel, Y.~I. and Kontorova, T. (1938),  ``On the Theory of Plastic
  Deformation and Twinning'', {\em Zh. Eksp. Teor. Fiz.}  {\bf 8}, 1340.

\item[]
Gao, J., Luedtke, W.~D., and Landman, U. (1995),  ``Nano-Elastohydrodynamics:
  Structure, Dynamics, and Flow in Non-uniform Lubricated Junctions'', {\em
  Science}  {\bf 270}, 605--608.

\item[]
Gao, J., Luedtke, W.~D., and Landman, U. (1996),  ``Nano-Elastohydrodynamics:
  Structure, Dynamics and Flow in Nonuniform Lubricated Junctions'', in {\em
  Physics of Sliding Friction} (Persson, B. N.~J. and Tosatti, E., eds.), pp.
  325--348, Kluwer, Dordrecht.

\item[]
Gao, J., Luedtke, W.~D., and Landman, U. (1997a),  ``Origins of Solvation
  Forces in Confined Films'', {\em J. Phys. Chem. B}  {\bf 101}, 4013--4023.

\item[]
Gao, J., Luedtke, W.~D., and Landman, U. (1997b),  ``Structure and Solvation
  Forces in Confined Films: Linear and Branched Alkanes'', {\em J. Chem. Phys.}
   {\bf 106}, 4309--4318.

\item[]
Gee, M.~L., McGuiggan, P.~M., Israelachvili, J.~N., and Homola, A.~M. (1990),
  ``Liquid to solid transitions of molecularly thin films under shear'', {\em
  J. Chem. Phys.}  {\bf 93}, 1895--1906.

\item[]
Georges, J.~M., Millot, S., Loubet, J.~L., Touck, A., and Mazuyer, D. (1993),
  ``Surface roughness and squeezed films at molecular level'', in {\em Thin
  Films in Tribology} (Dowson, D., Taylor, C.~M., Childs, T. H.~C., Godet, M.,
  and Dalmaz, G., eds.), pp. 443--452, Elsevier, Amsterdam.

\item[]
Glosli, J.~N. and McClelland, G. (1993),  ``Molecular dynamics study of sliding
  friction of ordered organic monolayers'', {\em Phys. Rev. Lett.}  {\bf 70},
  1960--1963.

\item[]
Gopal, A.~D. and Durian, D.~J. (1995),  ``Nonlinear Bubble Dynamics in a Slowly
  Driven Foam'', {\em Phys. Rev. Lett.}  {\bf 75}, 2610--2614.

\item[]
Granick, S. (1992),  ``Motions and Relaxations of Confined Liquids'', {\em
  Science}  {\bf 253}, 1374--1379.

\item[]
Greenwood, J.~A. and Williamson, J. B.~P. (1966),  ``Contact of nominally flat
  surfaces'', {\em Proc. Roy. Soc. A}  {\bf 295}, 300--319.

\item[]
Grest, G.~S. and Kremer, K. (1986),  ``Molecular dynamics simulations for
  polymers in the presence of a heat bath'', {\em Phys. Rev. A}  {\bf 33},
  3628--3631.

\item[]
Gr\"uner, G. (1988),  ``The dynamics of charge-density waves'', {\em Rev. Mod.
  Phys.}  {\bf 60}, 1129--1181.

\item[]
Gr\"uner, G., Zawadowski, A., and Chaikin, P.~M. (1981),  ``Nonlinear
  Conductivity and Noise due to Charge-Density-Wave Depinning in NbSe$_3$'',
  {\em Phys. Rev. Lett.}  {\bf 46}, 511--517.

\item[]
Gyalog, T., Bammerlin, M., L\"uthi, R., Meyer, E., and Thomas, H. (1995),
  ``Mechanism of Atomic Friction'', {\em Europhys. Lett.}  {\bf 31}, 269--274.

\item[]
Hammerberg, J.~E., Holian, B.~L., R\"oder, J., Bishop, A.~R., and Zhou, J.~J.
  (1998),  ``Nonlinear dynamics and the problem of slip at material
  interfaces'', {\em Physica D}  {\bf 123}, 330--340.

\item[]
Hannon, L., Lie, G.~C., and Clementi, E. (1988),  ``Micro-Hydrodynamics'', {\em
  J. Stat. Phys.}  {\bf 51}, 965--979.

\item[]
Harrison, J.~A. and Brenner, D.~W. (1994),  ``Simulated Tribochemistry: An
  Atomic-Scale View of the Wear of Diamond'', {\em J. Am. Chem. Soc.}  {\bf
  116}, 10399--10402.

\item[]
Harrison, J.~A., Brenner, D.~W., White, C.~T., and Colton, R.~J. (1991),
  ``Atomistic Mechanisms of Adhesion and Compression of Diamond Surfaces'',
  {\em Thin Solid Films}  {\bf 206}, 213--219.

\item[]
Harrison, J.~A. and Perry, S.~S. (1998),  ``Friction in the Presence of
  Molecular Lubricants and Solid/Hard Coatings'', {\em MRS Bull.}  {\bf 23}(6),
  27--31.

\item[]
Harrison, J.~A., Stuart, S.~J., and Brenner, D.~W. (1999),  ``Atomic-Scale
  Simulation of Tribological and Related Phenomena'', in {\em Handbook of
  Micro/Nanotribology} (Bhushan, B., ed.), pp. 525--594. CRC Press, Boca Raton.

\item[]
Harrison, J.~A., White, C.~T., Colton, R.~J., and Brenner, D.~W. (1992a),
  ``Nanoscale Investigation of Indentation, Adhesion and Fracture of Diamond
  (111) Surfaces'', {\em Surf. Sci.}  {\bf 271}, 57--67.

\item[]
Harrison, J.~A., White, C.~T., Colton, R.~J., and Brenner, D.~W. (1992b),
  ``Molecular-Dynamic Simulations of Atomic-Scale Friction of Diamond
  Surfaces'', {\em Phys. Rev. B}  {\bf 46}, 9700--9708.

\item[]
Harrison, J.~A., White, C.~T., Colton, R.~J., and Brenner, D.~W. (1993),
  ``Effects of Chemically-Bound, Flexible Hydrocarbon Species on the Frictional
  Properties of Diamond Surfaces'', {\em J. Phys. Chem.}  {\bf 97}, 6573--6576.

\item[]
Hayward, I.~P. (1991),  ``Friction and Wear Properties of Diamonds and Diamond
  Coatings'', {\em Surf. Coat. Tech.}  {\bf 49}, 554--559.

\item[]
He, G., M\"user, M.~H., and Robbins, M.~O. (1999),  ``Adsorbed Layers and the
  Origin of Static Friction'', {\em Science}  {\bf 284}, 1650--1652.

\item[]
He, G. and Robbins, M.~O., unpublished.

\item[]
Heinbuch, U. and Fischer, J. (1989),  ``Liquid flow in pores: Slip, no-slip, or
  multilayer sticking'', {\em Phys. Rev. A}  {\bf 40}, 1144--1146.

\item[]
Heslot, F., Baumberger, T., Perrin, B., Caroli, B., and Caroli, C. (1994),
  ``Creep, stick-slip, and dry-friction dynamics: Experiments and a heuristic
  model'', {\em Phys. Rev. E}  {\bf 49}, 4973--4988.

\item[]
Hirano, M. and Shinjo, K. (1990),  ``Atomistic locking and friction'', {\em
  Phys. Rev. B}  {\bf 41}, 11837--11851.

\item[]
Hirano, M. and Shinjo, K. (1993),  ``Superlubricity and frictional
  anisotropy'', {\em Wear}  {\bf 168}, 121--125.

\item[]
Hirano, M., Shinjo, K., Kaneko, R., and Murata, Y. (1991),  ``Anisotropy of
  Frictional Forces in Muscovite Mica'', {\em Phys. Rev. Lett.}  {\bf 67},
  2642--2645.

\item[]
Hirano, M., Shinjo, K., Kaneko, R., and Murata, Y. (1997),  ``Observation of
  superlubricity by scanning tunneling microscopy'', {\em Phys. Rev. Lett.}
  {\bf 78}, 1448--1451.

\item[]
H\"olscher, H., Schwarz, U.~D., and Wiesendanger, R. (1997),  ``Simulation of
  the scan process in friction force microscopy'', in {\em Materials Research
  Society Symposia Proceedings} (Bhushan, B., ed.), pp. 379--384, Kluwer
  Academic Publishers, Netherlands.

\item[]
Horn, R.~G. and Israelachvili, J.~N. (1981),  ``Direct measurement of
  structural forces between two surfaces in a nonpolar liquid'', {\em J. Chem.
  Phys.}  {\bf 75}, 1400--1412.

\item[]
Hu, H.-W., Carson, G.~A., and Granick, S. (1991),  ``Relaxation Time of
  Confined Liquids under Shear'', {\em Phys. Rev. Lett.}  {\bf 66}, 2758--2761.

\item[]
Hu, Y.-Z., Wang, H., Guo, Y., Shen, Z.-J., and Zheng, L.-Q. (1996),
  ``Simulations of Lubricant Rheology in Thin Film Lubrication Part II:
  Simulation of Couette Flow'', {\em Wear}  {\bf 196}, 249--253.

\item[]
Huh, C. and Scriven, L.~E. (1971),  ``Hydrodynamic Model of Steady Movement of
  a Solid/Liquid/Fluid Contact Line'', {\em J. Colloid. Interface Sci.}  {\bf
  35}, 85--101.

\item[]
Israelachvili, J.~N. (1986),  ``Measurement of the Viscosity of Liquids in Very
  Thin Films'', {\em J. Colloid Interface Sci.}  {\bf 110}, 263--271.

\item[]
Israelachvili, J.~N. (1991), {\em Intermolecular and Surface Forces, 2nd ed.},
  Academic Press, London.

\item[]
Jacobsen, K.~W., Norskov, J.~K., and Puska, M.~J. (1987),  ``Interatomic
  Interactions in the effective-medium theory'', {\em Phys. Rev. B}  {\bf 35},
  7423--7442.

\item[]
Jaeger, H.~M., Nagel, S.~R., and Behringer, R.~P. (1996),  ``Granular solids,
  liquids, and gases'', {\em Rev. Mod. Phys.}  {\bf 68}, 1259--1273.

\item[]
Joanny, J.~F. and Robbins, M.~O. (1990),  ``Motion of a contact line on a
  heterogeneous surface'', {\em J. Chem. Phys.}  {\bf 92}, 3206--3212.

\item[]
Kawaguchi, T. and Matsukawa, H. (1998),  ``Anomalous pinning behavior in an
  incommensurate two-chain model of friction'', {\em Phys. Rev. B}  {\bf 58},
  15866--15877.

\item[]
Khare, R., de~Pablo, J.~J., and Yethiraj, A. (1996),  ``Rheology of Confined
  Polymer Melts'', {\em Macromolecules}  {\bf 29}, 7910--7918.

\item[]
Klein, J. and Kumacheva, E. (1995),  ``Confinement-Induced Phase Transitions in
  Simple Liquids'', {\em Science}  {\bf 269}, 816--819.

\item[]
Koike, A. (1999),  ``Molecular Dynamics Study of Tribological Behavior of
  Confined Branched and Linear Perfluoropolyethers'', {\em J. Phys. Chem. B}
  {\bf 103}, 4578--4589.

\item[]
Koike, A. and Yoneya, M. (1998),  ``Chain Length Effects on Frictional Behavior
  of Confined Ultrathin Films of Linear Alkanes Under Shear'', {\em J. Phys.
  Chem. B}  {\bf 102}, 3669--3675.

\item[]
Koplik, J., Banavar, J.~R., and Willemsen, J.~F. (1988),  ``Molecular Dynamics
  of Poiseuille Flow and Moving Contact Lines'', {\em Phys. Rev. Lett.}  {\bf
  60}, 1282--1285.

\item[]
Koplik, J., Banavar, J.~R., and Willemsen, J.~F. (1989),  ``Molecular dynamics
  of fluid flow at solid surfaces'', {\em Phys. Fluids A}  {\bf 1}, 781--794.

\item[]
Kremer, K. and Grest, G.~S. (1990),  ``Dynamics of Entangled Linear Polymer
  Melts: A Molecular-Dynamics Simulation'', {\em J. Chem. Phys.}  {\bf 92},
  5057--5086.

\item[]
Krim, J., Solina, D.~H., and Chiarello, R. (1991),  ``Nanotribology of a Kr
  Monolayer: A Quartz-Crystal Microbalance Study of Atomic-Scale Friction'',
  {\em Phys. Rev. Lett.}  {\bf 66}, 181--184.

\item[]
Krim, J., Watts, E.~T., and Digel, J. (1990),  ``Slippage of simple liquid
  films adsorbed on silver and gold substrates'', {\em J. Vac. Sci. Technol. A}
   {\bf 8}, 3417--3420.

\item[]
Krim, J. and Widom, A. (1988),  ``Damping of a crystal oscillator by an
  adsorbed monolayer and its relation to interfacial viscosity'', {\em Phys.
  Rev. B}  {\bf 38}, 12184--12189.

\item[]
Kr\"oger, M., Loose, W., and Hess, S. (1993),  ``Rheology and
  Structural-Changes of Polymer Melts via Nonequilibrium Molecular Dynamics'',
  {\em J. Rheology}  {\bf 37}, 1057--1079.

\item[]
Landman, U. and Luedtke, W.~D. (1989),  ``Dynamics of Tip-Substrate
  Interactions in Atomic Force Microscopy'', {\em Surf. Sci. Lett.}  {\bf 210},
  L117--L184.

\item[]
Landman, U. and Luedtke, W.~D. (1991),  ``Nanomechanics and Dynamics of
  Tip-Substrate Interactions'', {\em J. Vac. Sci. Technol. B}  {\bf 9},
  414--423.

\item[]
Landman, U., Luedtke, W.~D., Burnham, N.~A., and Colton, R.~J. (1990),
  ``Atomistic Mechanisms and Dynamics of Adhesion, Nanoindentation, and
  Fracture'', {\em Science}  {\bf 248}, 454--461.

\item[]
Landman, U., Luedtke, W.~D., and Gao, J. (1996),  ``Atomic-Scale Issues in
  Tribology: Interfacial Junctions and Nano-elastohydrodynamics'', {\em
  Langmuir}  {\bf 12}, 4514--4528.

\item[]
Landman, U., Luedtke, W.~D., and Ribarsky, M.~W. (1989),  ``Structural and
  dynamical consequences of interactions in interfacial systems'', {\em J. Vac.
  Sci. Technol. A}  {\bf 7}, 2829--2839.

\item[]
Landman, U., Luedtke, W.~D., and Ringer, E.~M. (1992),  ``Atomistic mechanisms
  of adhesive contact formation and interfacial processes'', {\em Wear}  {\bf
  153}, 3--30.

\item[]
Liebsch, A., Gon\c{c}alves, S., and Kiwi, M. (1999),  ``Electronic versus
  Phononic Friction of Xenon on Silver'', {\em Phys. Rev. B}  {\bf 60},
  5034--5043.

\item[]
Loose, W. and Ciccotti, G. (1992),  ``Temperature and temperature control in
  nonequilibrium-molecular-dynamics simulations of the shear flow of dense
  liquids'', {\em Phys. Rev. A}  {\bf 45}, 3859--3866.

\item[]
Lupowski, M. and van Swol, F. (1991),  ``Ultrathin films under shear'', {\em J.
  Chem. Phys.}  {\bf 95}, 1995--1998.

\item[]
Magda, J., Tirrell, M., and Davis, H.~T. (1985),  ``Molecular Dynamics of
  Narrow, Liquid-Filled Pores'', {\em J. Chem. Phys.}  {\bf 83}, 1888--1901.

\item[]
Mak, C. and Krim, J. (1998),  ``Quartz-crystal microbalance studies of the
  velocity dependence of interfacial friction'', {\em Phys. Rev. B}  {\bf 58},
  5157--5159.

\item[]
Manias, E., Bitsanis, I., Hadziioannou, G., and Brinke, G.~T. (1996),  ``On the
  nature of shear thinning in nanoscopically confined films'', {\em Europhys.
  Lett.}  {\bf 33}, 371--376.

\item[]
Matsukawa, H. and Fukuyama, H. (1994),  ``Theoretical study of friction:
  One-dimensional clean surfaces'', {\em Phys. Rev. B}  {\bf 49}, 17286--17292.

\item[]
Maxwell, J.~C. (1867), {\em Philos. Trans. R. Soc. London Ser. A}  {\bf 170},
  231.

\item[]
McClelland, G.~M. (1989),  ``Friction at Weakly Interacting Interfaces'', in
  {\em Adhesion and Friction} (Grunze, M. and Kreuzer, H.~J., eds.), volume~17,
  pp. 1--16, Springer Verlag, Berlin.

\item[]
McClelland, G.~M. and Cohen, S.~R. (1990),  ``Tribology at the Atomic Scale'',
  in {\em Chemistry \& Physics of Solid Surfaces VII} (Vanselow, R. and Rowe,
  R., eds.), pp. 419--445, Springer Verlag, Berlin.

\item[]
McClelland, G.~M. and Glosli, J.~N. (1992),  ``Friction at the Atomic Scale'',
  in {\em Fundamentals of Friction: Macroscopic and Microscopic Processes}
  (Singer, I.~L. and Pollock, H.~M., eds.), pp. 405--422, Kluwer, Dordrecht.

\item[]
McCumber, D.~E. (1968),  ``Effect of ac Impedance on the dc Voltage-Current
Characteristics of Superconductor Weak-Link Junctions'', {\em J. App. Phys.}
{\bf 39}(7), 3113--3118.

\item[]
Mundy, C.~J., Balasubramanian, S., and Klein, M.~L. (1996),  ``Hydrodynamic
  boundary conditions for confined fluids via a nonequilibrium molecular
  dynamics simulation'', {\em J. Chem. Phys.}  {\bf 105}, 3211--3214.

\item[]
M\"user, M.~H. and Robbins, M.~O., to be published.

\item[]
M\"user, M.~H. and Robbins, M.~O. (1999),  ``Condition for static friction
  between flat crystalline surfaces'', {\em Phys. Rev. B}  {\bf 60}.

\item[]
Nasuno, S., Kudrolli, A., and Gollub, J.~P. (1997),  ``Friction in Granular
  Layers: Hysteresis and Precursors'', {\em Phys. Rev. Lett.}  {\bf 79},
  949--952.

\item[]
Nieminen, J.~A., Sutton, A.~P., and Pethica, J.~B. (1992),  ``Static Junction
  Growth During Frictional Sliding of Metals'', {\em Acta Metall. Mater.}  {\bf
  40}, 2503--2509.

\item[]
Nordholm, S. and Haymet, A. D.~J. (1980),  ``Generalized van der Waals Theory.
  I Basic Formulation and Application to Uniform Fluids'', {\em Aust. J. Chem.}
   {\bf 33}, 2013--2027.

\item[]
Nos\'e, S. (1991),  ``Constant Temperature Molecular Dynamics Methods'', {\em
  Prog. Theor. Phys. Supp.}  {\bf 103}, 1--46.

\item[]
Ohzono, T., Glosli, J.~N., and Fujihara, M. (1998),  ``Simulations of wearless
  friction at a sliding interface between ordered organic monolayers'', {\em
  Jpn. J. Appl. Physics}  {\bf 37}, 6535--6543.

\item[]
Paliy, M., Braun, O.~M., Dauxois, T., and Hu, B. (1997),  ``Dynamical phase
  diagram of the dc-driven underdamped Frenkel-Kontorova Chain'', {\em Phys.
  Rev. E}  {\bf 56}, 4025--4030.

\item[]
Paul, W., Yoon, D.~Y., and Smith, G.~D. (1995),  ``An optimized united atom
  model for simulations of polymethylene melts'', {\em J. Chem. Phys}  {\bf
  103}, 1702--1709.

\item[]
Perry, M.~D. and Harrison, J.~A. (1996),  ``Molecular Dynamics Investigations
  of the Effects of Debris Molecules on the Friction and Wear of Diamond'',
  {\em Thin Solid Films}  {\bf 290-291}, 211--215.

\item[]
Perry, M.~D. and Harrison, J.~A. (1997),  ``Friction between Diamond Surfaces
  in the Presence of Small Third-Body Molecules'', {\em J. Phys. Chem. B}  {\bf
  101}, 1364--1373.

\item[]
Persson, B. N.~J. (1991),  ``Surface resistivity and vibrational damping in
  adsorbed layers'', {\em Phys. Rev. B}  {\bf 44}, 3277--3296.

\item[]
Persson, B. N.~J. (1993a),  ``Theory of friction and boundary lubrication'',
  {\em Phys. Rev. B}  {\bf 48}, 18140--18158.

\item[]
Persson, B. N.~J. (1993b),  ``Theory and Simulation of Sliding Friction'', {\em
  Phys. Rev. Lett.}  {\bf 71}, 1212--1215.

\item[]
Persson, B. N.~J. (1995),  ``Theory of Friction: Dynamical Phase Transitions in
  Adsorbed Layers'', {\em J. Chem. Phys.}  {\bf 103}, 3849--3860.

\item[]
Persson, B. N.~J. (1998), {\em Sliding Friction: Physical Principles and
  Applications}, Springer, Berlin.

\item[]
Persson, B. N.~J. and Ballone, P., unpublished.

\item[]
Persson, B. N.~J. and Nitzan, A. (1996),  ``Linear sliding friction: On the
  origin of the microscopic friction for Xe on silver'', {\em Surf. Sci.}  {\bf
  367}, 261--275.

\item[]
Persson, B. N.~J. and Tosatti, E. (1996),  ``Theory of Friction: Elastic
  Coherence Length and Earthquake Dynamics'', in {\em Physics of Sliding
  Friction} (Persson, B. N.~J. and Tosatti, E., eds.), pp. 179--189, Kluwer,
  Dordrecht.

\item[]
Persson, B. N.~J. and Volokitin, A.~I. (1995),  ``Electronic friction of
  physisorbed molecules'', {\em J. Chem. Phys.}  {\bf 103}, 8679--8683.

\item[]
Plischke, M. and Henderson, D. (1986),  ``Density Profiles and Pair Correlation
  - Functions of Lennard-Jones Fluids Near a Hard-Wall'', {\em J. Chem. Phys.}
  {\bf 84}, 2846--2852.

\item[]
Rabin, Y. and Hersht, I. (1993),  ``Thin Liquid Layers in Shear-Non-Newtonian
  Effects'', {\em Physica A}  {\bf 200}, 708--712.

\item[]
Rabinowicz, E. (1965), {\em Friction and Wear of Materials}, Wiley, New York.

\item[]
Raffi-Tabar, H., Pethica, J.~B., and Sutton, A.~P. (1992),  ``Influence of
  Adsorbate Monolayer on the Nano-Mechanics of Tip-Substrate Interactions'',
  {\em Mater. Res. Soc. Symp. Proc.}  {\bf 239}, 313--318.

\item[]
Rajasekaran, E., Zeng, X.~C., and Diestler, D.~J. (1997),  ``Frictional
  anisotropy and the role of lattice relaxation in molecular tribology of
  crystalline interfaces'', in {\em Materials Research Society Symposia
  Proccedings} (Bhushan, B., ed.), pp. 379--384, Kluwer, Netherlands.

\item[]
Raphael, E. and deGennes, P.~G. (1989),  ``Dynamics of Wetting with Non-ideal
  Surfaces - The Single Defect Problem'', {\em J. Chem. Phys.}  {\bf 90},
  7577--7584.

\item[]
Ribarsky, M.~W. and Landman, U. (1992),  ``Structure and dynamics of
  normal-alkanes confined by Solid-Surfaces. 1. Stationary Crystalline
  Boundaries'', {\em J. Chem. Phys.}  {\bf 97}, 1937--1949.

\item[]
Rigney, D.~A. and Hammerberg, J.~E. (1998),  ``Unlubricated Sliding Behavior of
  Metals'', {\em MRS Bull.}  {\bf 23}(6), 32--36.

\item[]
Robbins, M.~O. and Baljon, A. R.~C. (2000),  ``Response of Thin Oligomer Films
  to Steady and Transient Shear'', in {\em Microstructure and Microtribology of
  Polymer Surfaces} (Tsukruk, V.~V. and Wahl, K.~J., eds.), pp. 91--117.
  American Chemical Society, Washington DC.

\item[]
Robbins, M.~O. and Krim, J. (1998),  ``Energy Dissipation in Interfacial
  Friction'', {\em MRS Bull.}  {\bf 23}(6), 23--26.

\item[]
Robbins, M.~O. and Mountain, R.~D., unpublished.

\item[]
Robbins, M.~O. and Smith, E.~D. (1996),  ``Connecting Molecular-Scale and
  Macroscopic Tribology'', {\em Langmuir}  {\bf 12}, 4543--4547.

\item[]
Robbins, M.~O. and Thompson, P.~A. (1991),  ``Critical Velocity of Stick-Slip
  Motion'', {\em Science}  {\bf 253}, 916.

\item[]
Robbins, M.~O. (2000) ``Jamming, Friction, and Unsteady Rheology'', in
{\em Jamming and Rheology: Constrained Dynamics on
Microscopic and Macroscopic Scales},
(Liu, A. J. and Nagel, S. R., eds.) Taylor and Francis, London.

\item[]
Rozman, M.~G., Urbakh, M., and Klafter, J. (1996),  ``Stick-Slip Motion and
  Force Fluctuations in a Driven Two-Wave Potential'', {\em Phys. Rev. Lett.}
  {\bf 77}, 683--686.

\item[]
Rozman, M.~G., Urbakh, M., and Klafter, J. (1997),  ``Stick-slip dynamics as a
  probe of frictional forces'', {\em Europhys. Lett.}  {\bf 39}, 183--188.

\item[]
Rozman, M.~G., Urbakh, M., Klafter, J., and Elmer, F.-J. (1998),  ``Atomic
  Scale Friction and Different Phases of Motion of Embedded Molecular
  Systems'', {\em J. Phys. Chem. B}  {\bf 102}, 7924--7930.

\item[]
Ruina, A. (1983),  ``Slip Instability and State Variable Friction Laws'', {\em
  J. Geophys. Res.}  {\bf 88}, 10359--10370.

\item[]
Ryckaert, J.~P. and Bellemans, A. (1978),  ``Molecular Dynamics of Liquid
  Alkanes'', {\em Discuss. Faraday Soc.}  {\bf 66}, 95--106.

\item[]
Sarman, S.~S., Evans, D.~J., and Cummings, P.~T. (1998),  ``Recent developments
  in non-Newtonian molecular dynamics'', {\em Phys. Rep.}  {\bf 305}, 1--92.

\item[]
Schaich, W.~L. and Harris, J. (1981),  ``Dynamic Corrections to van der Waals
  Potentials'', {\em J. Phys. F: Met. Phys.}  {\bf 11}, 65--78.

\item[]
Schneider, T. and Stoll, E. (1978),  ``Molecular Dynamics Study of a
  Three-Dimensional One-Component Model for Distortive Phase Transitions'',
  {\em Phys. Rev. B}  {\bf 17}, 1302--1322.

\item[]
Schoen, M., Cushman, J.~H., Diestler, D.~J., and Rhykerd, C.~L. (1988),
  ``Fluids in Micropores II: Self-Diffusion in a Simple Classical Fluid in a
  Slit-Pore'', {\em J. Chem. Phys.}  {\bf 88}, 1394--1406.

\item[]
Schoen, M., Rhykerd, C.~L., Diestler, D.~J., and Cushman, J.~H. (1987),
  ``Fluids in Micropores. I. Structure of a Simple Classical Fluid in a
  Slit-Pore'', {\em J. Chem. Phys.}  {\bf 87}, 5464--5476.

\item[]
Schoen, M., Rhykerd, C.~L., Diestler, D.~J., and Cushman, J.~H. (1989),
  ``Shear Forces in Molecularly Thin Films'', {\em Science}  {\bf 245},
  1223--1225.

\item[]
Shinjo, K. and Hirano, M. (1993),  ``Dynamics of Friction: Superlubric State'',
  {\em Surface Science}  {\bf 283}, 473--478.

\item[]
Smith, E.~D., Cieplak, M., and Robbins, M.~O. (1996),  ``The Friction on
  Adsorbed Monolayers'', {\em Phys. Rev. B.}  {\bf 54}, 8252--8260.

\item[]
Sneddon, L., Cross, M.~C., and Fisher, D.~S. (1982),  ``Sliding Conductivity of
  Charge-Density Waves'', {\em Phys. Rev. Lett.}  {\bf 49}, 292--295.

\item[]
Snook, I.~K. and van Megen, W. (1980),  ``Solvation in simple dense fluids.
  I'', {\em J. Chem. Phys.}  {\bf 72}, 2907--2914.

\item[]
Sokoloff, J.~B. (1990),  ``Theory of energy dissipation in sliding crystal
  surfaces'', {\em Phys. Rev. B}  {\bf 42}, 760--765.

\item[]
S$\o$rensen, M.~R., Jacobsen, K.~W., and Stoltze, P. (1996),  ``Simulations of
  atomic-scale sliding friction'', {\em Phys Rev. B}  {\bf 53}, 2101--2113.

\item[]
Steele, W.~A. (1973),  ``The physical interaction of gases with crystalline
  solids. I. Gas-solid energies and properties of isolated absorbed atoms'',
  {\em Surf. Sci.}  {\bf 36}, 317--352.

\item[]
Stevens, M.~J., Mondello, M., Grest, G.~S., Cui, S.~T., Cochran, H.~D., and
  Cummings, P.~T. (1997),  ``Comparison of shear flow of hexadecane in a
  confined geometry and in bulk'', {\em J. Chem. Phys.}  {\bf 106}, 7303--7314.

\item[]
Stevens, M.~J. and Robbins, M.~O. (1993),  ``Simulations of shear-induced
  melting and ordering'', {\em Phys. Rev. E}  {\bf 48}, 3778--3792.

\item[]
Stevens, M.~J., Robbins, M.~O., and Belak, J.~F. (1991),  ``Shear-Melting of
  Colloids: A Non-Equilibrium Phase Diagram'', {\em Phys. Rev. Lett.}  {\bf
  66}, 3004--3007.

\item[]
Suh, N.~P. (1986), {\em Tribophysics}, Prentice-Hall, Englewood Cliffs.

\item[]
Taub, H., Torzo, G., Lauter, H.~J., and S.~C.~Fain, J. (1991), {\em Phase
  Transitions in Surface Films 2}, Plenum Press, New York.

\item[]
Thompson, P.~A. and Grest, G.~S. (1991),  ``Granular Flow: Friction and the
  Dilatancy Transition'', {\em Phys. Rev. Lett.}  {\bf 67}, 1751--1754.

\item[]
Thompson, P.~A., Grest, G.~S., and Robbins, M.~O. (1992),  ``Phase Transitions
  and Universal Dynamics in Confined Films'', {\em Phys. Rev. Lett.}  {\bf 68},
  3448--3451.

\item[]
Thompson, P.~A. and Robbins, M.~O. (1989),  ``Simulations of Contact-Line
  Motion: Slip and the Dynamic Contact Angle'', {\em Phys. Rev. Lett.}  {\bf
  63}, 766--769.

\item[]
Thompson, P.~A. and Robbins, M.~O. (1990a),  ``Shear flow near solids:
  Epitaxial order and flow boundary conditions'', {\em Phys. Rev. A}  {\bf 41},
  6830--6837.

\item[]
Thompson, P.~A. and Robbins, M.~O. (1990b),  ``Origin of Stick-Slip Motion in
  Boundary Lubrication'', {\em Science}  {\bf 250}, 792--794.

\item[]
Thompson, P.~A., Robbins, M.~O., and Grest, G.~S. (1993),  ``Simulations of
  Lubricant Behavior at the Interface with Bearing Solids'', in {\em Thin Films
  in Tribology} (Dowson, D., Taylor, C.~M., Childs, T. H.~C., Godet, M., and
  Dalmaz, G., eds.), pp. 347--360, Elsevier, Amsterdam.

\item[]
Thompson, P.~A., Robbins, M.~O., and Grest, G.~S. (1995),  ``Structure and
  Shear Response in Nanometer-Thick Films'', {\em Israel J. of Chem.}  {\bf
  35}, 93--106.

\item[]
Thompson, P.~A. and Troian, S.~M. (1997),  ``A general boundary condition for
  liquid flow at solid surfaces'', {\em Nature}  {\bf 389}, 360--363.

\item[]
Tomagnini, O., Ercolessi, F., and Tosatti, E. (1993),  ``Microscopic
  Interaction between a Gold Tip and a Pb(110) Surface'', {\em Surf. Sci.}
  {\bf 287/288}, 1041--1045.

\item[]
Tomassone, M.~S., Sokoloff, J.~B., Widom, A., and Krim, J. (1997),  ``Dominance
  of Phonon Friction for a Xenon Film on a Silver (111) Surface'', {\em Phys.
  Rev. Lett.}  {\bf 79}, 4798--4801.

\item[]
Tomlinson, G.~A. (1929),  ``A molecular theory of friction'', {\em Phil. Mag.
  Series}  {\bf 7}, 905--939.

\item[]
Toxvaerd, S. (1981),  ``The structure and thermodynamics of a solid-fluid
  interface'', {\em J. Chem. Phys.}  {\bf 74}, 1998--2008.

\item[]
Tsch\"op, W., Kremer, K., Batoulis, J., B\"urger, T., and Hahn, O. (1998a),
  ``Simulation of polymer melts. I. Coarse graining procedure for
  polycarbonates'', {\em Acta Polym.}  {\bf 49}, 61--74.

\item[]
Tsch\"op, W., Kremer, K., Batoulis, J., B\"urger, T., and Hahn, O. (1998b),
  ``Simulation of polymer melts. II. From coarse grained models back to
  atomistic description'', {\em Acta Polym.}  {\bf 49}, 75--79.

\item[]
Tutein, A.~B., Stuart, S.~J., and Harrison, J.~A. (2000),  ``Indentation
  Analysis of Linear-Chain Hydrocarbon Monolayers Anchored to Diamond'', {\em
  J. Phys. Chem. B}  {\bf YY}, xx--xx.

\item[]
Urbakh, M., Daikhin, L., and Klafter, J. (1995),  ``Dynamics of Confined
  Liquids Under Shear'', {\em Phys. Rev. E}  {\bf 51}, 2137--2141.

\item[]
Veje, C.~T., Howell, D.~W., and Behringer, R.~P. (1999),  ``Kinematics of a
  two-dimensional granular Couette experiment at the transition to shearing'',
  {\em Phys. Rev. E}  {\bf 59}, 739--745.

\item[]
Volmer, A. and Natterman, T. (1997),  ``Towards a Statistical Theory of Solid
  Dry Friction'', {\em Z. Phys. B}  {\bf 104}, 363--371.

\item[]
Weiss, M. and Elmer, F.-J. (1996),  ``Dry Friction in the
  Frenkel-Kontorova-Tomlinson Model: Static Properties'', {\em Phys. Rev. B}
  {\bf 53}, 7539--7549.

\item[]
Xia, T.~K., Ouyang, J., Ribarsky, M.~W., and Landman, U. (1992),  ``Interfacial
  Alkane Films'', {\em Phys. Rev. Lett.}  {\bf 69}, 1967--1970.
\item[]
Yoshizawa, H. and Israelachvili, J.~N. (1993),  ``Fundamental Mechanisms of
  Interfacial Friction. 1. Stick-Slip Friction of Spherical and Chain
  Molecules'', {\em J. Phys. Chem.}  {\bf 97}, 11300--11313.
\end{description}
\end{document}